\documentclass[epj]{svjour} 
\usepackage{
  cite,
  graphicx, 
  dsfont,
  url
}

\title{Stochastic analysis of different rough surfaces}

\author{M.~Waechter\inst{1}\thanks{\email{matthias.waechter@uni-oldenburg.de}}
  \and F.~Riess\inst{1}
  \and Th.~Schimmel\inst{2}
  \and U.~Wendt\inst{3}
  \and J.~Peinke\inst{1}\thanks{\email{peinke@uni-oldenburg.de}}
}

\institute{
  Institute of Physics, Carl-von-Ossietzky University, 
  D-26111 Oldenburg, Germany
  \and
  Institute of Applied Physics,
  University of Karlsruhe,
  D-76128 Karlsruhe, Germany
  \and
  School of Materials Science,
  Otto-von-Guericke University,
  D-39016 Magdeburg, Germany
}

\date{\today}

\abstract{  
  This paper shows in detail the application of a new stochastic approach for
  the characterization of surface height profiles, which is based on the
  theory of Markov processes. 
  With this analysis we achieve a characterization of the scale dependent
  complexity of surface roughness by means of a Fokker-Planck or Langevin
  equation, providing the complete stochastic information of multiscale joint
  probabilities.
  The method is applied to several surfaces with different properties, for the
  purpose of showing the utility of this method in more detail. In particular
  we show evidence of the Markov properties, and we estimate the parameters of
  the Fokker-Planck equation by pure, parameter-free data analysis. The
  resulting Fokker-Planck equations are verified by numerical reconstruction
  of the conditional probability density functions.
  The results are compared with those from the analysis of multi-affine and
  extended multi-affine scaling properties which is often used for surface
  topographies. The different surface structures analysed here show in detail
  the advantages and disadvantages of these methods.
}

\newcommand{\un}[1]{\ensuremath{\,\mathrm{#1}}}
\newcommand{\mum}{\ensuremath{\,\mathrm{\mu m}}}
\newcommand{\Tik}{\rule[-3pt]{1pt}{4pt}}
\newcommand{\balken}[2]{%
  \raisebox{-2ex}{\makebox[0pt][l]{\hspace{0.5em}\footnotesize #2}}%
  \Tik\rule{#1}{1pt}\Tik%
}
\begin{document}
\maketitle

\section{Introduction}
\label{sec:intro}

Among the great variety of complex and disordered systems the complexity of
surface roughness is attracting a great deal of scientific interest
\cite{Sayles1978,Vicsek1992-e,Barabasi1995,Davies1999,Wendt2002a}. The physical
and chemical properties of surfaces and interfaces are to a significant degree
determined by their topographic structure. Thus a comprehensive
characterization of their topography is of vital interest from a scientific
point of view as well as for many applications
\cite{Dharmadhikari1999,Saitou2001,Sydow2003}.

Most popular methods used today for the characterization of surface roughness
are based on the concepts of self-affinity and multi-affinity, where the
multifractal $f(\alpha)$ spectrum has been regarded as the most complete
characterization of a surface
\cite{Feder1988,Family1991,Vicsek1992-e,Barabasi1995}. One example of a
measure of roughness which is commonly used in this context is the rms surface
width $w_r(x)=\langle (h(\tilde{x})-\bar{h})^2 \rangle^{1/2}_r$, where
$h(\tilde{x})$ is the measured height at point $\tilde{x}$,
$\langle\,\cdot\,\rangle_r$ denotes the average over an interval of length $r$
around the selected point $x$, and $\bar{h}$ the mean value of $h(\tilde{x})$
in that interval. Thus the roughness is measured at a specific location $x$
and over a specific scale $r$. Then a scaling regime of the ensemble average
$\langle w_r \rangle$ in $r$, if it exists, is analyzed according to $\langle
w_r^\alpha \rangle \sim r^{\xi_\alpha}$, usually $\alpha\in\mathds{Q}$. Here,
$\langle\,\cdot\,\rangle$ denotes the mean over the available range in $x$.
For a more thorough introduction into scaling concepts we refer the reader to
the literature, e.g.\ \cite{Feder1988,Family1991,Vicsek1992-e,Barabasi1995}.
Terms concerning scaling concepts which are used in this paper are rapidly
introduced in section~\ref{sec:scaling}. Here, we have to note the following
points which concern stochastic aspects of roughness analysis: First, the
ensemble average $\langle w_r \rangle$ must obey a scaling law as mentioned
above, and second, the statistics of $w_r(x)$ are investigated over distinct
length scales $r$, thus possible correlations between $w_r(x)$ and $w_{r'}(x)$
on different scales $r,r'$ are not examined.

In this paper we want to give a deeper introduction into a new approach to
surface roughness analysis which has recently been introduced by us
\cite{Friedrich1998a,Waechter2003_plus_preprint} and by \cite{Jafari2003}.
This method is based on stochastic processes which should grasp the scale
dependency of surface roughness in a most general way. No scaling feature is
explicitly required, and especially the correlations between different scales
$r$ and $r'$ are investigated. To this end we present a systematic procedure
as to how the explicit form of a stochastic process for the $r$-evolution of a
roughness measure similar to $w_r(x)$ can be extracted directly from the
measured surface topography. This stochastic approach turns out to be a
promising tool also for other systems with scale dependent complexity such as
turbulence \cite{Friedrich1998a,Renner2001,Lueck1999}, financial data
\cite{Friedrich2000a,Ausloos2003}, and cosmic background radiation
\cite{Ghasemi2003}. Also this stochastic approach has recently
enabled the numerical reconstruction of surface topographies
\cite{Jafari2003}.

Here we demonstrate our ansatz by analysing a number of data sets from
different surfaces.  The purpose is to show extensively the utility of this
method for a wide class of rough surfaces. The examples show different kinds of
scaling properties which are, in addition, briefly analysed.  Among these
examples is a collection of road surfaces measured with a specially designed
profile scanner. Preliminary results of the analysis of one of these surfaces
have already been presented \cite{Waechter2003_plus_preprint}. AFM measurement
data from an evaporated gold film have already been analysed in an earlier
stage of the method \cite{Friedrich1998a}. Since then, the method has been
significantly refined and extended. Measurements of a steel crack surface were
taken by confocal laser scanning microscopy (CLSM) \cite{Wendt2002a}.

As a measure of surface roughness we use the \emph{height increment}
\cite{CenteredInc}
\begin{equation}
  \label{eq:h_r}
  h_r(x) := h(x+r/2) - h(x-r/2)
\end{equation}
depending on the length scale $r$. For other scale dependent roughness
measures, see \cite{Family1991,Barabasi1995}.  The height increment $h_r$ is
used because its moments, which are well-known as structure functions (see
section~\ref{sec:scaling}), are closely connected with spatial correlation
functions.  Nevertheless, it should be pointed out that our method presented in
the following could be easily generalized to any scale dependent measure,
\emph{e.g.} the above-mentioned $w_r(x)$ or wavelet functions \cite{Haase2003}.
As a new ansatz, $h_r$ is regarded as a \emph{stochastic variable in $r$}.
Without loss of generality we consider the process as being directed from
larger to smaller scales.  The focus of our method is the investigation of how
the surface roughness is linked between different length scales.

In the remainder of this paper we will first summarize in
section~\ref{sec:Markov_theory} some central aspects of the theory of Markov
processes which form the basis of our analysis procedure. Details concerning
the measurement data are presented in section~\ref{sec:measurement_data},
their scaling properties are analyzed in section~\ref{sec:scaling}. The Markov
properties of our examples are investigated in section \ref{sec:markov_props}.
In section~\ref{sec:Dk_estimation} we estimate for each data set the
parameters of a Fokker-Planck equation. The ability of this equation to
describe the statistics of $h_r$ in the scale variable $r$ is then examined
in section~\ref{sec:veri_coeff}, followed by concluding remarks in
section~\ref{sec:conclusions}.

\section{Surface roughness as a Markov process}
\label{sec:Markov_theory}

Complete information about the stochastic process would be available from the
knowledge of all possible $n$-point, or more precisely $n$-scale,
joint probability density functions (pdf) %
$p(h_1, r_1; h_2, r_2; \ldots ; h_n, r_n)$ %
describing the probability of finding simultaneously the increments $h_1$ on
the scale $r_1$, $h_2$ on the scale $r_2$, and so forth up to $h_n$ on the
scale $r_n$. Here we use the notation $h_i(x)=h_{r_i}(x)$, see
eq.~(\ref{eq:h_r}). Without loss of generality we take $r_1<r_2<\ldots<r_n$.
As a first question one has to ask for a suitable simplification. In any case
the $n$-scale joint pdf can be expressed by multiconditional pdf
\begin{eqnarray} \label{eq:condpdf}
  \lefteqn{p(h_1,r_1;\ldots;h_n,r_n) = }\nonumber\\
  &&p(h_1,r_1 | h_2,r_2;\ldots;h_n,r_n)\cdot p(h_2,r_2|h_3,r_3;\ldots;h_n,r_n)
  \nonumber\\
  &&{}\cdot\ldots\cdot p(h_{n-1},r_{n-1}|h_n,r_n)\cdot p(h_n,r_n)\,.
\end{eqnarray}
Here, $p(h_i,r_i| h_j,r_j)$ denotes a \emph{conditional probability} of
finding the increment $h_i$ on the scale $r_i$ under the condition that
simultaneously, \emph{i.e.}\ at the same location $x$, on a larger scale $r_j$
the value $h_j$ was found. 
It is defined with the help of the joint probability $p(h_i,r_i; h_j,r_j)$ by
\begin{equation}
  \label{eq:condpdf_def}
  p(h_i,r_i| h_j,r_j) = \frac{p(h_i,r_i; h_j,r_j)}{p(h_j,r_j)}\;.
\end{equation}

An important simplification arises if
\begin{equation}
   \label{eq:markov_straight}
   p(h_i, r_i | h_{i+1}, r_{i+1}; \ldots; h_n, r_n) =
   p(h_i, r_i | h_{i+1}, r_{i+1})\;.
\end{equation}
This property is the defining feature of a Markov process evolving from
$r_{i+1}$ to $r_i$. Thus for a Markov process the $n$-scale joint pdf
factorize into $n$ conditional pdf
\begin{eqnarray}
   \label{eq:markov1}
       \lefteqn{p(h_1,r_1;\ldots;h_n,r_n) = p(h_1,r_1\, | h_2,r_2)}\nonumber\\
       &&{}\cdot \ldots \cdot p(h_{n-1},r_{n-1}\, | h_n,r_n)
           \cdot p(h_n,r_n) \;.
\end{eqnarray}
The Markov property implies that the $r$-dependence of $h_r$ can be regarded
as a stochastic process evolving in $r$, driven by deterministic and random
forces. 
Here it should be noted that if condition (\ref{eq:markov_straight}) holds
this is true for a process evolving in $r$ from large down to small scales as
well as the reverse from small to large scales \cite{Renner2001Diss}.
Equation (\ref{eq:markov1}) also emphasizes the fundamental meaning of
conditional probabilities for Markov processes since they determine any
$n$-scale joint pdf and thus the complete statistics of the process.

For any Markov process a Kramers-Moyal expansion of the governing master
equation exists \cite{Risken1984}. For our height profiles it takes the form
\begin{eqnarray}
   \label{eq:KME}
     \lefteqn{-r\,\frac{\partial}{\partial r}\;p(h_r,r|h_0,r_0)\,=}\nonumber\\
     &&\sum_{k=1}^{\infty}
     \left( -\frac{\partial}{\partial h_r} \right)^k
        D^{(k)}(h_r,r)\, p(h_r,r|h_0,r_0) \,.
\end{eqnarray}
The minus sign on the left side of eq.~(\ref{eq:KME}) expresses the direction
of the process from larger to smaller scales, furthermore the factor $r$
corresponds to a logarithmic variable $\rho=\ln r$ which leads to simpler
results in the case of the scaling behaviour \cite{FPE}. 
To derive the Kramers-Moyal coefficients $D^{(k)}(h_r,r)$, the limit $\Delta r
\rightarrow 0$ of the conditional moments has to be performed:
\begin{equation}
  \label{eq:Dk_def}
    D^{(k)}(h_r,r) = \lim_{\Delta r \rightarrow 0} M^{(k)}(h_r,r,\Delta r)\,, 
    \qquad\mbox{where}
\end{equation}
\begin{eqnarray}
   \label{eq:Mk_def}
   \lefteqn{M^{(k)}(h_r,r,\Delta r) =}\nonumber\\
     &&\frac{r}{k!\Delta r}
     \int_{\scriptscriptstyle-\infty}^{\scriptscriptstyle+\infty}
     (\tilde{h}-h_r)^k \, p(\tilde{h},r-\Delta r | h_r,r) \,  d\tilde{h} \,.
\end{eqnarray}
The moments $M^{(k)}(h_r,r,\Delta r)$ characterize the alteration of the
conditional probability $p(h_r,r|h_0,r_0)$ over a finite step size 
$\Delta r=r_0-r$ and are thus also called ``transitional moments''.

A second major simplification is valid if the noise included in the process is
Gaussian distributed. In this case the coefficient $D^{(4)}$ vanishes (from
eqs.~(\ref{eq:Dk_def}) and (\ref{eq:Mk_def}) it can be seen that $D^{(4)}$ is
a measure of non-gaussianity of the included noise).  According to Pawula's
theorem, together with $D^{(4)}$ all the $D^{(k)}$ with $k\geq 3$ disappear
and the Kramers-Moyal expansion (\ref{eq:KME}) collapses to a Fokker-Planck
equation \cite{Risken1984}, also known as Kolmogorov equation
\cite{Kolmogorov1931}:
\begin{eqnarray}
   \label{eq:FPE1}
     \lefteqn{-r\,\frac{\partial}{\partial r}\;p(h_r,r|h_0,r_0)\,=}\\
     &&\left\{ -\frac{\partial}{\partial h_r} D^{(1)}(h_r,r)
       + \frac{\partial^2}{\partial h_r^2} D^{(2)}(h_r,r)
     \right\} p(h_r,r|h_0,r_0) 
     \nonumber
\end{eqnarray}
The Fokker-Planck equation then describes the evolution of the conditional
probability density function from larger to smaller length scales and thus also
the complete $n$-scale statistics. The term $D^{(1)}(h_r,r)$ is commonly
denoted as the drift term, describing the deterministic part of the process,
while $D^{(2)}(h_r,r)$ is designated as the diffusion term, determined by the
variance of a Gaussian, $\delta$-correlated noise (compare also
eqs.~(\ref{eq:Dk_def}) and (\ref{eq:Mk_def})).

By integrating over $h_0$ it can be seen that the Fokker-Planck equation
(\ref{eq:FPE1}) is also valid for the unconditional probabilities $p(h_r,r)$
(see also section \ref{sec:veri_coeff}). Thus it covers also the behaviour of
the moments $\langle h_r^n\rangle$ (also called structure functions) including
any possible scaling behaviour.
An equation for the moments can be obtained by additionally multiplying with
$h_r^n$ and integrating over $h_r$
\begin{eqnarray}
   \label{eq:moments}
   \lefteqn{-r \frac{\partial}{\partial r} \langle h_r^n \rangle =}\\
   && n \langle D^{(1)}(h_r,r) h_r^{n-1} \rangle
   + n(n-1) \langle D^{(2)}(h_r,r) h_r^{n-2} \rangle \;. \nonumber
\end{eqnarray}
For $D^{(1)}$ being purely linear in $h_r$ ($D^{(1)} = \alpha h_r$) and
$D^{(2)}$ purely quadratic ($D^{(2)}= \beta h_r^2$), the multifractal scaling
$\langle h_r^n\rangle \sim r^{\xi_n}$ with $\xi_n = n \alpha + n(n-1)\beta$ is
obtained from (\ref{eq:moments}). If in contrast $D^{(2)}$ is constant in
$h_r$, a monofractal scaling where $\xi_n$ are linear in $n$ may occur, see
\cite{Friedrich1998a}.

Lastly, we want to point out that the Fokker-Planck equation (\ref{eq:FPE1})
corresponds to the following Langevin equation (we use It\^{o}'s definition)
\cite{Risken1984}
\begin{equation}\label{eq:Langevin}
   -\frac{\partial h_r}{\partial r} =
   D^{(1)}(h_r,r)/r+\sqrt{D^{(2)}(h_r,r)/r}\,\Gamma(r)\;,
\end{equation}
where $\Gamma(r)$ is a Gaussian distributed, $\delta$-correlated noise. The
use of this Langevin model in the scale variable opens the possibility to
directly simulate surface profiles with given stochastic properties, similar
to \cite{Jafari2003}.

With this brief summary of the features of stochastic processes we have fixed
the scheme from which we will present our analysis of diverse rough surfaces.
There are three steps: First, the verification of the Markov property. Second,
the estimation of the drift and diffusion coefficients $D^{(1)}$ and $D^{(2)}$.
Third, the verification of the estimated coefficients by a numerical solution
of the corresponding Fokker-Planck equation, thus reconstructing the pdf which
are compared to the empirical ones.

\section{Measurement data}
\label{sec:measurement_data}

With the method outlined in section \ref{sec:Markov_theory} we analysed a
collection of road surfaces measured with a specially designed profile scanner
as well as two microscopic surfaces, namely an evaporated gold film and a
crack surface of a low-alloyed steel sample, as already mentioned in
section~\ref{sec:intro}.

The road surfaces have been measured with a specially designed surface profile
scanner. The Longitudinal resolution was 1.04\un{mm}, the profile length being
typically 20\un{m} or 19000 samples, respectively. Between ten and twenty
parallel profiles with a lateral distance of 10\un{mm} were taken for each
surface, see fig.~\ref{fig:data_road_scaling}. The vertical error was always
smaller than 0.5\un{mm} but in most cases approximately 0.1\un{mm}. Details can
be found in \cite{Waechter2002a}.

For the Au film data, the surface of four optical glass plates had been coated
with an Au layer of 60\un{nm} thickness by thermal evaporation
\cite{Friedrich1998a}. The topography of these films was measured by atomic
force microscopy at different resolutions, resulting in a set of images of
$256\times256$ pixels each, where every pixel specifies the surface height
relative to a reference plane, see fig.~\ref{fig:data_gold}. Out of these
images 99 could be used for the analysis presented here, resulting in about
$6.5\cdot 10^6$ data points. Sidelengths vary between 36\un{nm} and 2.8\un{\mu
  m}.

The sample of the crack measurements was a fracture surface of a low-alloyed
steel (german brand 10MnMoNi5-5). A detailed description of the measurements
can be found in \cite{Wendt2002a}. Three CLSM (Confocal Laser Scanning
Microscopy) images of size 512x512 pixels in different spatial resolutions were
available, see fig.~\ref{fig:Wendt_data}. Pixel sizes are 0.49, 0.98, and
1.95\mum, resulting in image widths of 251, 502, and 998\mum, repectively.
Unavoidable artefacts of the CLSM method were removed by simply omitting for
each image those data with the smallest and largest height value, similar to
\cite{Wendt2002a}. Nevertheless this cannot guarantee the detection of all the
artefacts. The possible consequences are discussed together with the results.

For the analysis in the framework of the theory of Markov processes, we will
normalize the measurement data by the quantity $\sigma_\infty$ defined by
\begin{equation}
  \label{eq:sigma_inf}
  \sigma_\infty^2 = \lim_{r \rightarrow \infty} \langle h_r^2 \rangle\,.
\end{equation}
Thus it is possible to obtain dimensionless data with a normalization
independent of the scale $r$, in contrast to e.g.\ $\sigma_r^2=\langle h_r^2
\rangle$. As a consequence the results, especially $M^{(k)}(h_r,r,\Delta
r)$ and $D^{(k)}(h_r,r)$ (cf.\ section~\ref{sec:Markov_theory}), will also be
dimensionless. From the definition it is easy to see that $\sigma_\infty$ can
be derived via 
$\sigma_\infty^2=2\sigma_x^2=2\langle(h(x)-\bar{h})^2\rangle$ if
$h(x)$ becomes uncorrelated for large distances $r$.

\section{Scaling analysis}
\label{sec:scaling}

In this paper a number of examples was selected from all data sets under
investigation. Because most popular methods of surface analysis are based on
scaling features of some topographical measure, the examples were chosen with
respect to their different scaling properties as well as their results from our
analysis based on the theory of Markov processes.

In the analysis presented here we use the well-known \emph{height increment}
$h_r(x)$, which has been defined in eq.~\ref{eq:h_r}, as a scale-dependent
measure of the complexity of rough surfaces \cite{CenteredInc}. Scaling
properties are reflected by the $r$-dependence of the so-called structure
functions
\begin{equation}
  \label{eq:S^n}
  S^n(r) = \langle |h_r^n| \rangle \,.
\end{equation}
If one then finds
\begin{equation}
  \label{eq:scaling_def}
  S^n(r) \sim r^{\xi_n}
\end{equation}
for a range of $r$, this regime is called the scaling range. In that range the
investigated profiles have \emph{self-affine} properties, i.e., they are
statistically invariant under an anisotropic scale transformation.  If
furthermore the dependence of the exponents $\xi_n$ on the order $n$ is
nonlinear, one speaks of \emph{multi-affine} scaling. Those properties are no
longer identified by a single scaling exponent, but an infinite set of
exponents. A detailed explanation of self- and multi-affine concepts is beyond
the scope of this article. Instead, we would like to refer the reader to the
literature \cite{Feder1988,Family1991,Vicsek1992-e,Barabasi1995}.
The power spectrum, which often is used to determine scaling properties, can
easily be derived from the second order structure function.  It is defined as
the Fourier transform of the autocorrelation function $R(r)$, which itself is
closely related to $S^2(r)$ by $R(r) = \langle h(x)^2\rangle -S^2(r)/2$,
by comparing eqs.~(\ref{eq:h_r}) and (\ref{eq:S^n}).

In addition to the $r$-dependence of the structure functions, a generalized
form of scaling behaviour can be determined analogously to the Extended Self
Similarity (ESS) method which is popular in turbulence research
\cite{Benzi1993}. When the $S^n(r)$ are plotted against a structure function of
specific order, say $S^3(r)$, in many cases an extended scaling regime is
found according to
\begin{equation}
  \label{eq:ESS}
  S^n(r) \sim \left(S^3(r)\right)^{\zeta_n}\,.
\end{equation}
Clearly, meaningful results are restricted to the regime where $S^3$ is
monotonous. It is easy to see that now the $\xi_n$ can be obtained by
\begin{equation}
  \label{eq:zeta_n}
  \xi_n=\zeta_n\cdot\xi_3\,, 
\end{equation}
cf.~\cite{Benzi1993}. While for turbulence it is widely accepted that by this
means experimental defiencies can be compensated to some degree, for surface
roughness the meaning of ESS lies merely in a generalized form of scaling
properties.

It should be noted that the results of any scaling analysis may be influenced
by the method of measurement, by the definition of the roughness measure, here
$h_r(x)$ (or $w_r(x)$ as mentioned in section \ref{sec:intro}), as well as by
the algorithms used for the analysis \cite{Wendt2002a, Alber1998a}.
Nevertheless, this problem is not addressed here as the main focus of our
investigations is the application of the theory of Markov processes to
experimental data.

\subsection{Surfaces with scaling properties}
\label{sec:scaling_with}

In fig.~\ref{fig:data_road_scaling} we present road surface data with
different kinds of scaling properties.  For each data set a short profile
section is shown. Structure functions of order one to six on double
logarithmic scale are presented in fig.~\ref{fig:Sn_road_scaling}. Following
the arguments in \cite{Renner2001}, higher order structure functions cannot be
evaluated with sufficient precision from the given amount of data points.
The worn asphalt pavement (Road~1) is an example of a comparably large scaling
regime over more than one order of magnitude in $r$. A surface with similar
features, namely a cobblestone road, has already been presented
in~\cite{Waechter2003_plus_preprint}.
Two separate scaling regions are found for a Y-shaped concrete stone pavement
(Road~2). Additionally a sharp notch can be seen in the structure functions at
$r=0.2\un{m}$, indicating a strong periodicity of the pavement caused by the
length of the individual stones.
The third example, a ``pebbled concrete'' pavement (Road~3), consists of
concrete stones with a top layer of washed pebbles. This material is also known
as ``exposed aggregate concrete''. Here, the scaling region of the structure
functions is only small.
For the basalt stone pavement (Road~4), being the fourth example, scaling
properties are poor. We have nevertheless marked a possible scaling range
and derived the respective scaling exponents for comparison with the other
examples.  Similar to the Y-shaped concrete stones, a periodicity can be found
at about 0.1\un{m} length scale.

\newlength{\breite}\setlength{\breite}{0.8\linewidth}
\begin{figure}[htbp]\centering
  \begin{picture}(0,0)\put(8,29){\makebox{\small Road~1}}\end{picture}%
  \includegraphics[width=\breite]{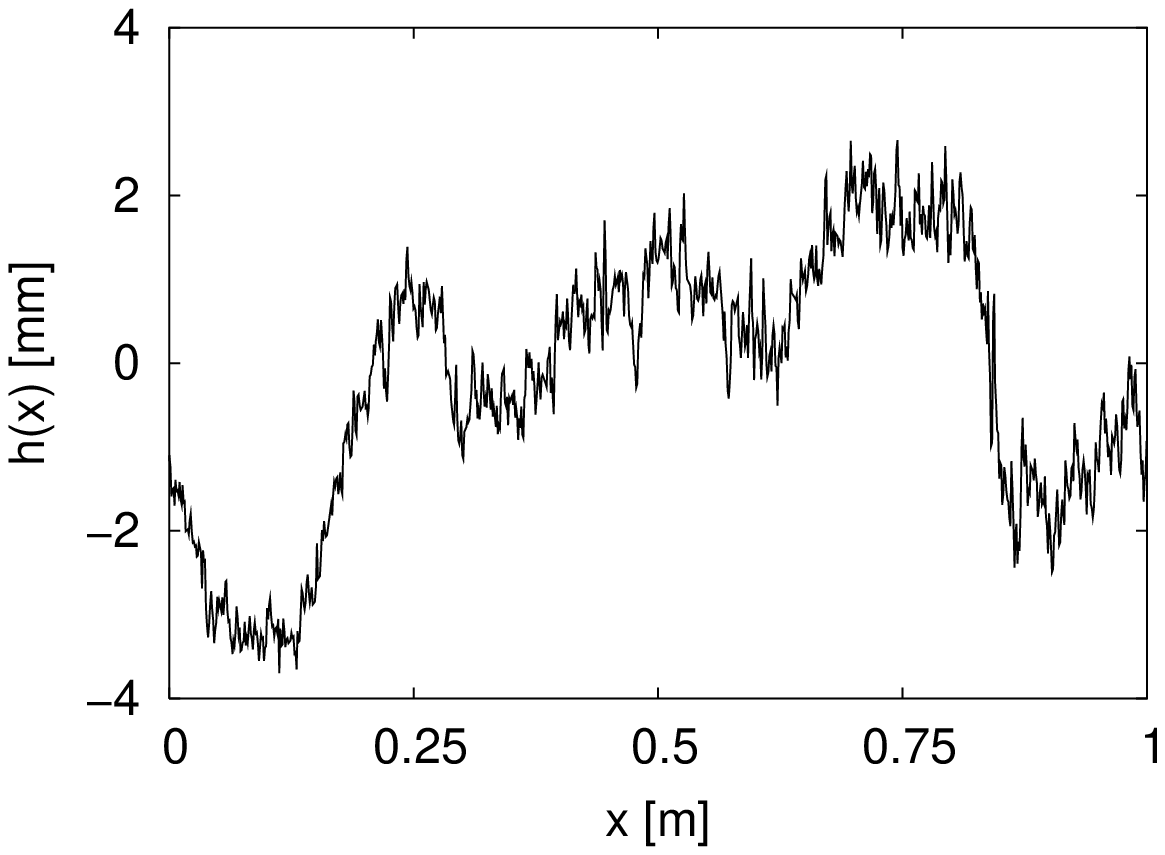}

  \begin{picture}(0,0)\put(8,29){\makebox{\small Road~2}}\end{picture}%
  \includegraphics[width=\breite]{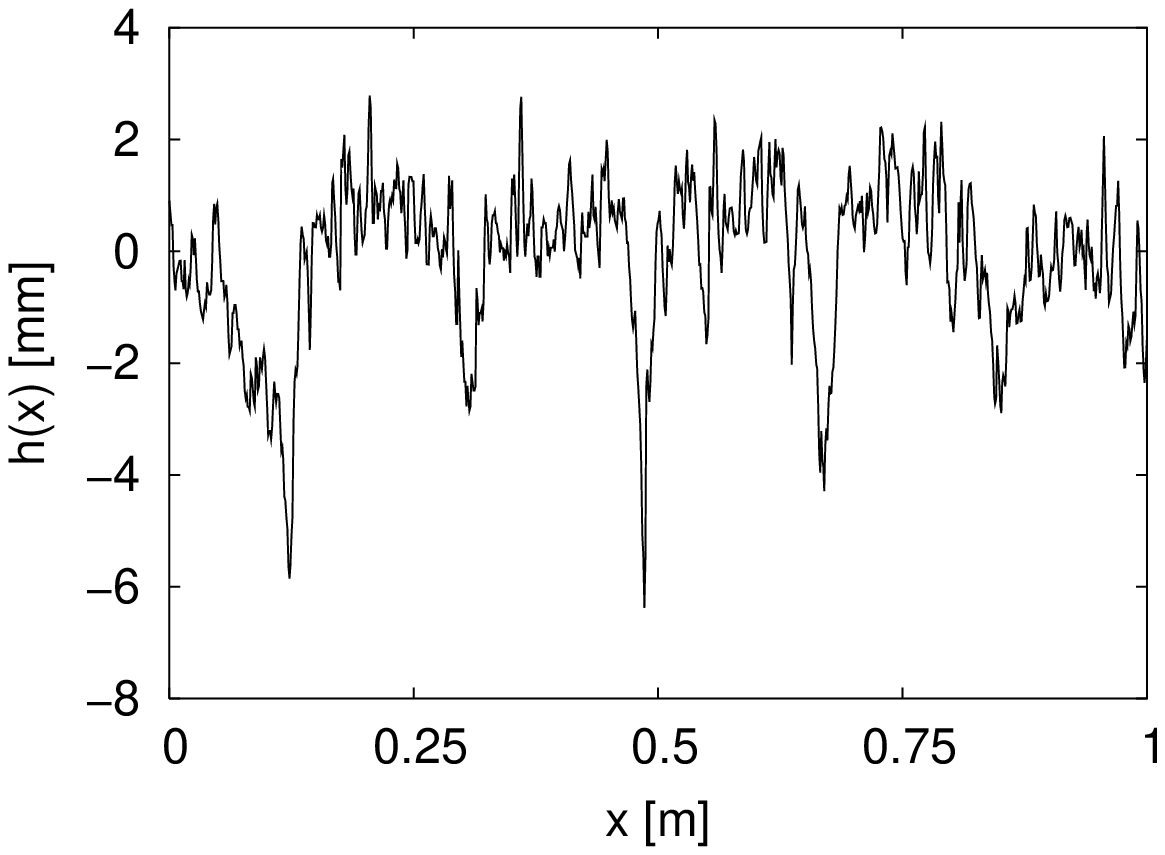}

  \begin{picture}(0,0)\put(8,29){\makebox{\small Road~3}}\end{picture}%
  \includegraphics[width=\breite]{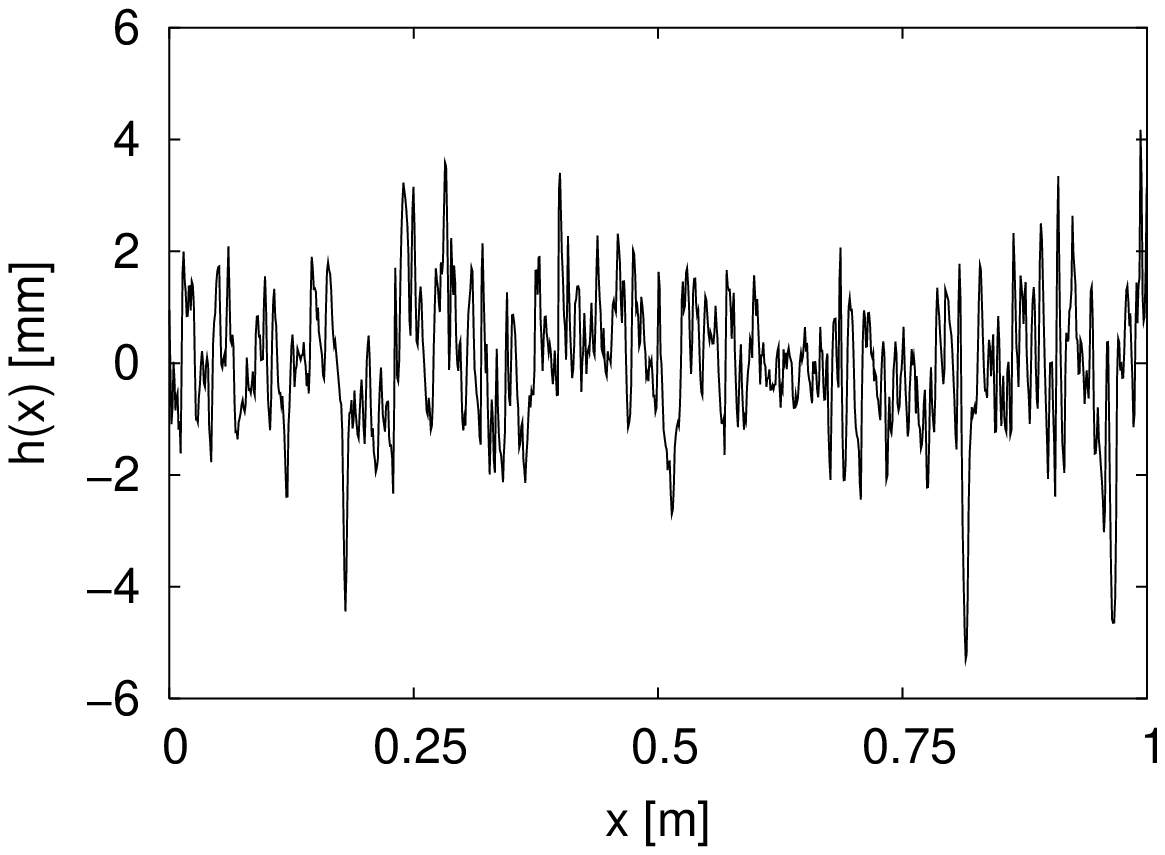}

  \begin{picture}(0,0)\put(8,29){\makebox{\small Road~4}}\end{picture}%
  \includegraphics[width=\breite]{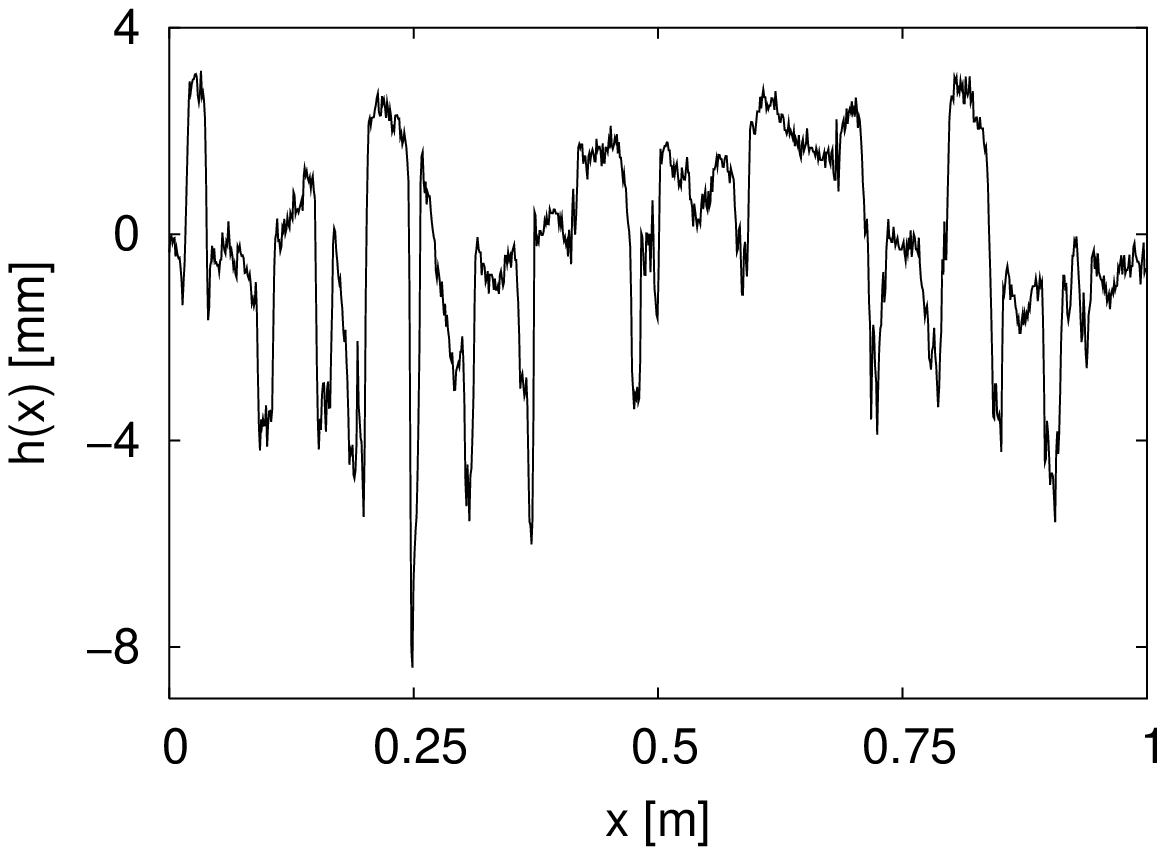}

  \caption{%
    Measurement data from selected road surfaces with different kinds of
    scaling properties.  %
    Pavements are worn asphalt (Road 1), Y-shaped concrete stones (Road 2),
    pebbled concrete stones (Road 3), and basalt stones (Road 4), from top to
    bottom. For each surface a short
    section of the respective height profile is shown.  }
  \label{fig:data_road_scaling}
\end{figure}

\begin{figure}[htbp]\centering
  \begin{picture}(0,0)\put(8,29){\makebox{\small Road~1}}\end{picture}%
  \includegraphics[width=\breite]{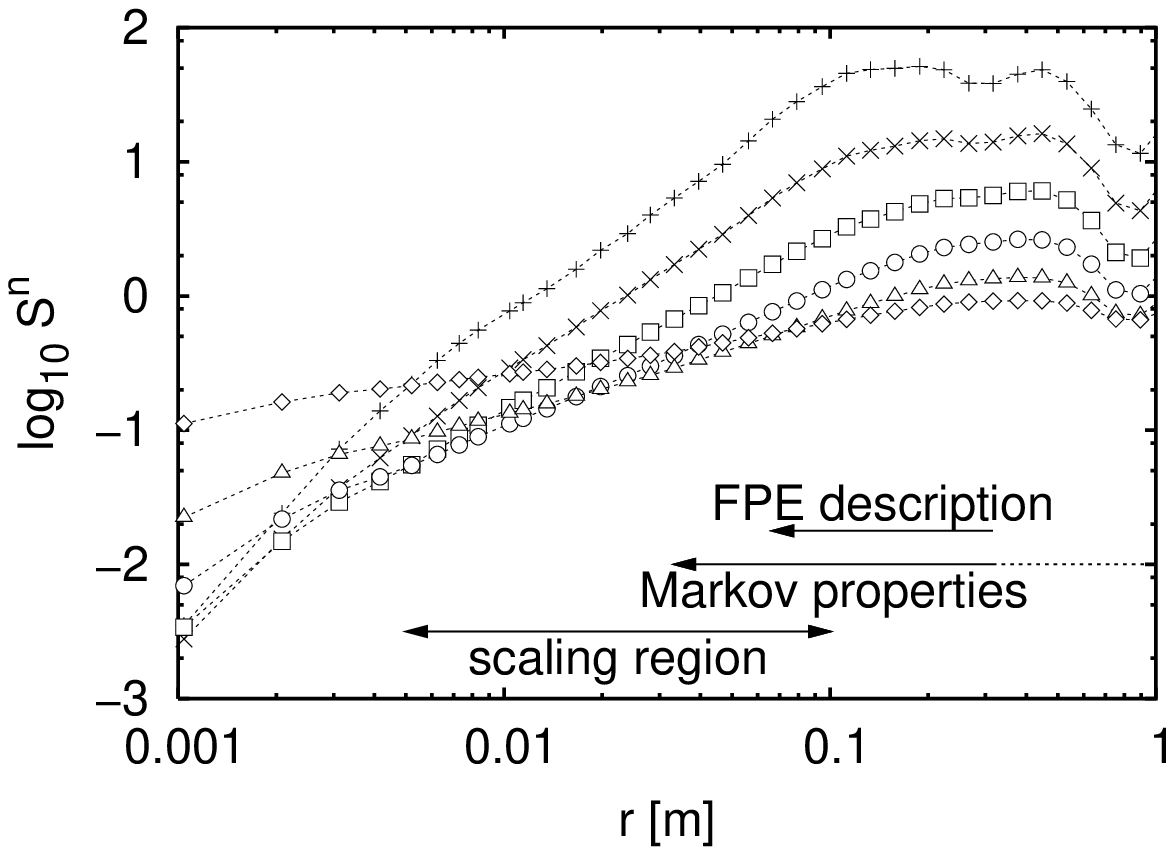}

  \begin{picture}(0,0)\put(8,29){\makebox{\small Road~2}}\end{picture}%
  \includegraphics[width=\breite]{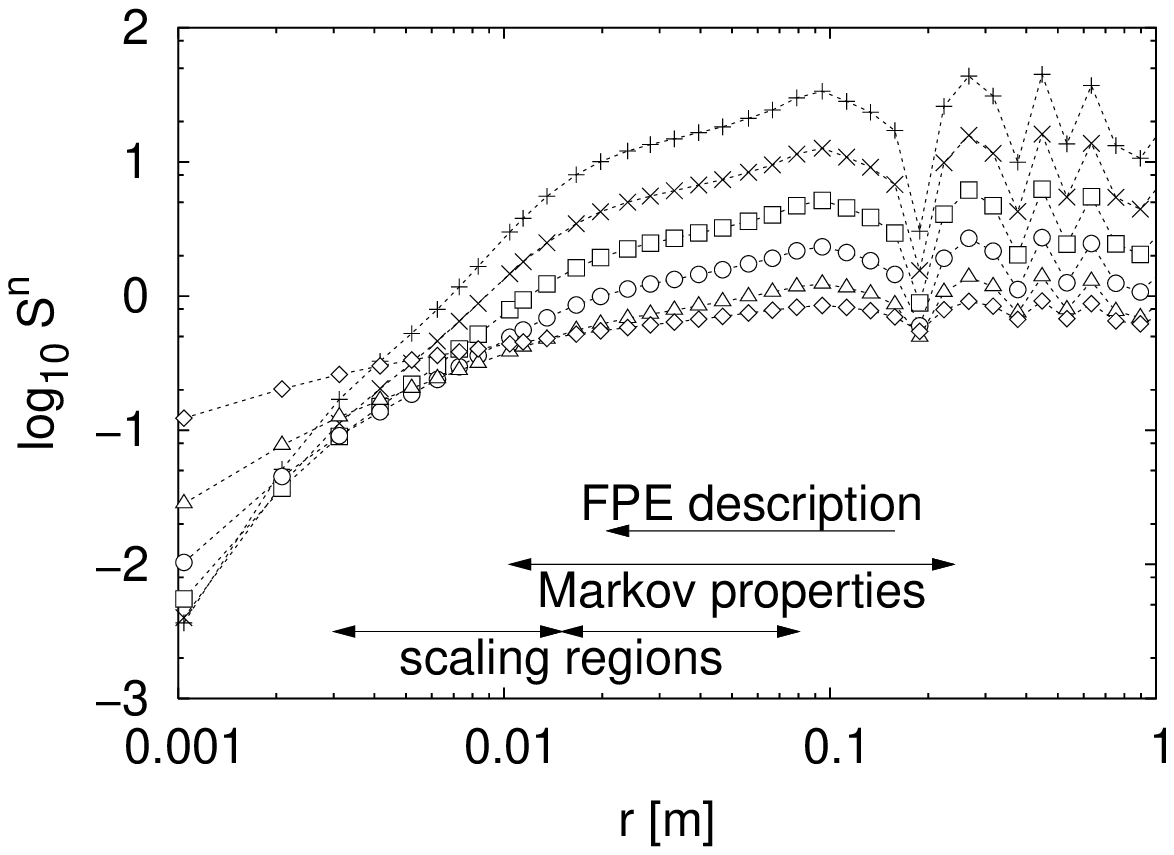}

  \begin{picture}(0,0)\put(8,29){\makebox{\small Road~3}}\end{picture}%
  \includegraphics[width=\breite]{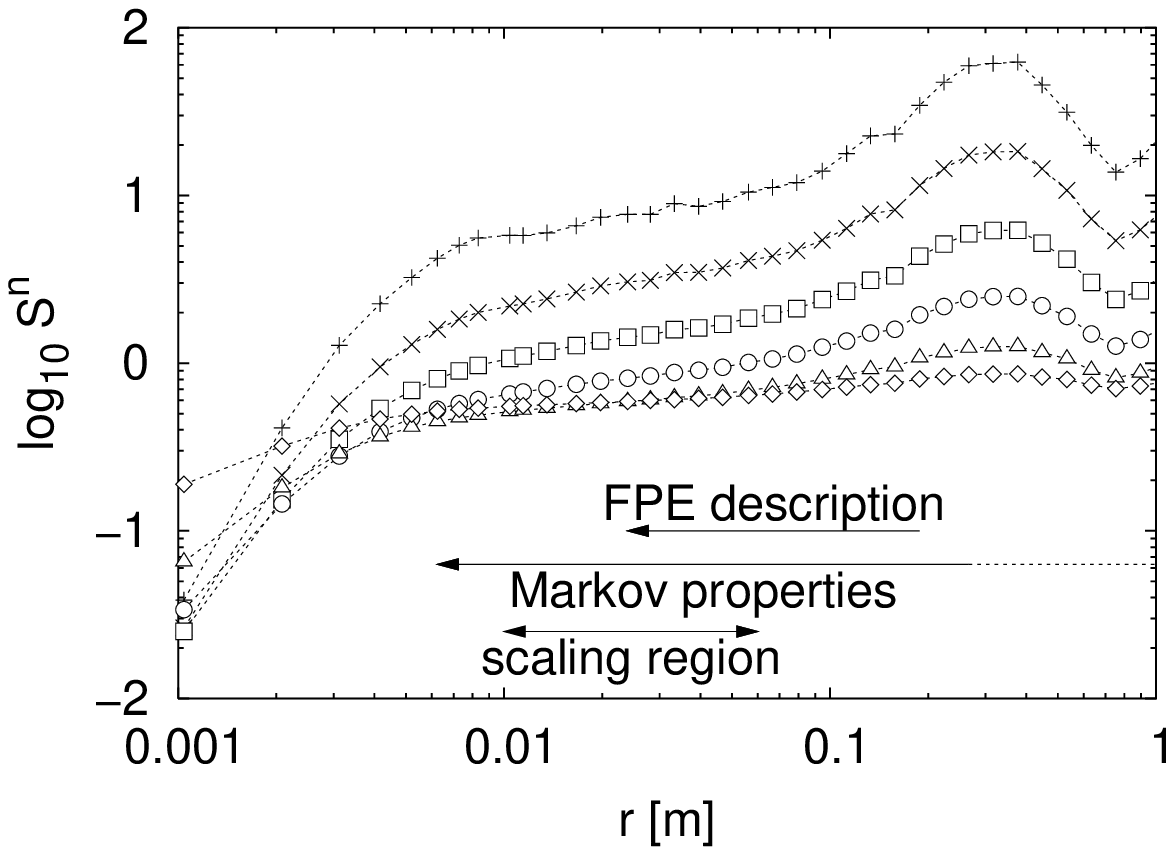}

  \begin{picture}(0,0)\put(8,29){\makebox{\small Road~4}}\end{picture}%
  \includegraphics[width=\breite]{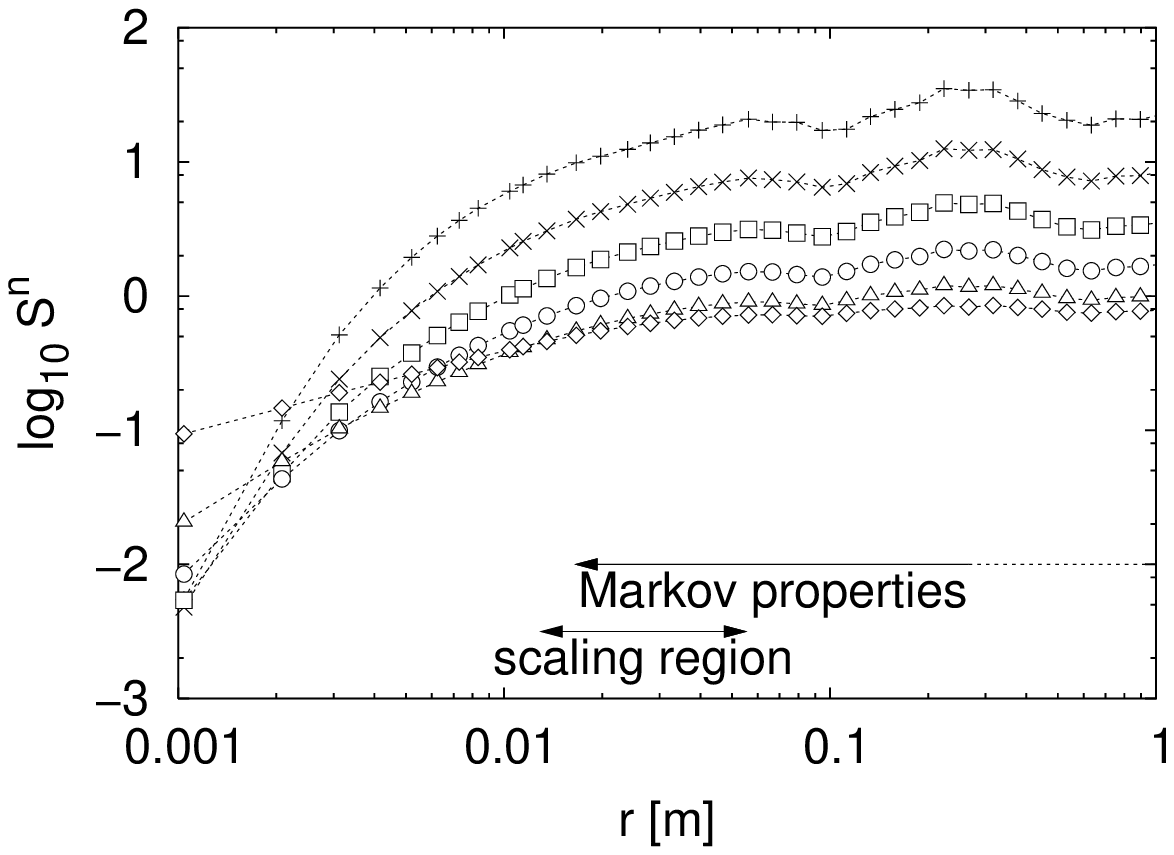}

  \caption{%
    Structure functions $S^n(r)$ of selected road surfaces (see
    fig.~\ref{fig:data_road_scaling}) with different kinds of scaling
    properties on a log-log scale (see text).
    Symbols correspond to orders $n$; diamonds ($n=1$), triangles
    ($n=2$), circles ($n=3$), squares ($n=4$), x signs ($n=5$), and plus signs
    ($n=6$).  }
  \label{fig:Sn_road_scaling}
\end{figure}

\begin{figure}[htbp]\centering
  \begin{picture}(0,0)\put(8,29){\makebox{\small Road~1}}\end{picture}%
  \includegraphics[width=\breite]{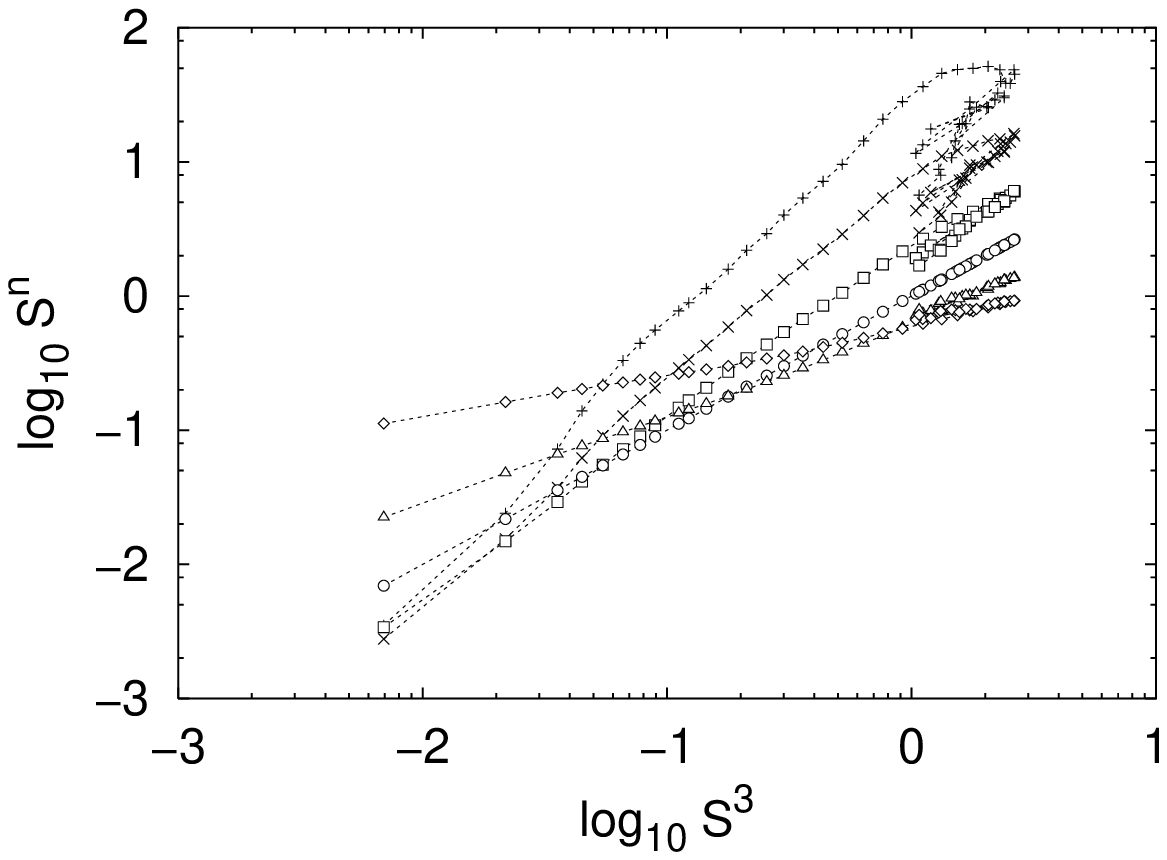}

  \begin{picture}(0,0)\put(8,29){\makebox{\small Road~2}}\end{picture}%
  \includegraphics[width=\breite]{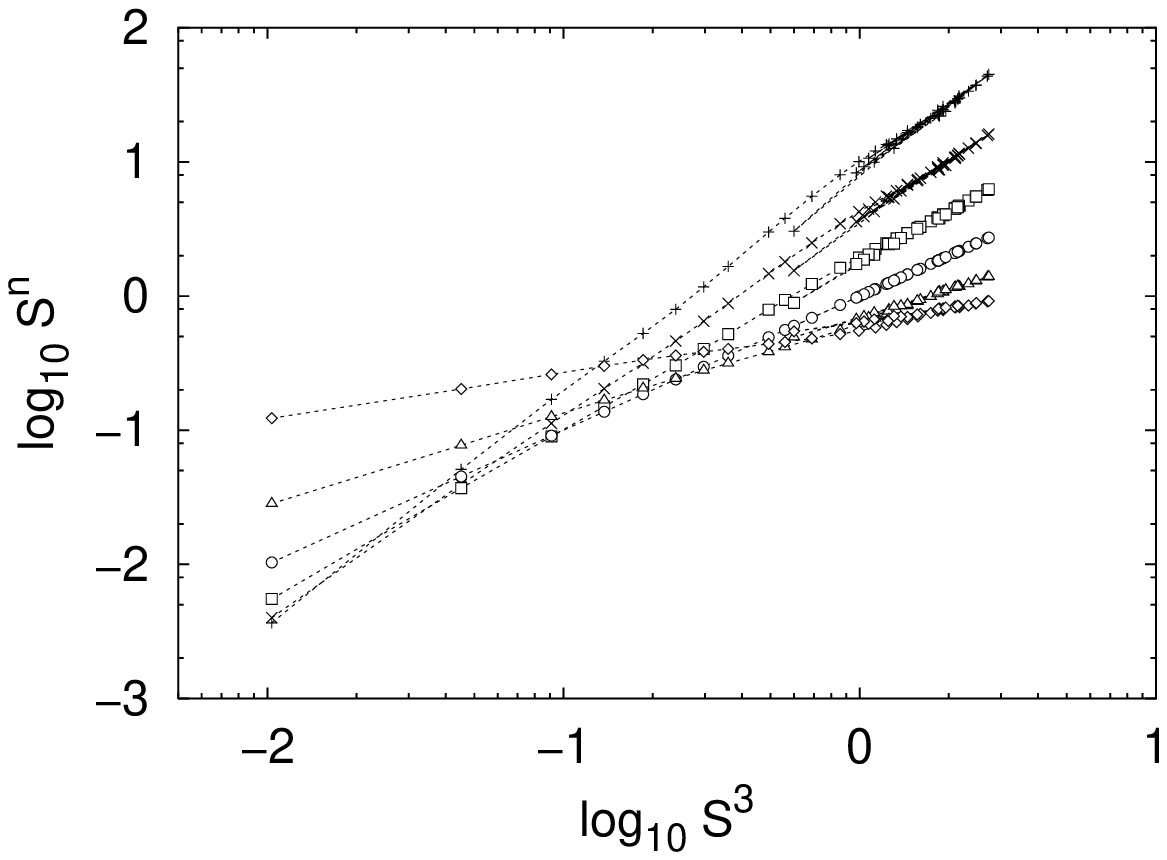}

  \begin{picture}(0,0)\put(8,29){\makebox{\small Road~3}}\end{picture}%
  \includegraphics[width=\breite]{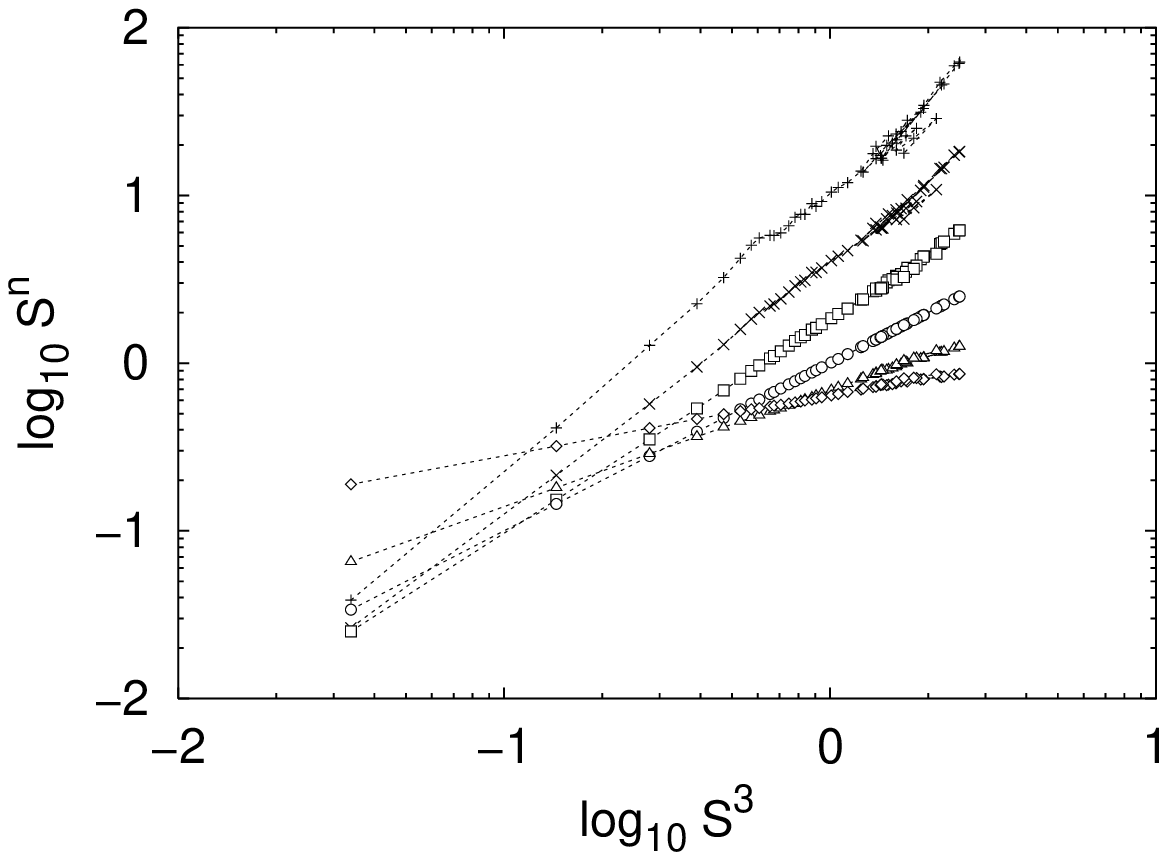}

  \begin{picture}(0,0)\put(8,29){\makebox{\small Road~4}}\end{picture}%
  \includegraphics[width=\breite]{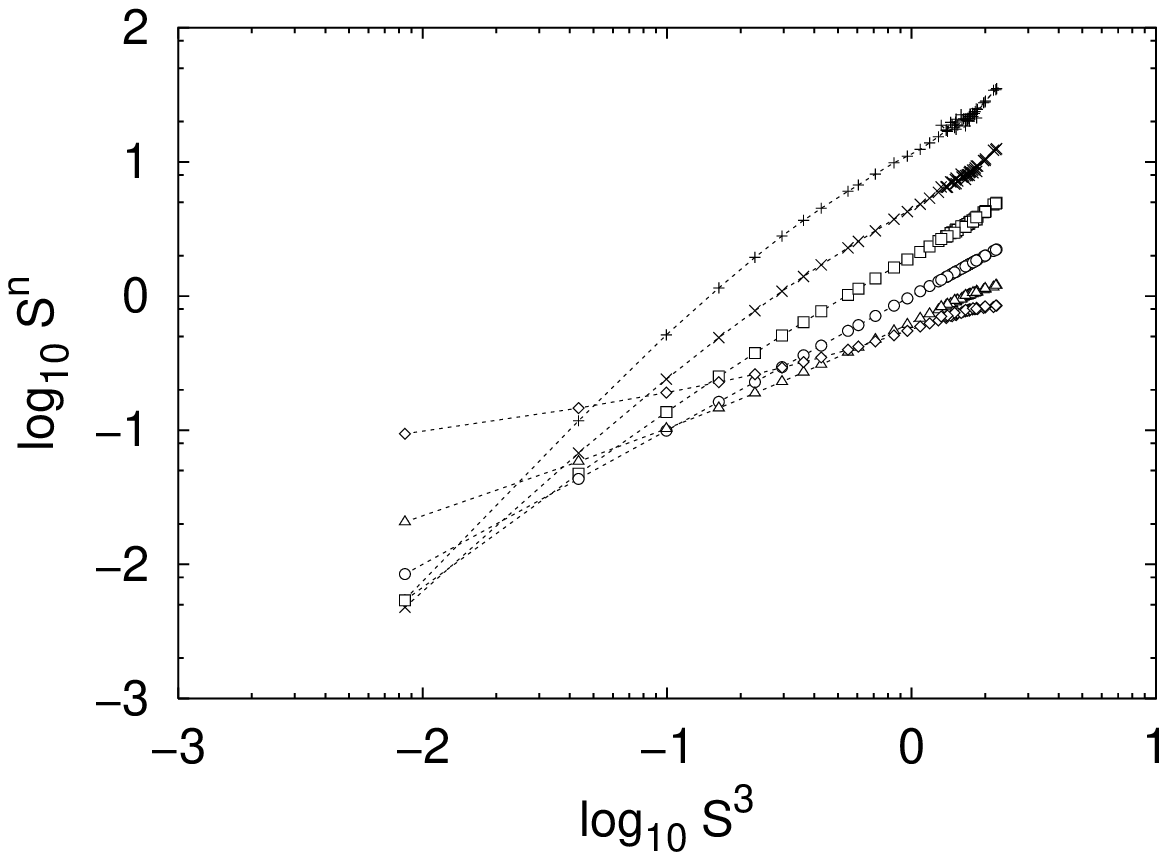}

  \caption[Road surface scaling]{%
    Generalized scaling analysis of the surfaces shown in
    fig.~\ref{fig:data_road_scaling}.
    Structure functions $S^n$ are displayed versus $S^3$ on a log-log scale.
    Symbols correspond to orders $n$ as in fig.~\ref{fig:Sn_road_scaling}.
  }
  \label{fig:data_road_scaling_ess}
\end{figure}

\begin{figure}[htbp]\centering
  \begin{picture}(0,0)\put(10,29){\makebox{\small Road~1}}\end{picture}%
  \includegraphics[width=\breite]{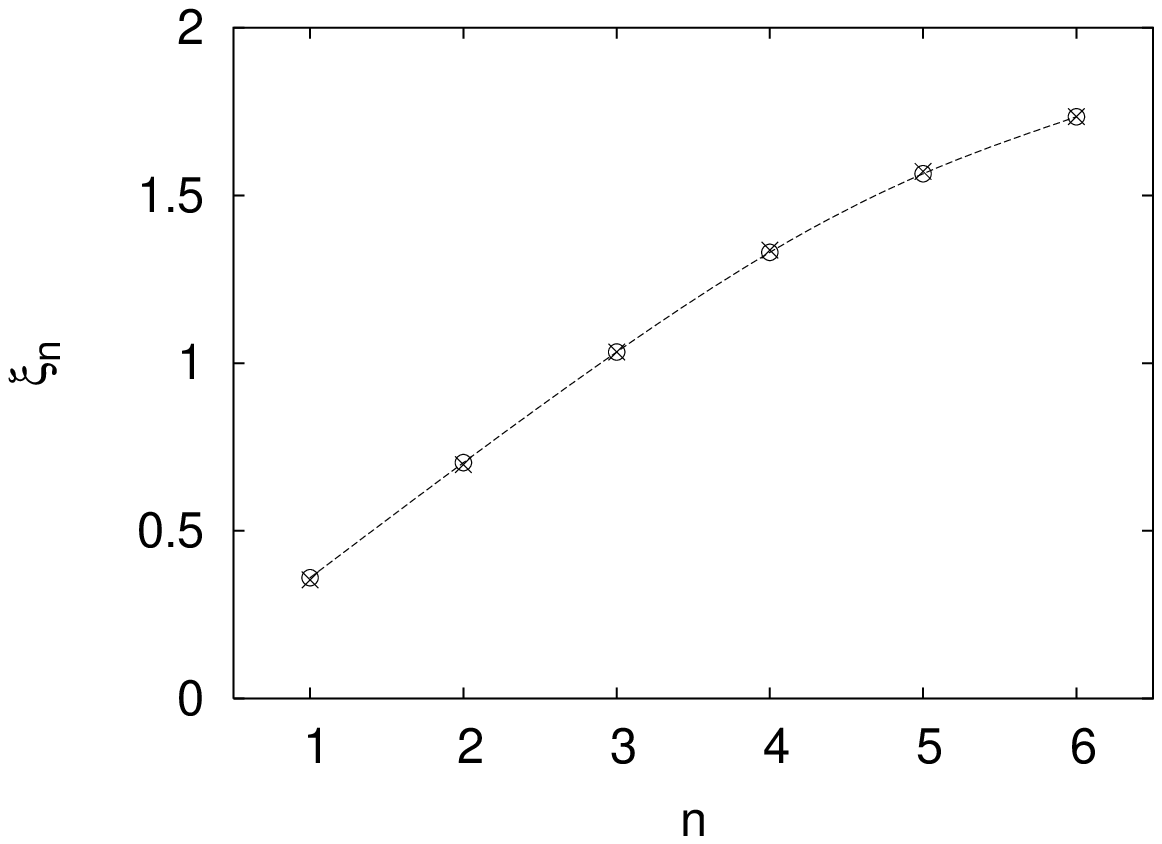}

  \begin{picture}(0,0)\put(10,29){\makebox{\small Road~2}}\end{picture}%
  \includegraphics[width=\breite]{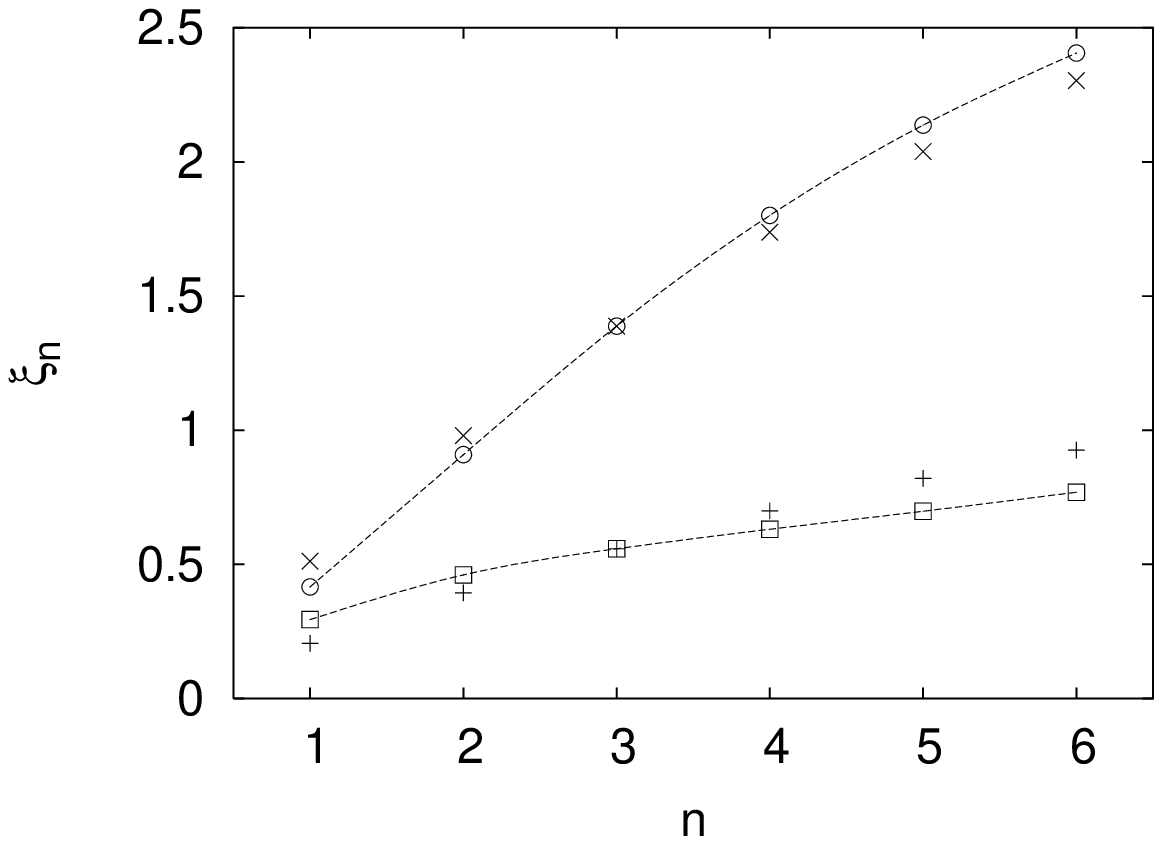}

  \begin{picture}(0,0)\put(10,29){\makebox{\small Road~3}}\end{picture}%
  \includegraphics[width=\breite]{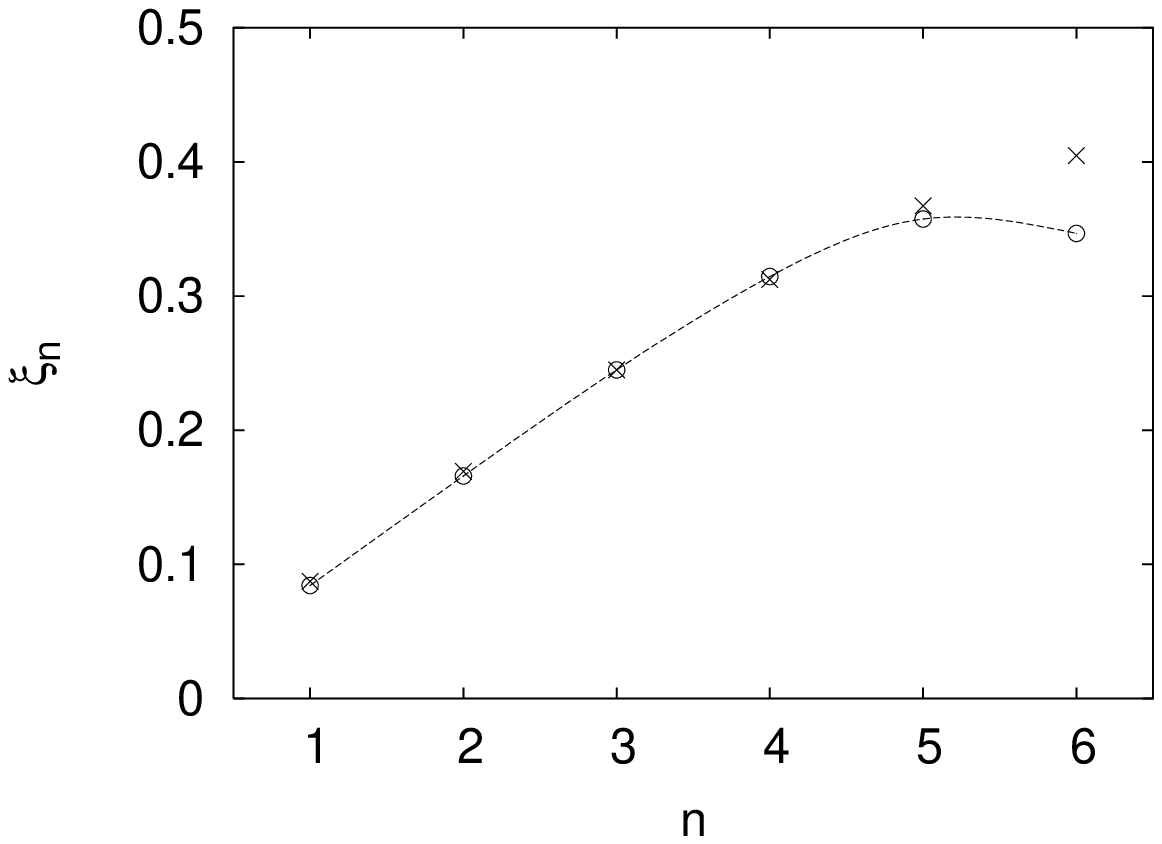}

  \begin{picture}(0,0)\put(10,29){\makebox{\small Road~4}}\end{picture}%
  \includegraphics[width=\breite]{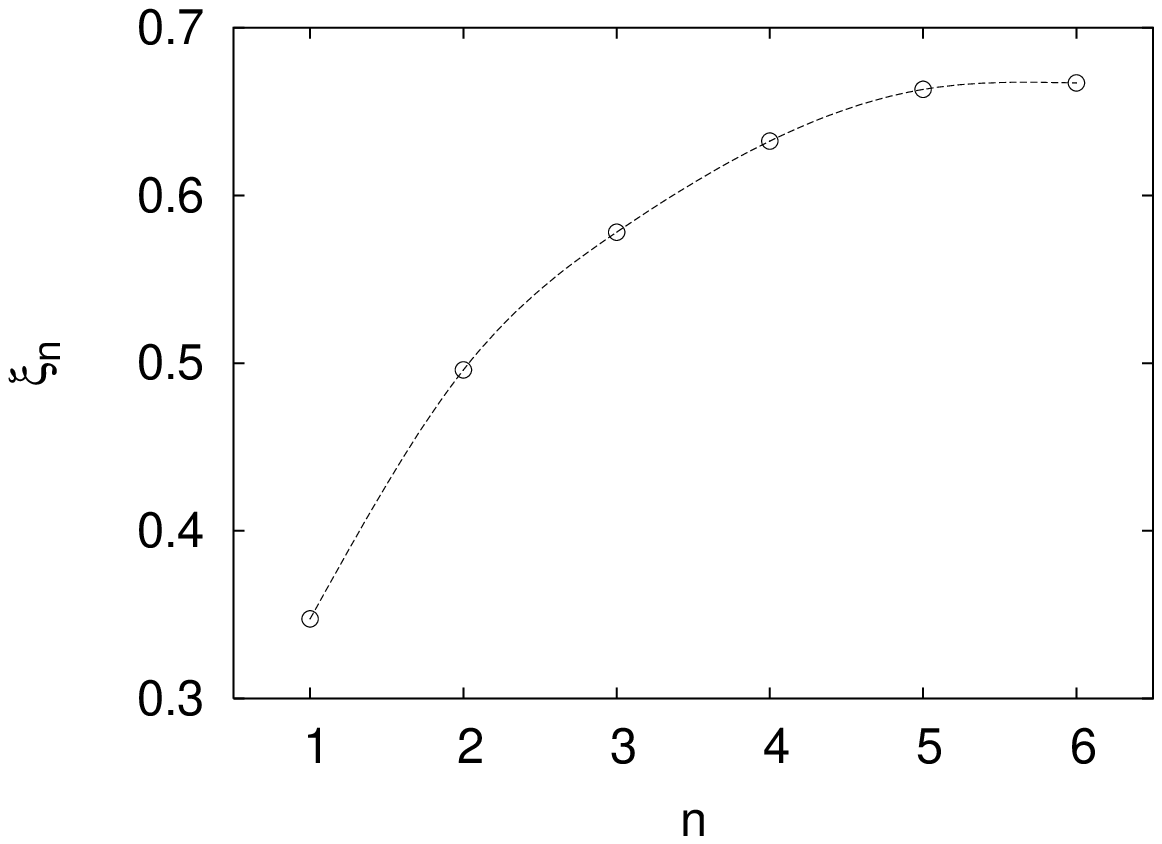}

  \caption[Road surface scaling]{%
    Scaling exponents $\xi_n$ of the surfaces shown in
    fig.~\ref{fig:data_road_scaling} achieved via eq.~(\ref{eq:scaling_def})
    (open symbols) and those obtained via $\zeta_n$ from eq.~(\ref{eq:zeta_n})
    (crosses).  For Road~2 two sets of exponents were obtained from the two
    scaling regimes found for $S^n(r)$ in fig.~\ref{fig:Sn_road_scaling}.}
  \label{fig:data_road_scaling_exp}
\end{figure}

The results for the generalized scaling behaviour according to
eq.~(\ref{eq:ESS}) are shown in fig.~\ref{fig:data_road_scaling_ess}. It can be
seen that indeed for three of the surfaces in fig.~\ref{fig:data_road_scaling}
an improved scaling behaviour is found by this method. Only for Road~4 do the
generalized scaling properties remain weak.
In fig.~\ref{fig:data_road_scaling_exp} the scaling exponents $\xi_n$ of the
structure functions within the marked scaling regimes in
fig.~\ref{fig:Sn_road_scaling} were determined and plotted against the order
$n$ as open symbols. Additionally, values of $\xi_n$ were derived according to
eqs.~(\ref{eq:ESS}) and (\ref{eq:zeta_n}) and added as crosses. For Road~2 two
sets of exponents correspond to the two distinct scaling regimes in
fig.~\ref{fig:Sn_road_scaling}. Even though there is only one set of $\zeta_n$
found in fig.~\ref{fig:data_road_scaling_ess}, two sets of $\xi_n$ are
obtained due to the two different $\xi_3$, see eqs.~(\ref{eq:scaling_def}) and
(\ref{eq:zeta_n}).

All surfaces from fig.~\ref{fig:data_road_scaling} show a more or less
nonlinear dependence of the $\xi_n$ on $n$, indicating multi-affine scaling
properties. The scaling exponents obtained via the generalized scaling
according to eq.~(\ref{eq:ESS}) are in good correspondence with the $\xi_n$
achieved by the application of eq.~(\ref{eq:scaling_def}). Deviations are seen
for Road~2 and at higher orders for Road~3, possibly caused by inaccuracies in
the fitting procedure. For Road~4 no generalized scaling is observed (compare
fig.~\ref{fig:data_road_scaling_ess}) and thus values of $\xi_n$ cannot be
derived from $\zeta_n$. From this we conclude that scaling properties for some
cases are questionable as a comprehensive tool to characterize the complexity
of a rough surface.

\begin{figure}[htbp]
    \setlength{\breite}{0.5\linewidth}
    \begin{picture}(0,0)%
      \put(0, 2){\balken{0.4545\breite}{50\un{nm}}}%
    \end{picture}%
    \begin{picture}(0,0)%
      \put(30, 2){\balken{0.4545\breite}{0.5\un{\mu m}}}%
    \end{picture}%
    \raisebox{2.5ex}{%
      \includegraphics[width=\breite]{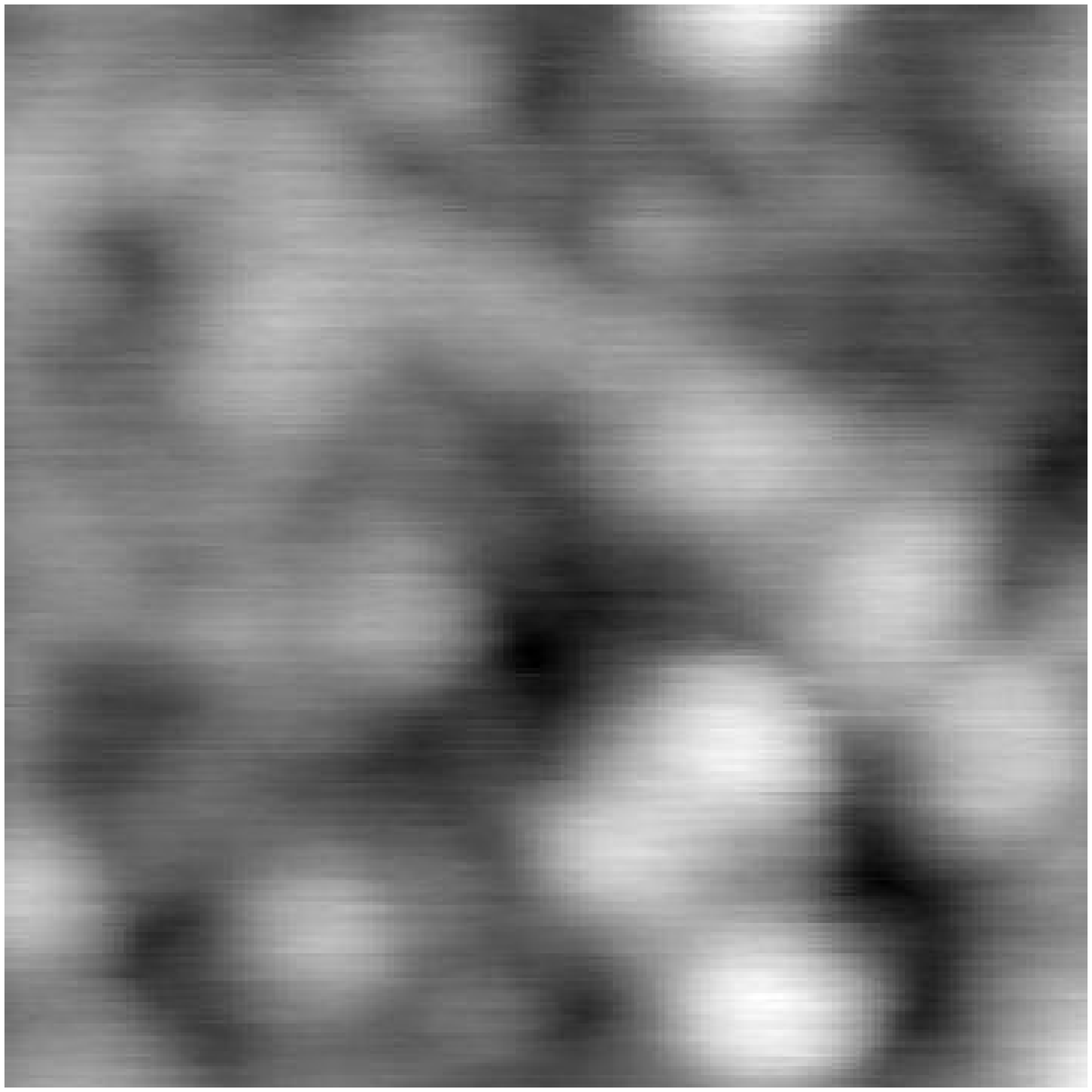}%
      \includegraphics[width=\breite]{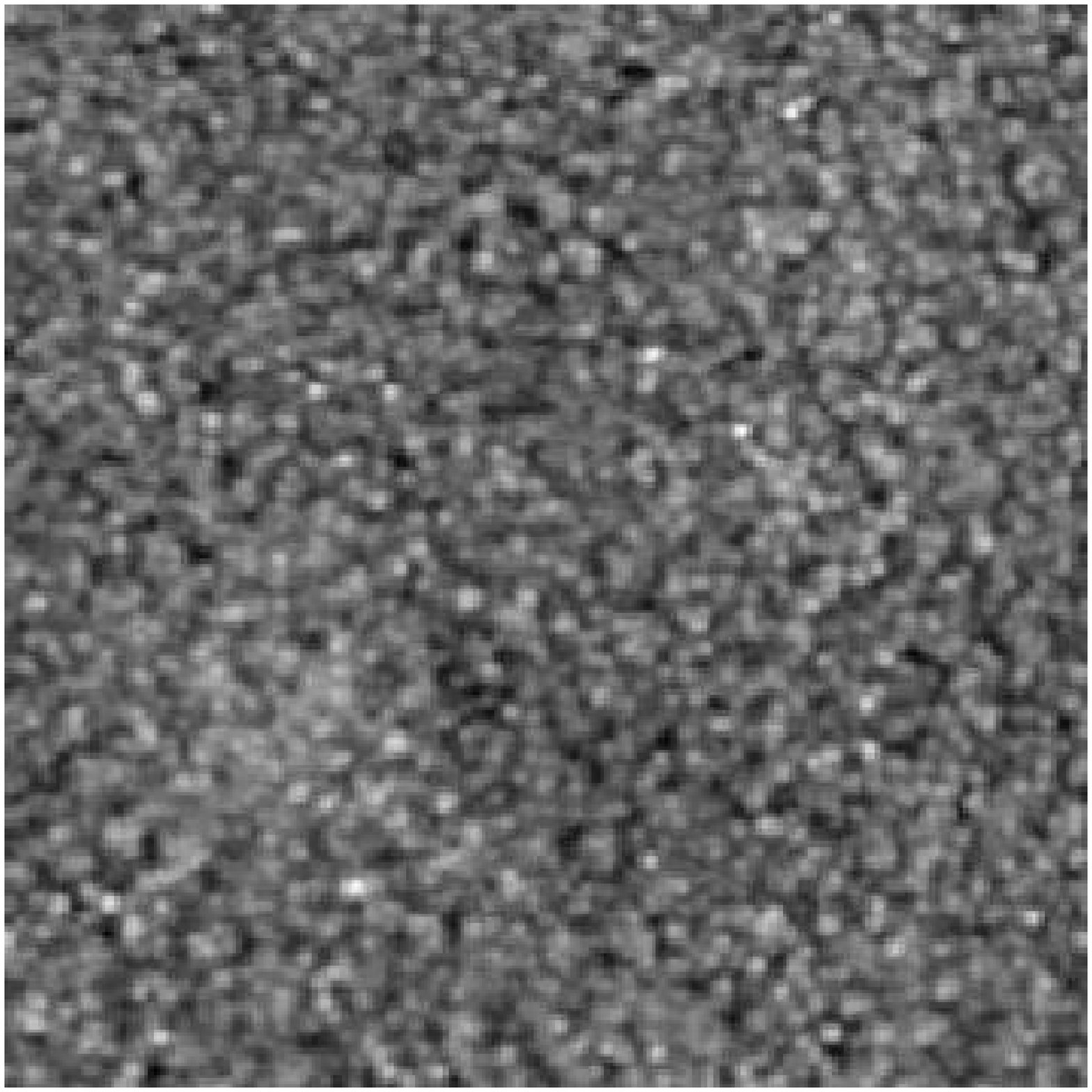}%
    }%
  \caption{%
    AFM images of the Au film surface. Sidelengths are 110\un{nm} and
    1.1\un{\mu m}. The relative surface height is represented as gray level.
    Maximum heights are 7.2\un{nm} and 13.3\un{nm}, respectively.
    }
  \label{fig:data_gold}
\end{figure}

\begin{figure}[htbp]\centering
  \setlength{\breite}{\linewidth}
  \includegraphics[width=0.8\breite]{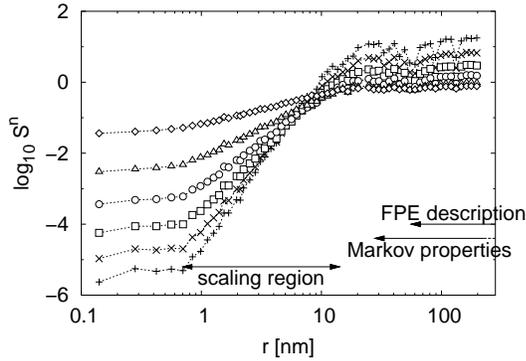}
  \caption{%
    Structure functions $S^n(r)$ of the Au film surface on a log-log scale.
    Symbols correspond to orders $n$ as in fig.~\ref{fig:Sn_road_scaling}.
  }
  \label{fig:Sn_gold}
\end{figure}

\begin{figure}[htbp]\centering
  \begin{picture}(0,0)\put(8,29){\makebox{\small (a)}}\end{picture}%
  \includegraphics[width=\breite]{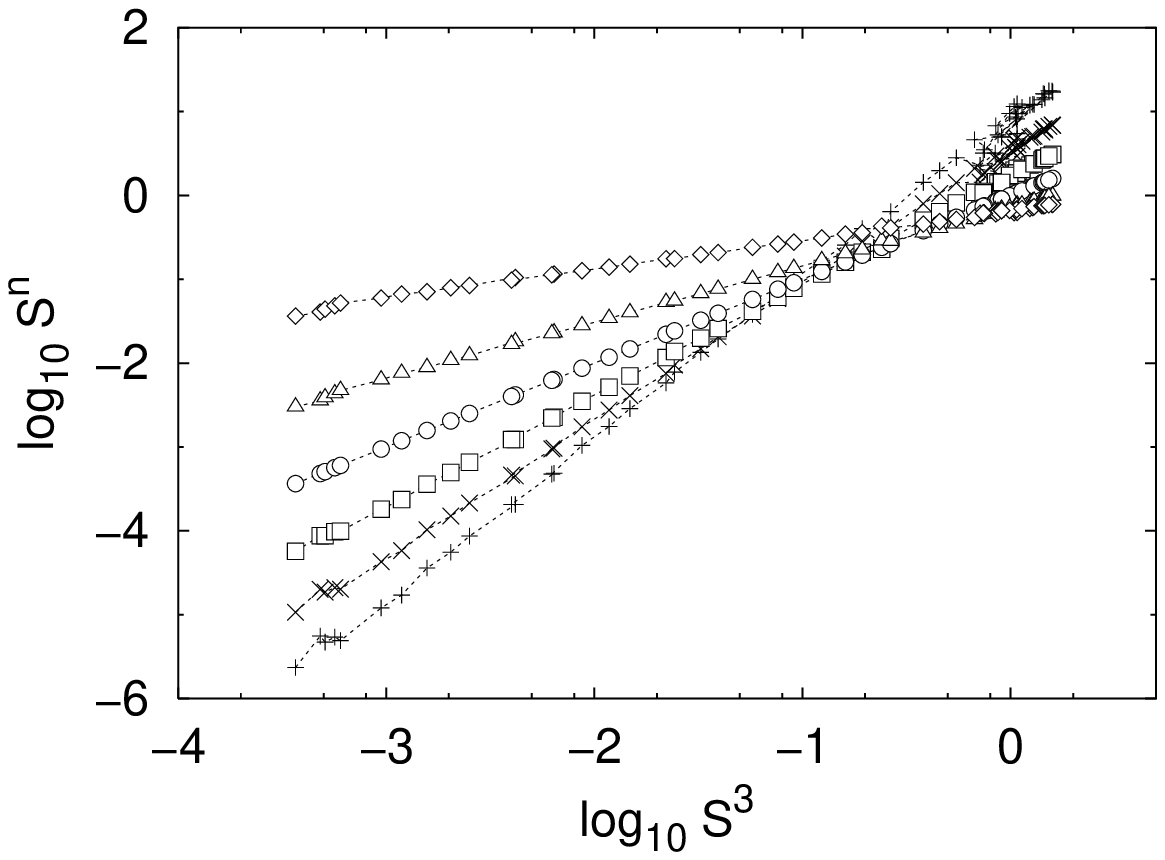}
  \begin{picture}(0,0)\put(8,29){\makebox{\small (b)}}\end{picture}%
  \includegraphics[width=\breite]{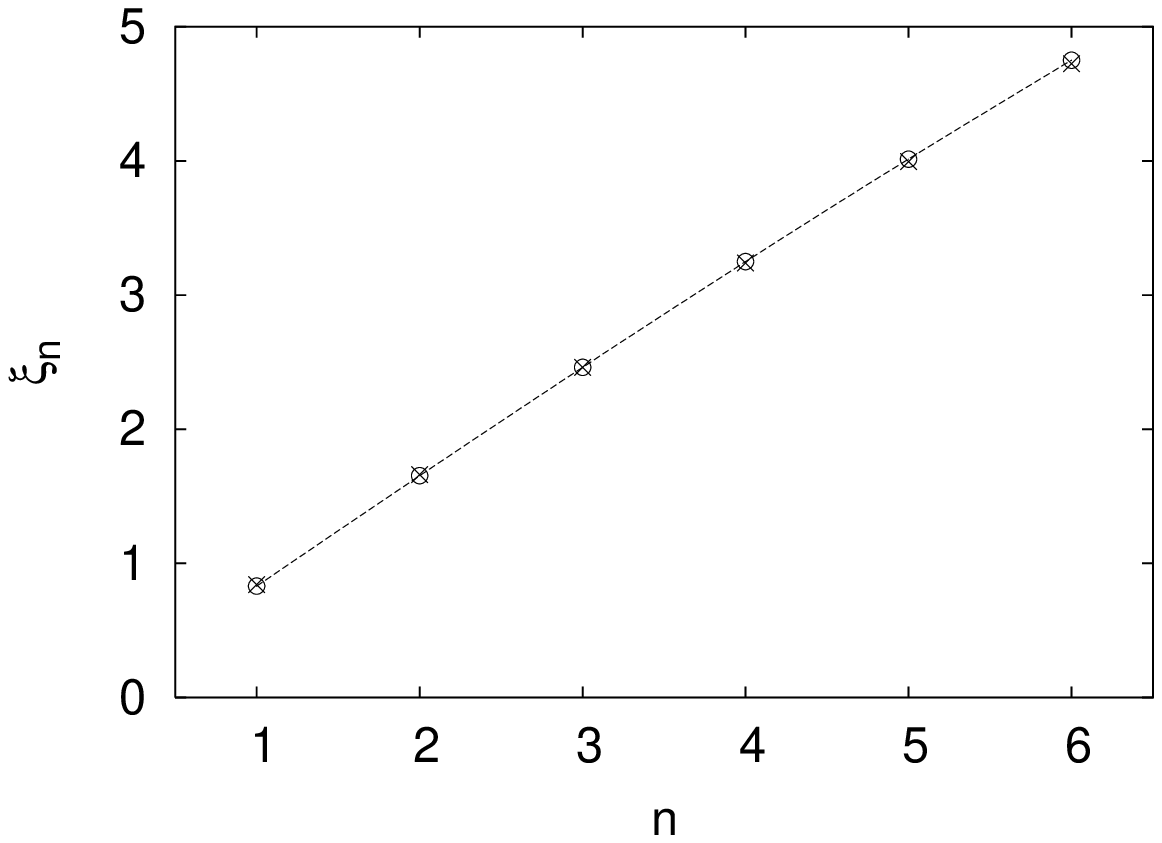}
  \caption{%
    Generalized scaling properties (a) and scaling exponents (b) of the
    Au film surface shown in fig.~\ref{fig:data_gold}. Scaling exponents
    $\xi_n$ achieved via eq.~(\ref{eq:ESS}) are marked by open symbols, those
    obtained via $\zeta_n$ from eq.~(\ref{eq:zeta_n}) by crosses. Compare also
    with fig.~\ref{fig:data_road_scaling_ess}. }
  \label{fig:data_gold_ess}
\end{figure}

An example for good scaling properties is the gold film surface (Au).  To
increase statistical accuracy, increments are evaluated here in the direction
of the rows of the images as well as the columns.
In fig.~\ref{fig:data_gold} two of the 99 images under investigation are
shown. Figure~\ref{fig:Sn_gold} presents the structure functions $S^n(r)$,
derived from all images. The surface is randomly covered with granules which
show no typical diameter. A scaling regime of more than one order of
magnitude in $r$ is found for the structure functions $S^n(r)$ in
fig.~\ref{fig:Sn_gold}. Generalized scaling behaviour is clearly present
as shown in fig.~\ref{fig:data_gold_ess}(a). The scaling exponents $\xi_n$
presented in part (b) of the same figure are nearly linear in $n$, thus this
surface can not be regarded as multi-affine, but appears to be self-affine.
Here, the $\xi_n$ achieved via eq.~(\ref{eq:zeta_n}) match perfectly those
obtained from eq.~(\ref{eq:ESS}).

\subsection{Surfaces without scaling properties}
\label{sec:scaling_without}

\begin{figure}[htbp]\centering
  \begin{picture}(0,0)\put(8,29){\makebox{\small (a)}}\end{picture}%
  \includegraphics[width=\breite]{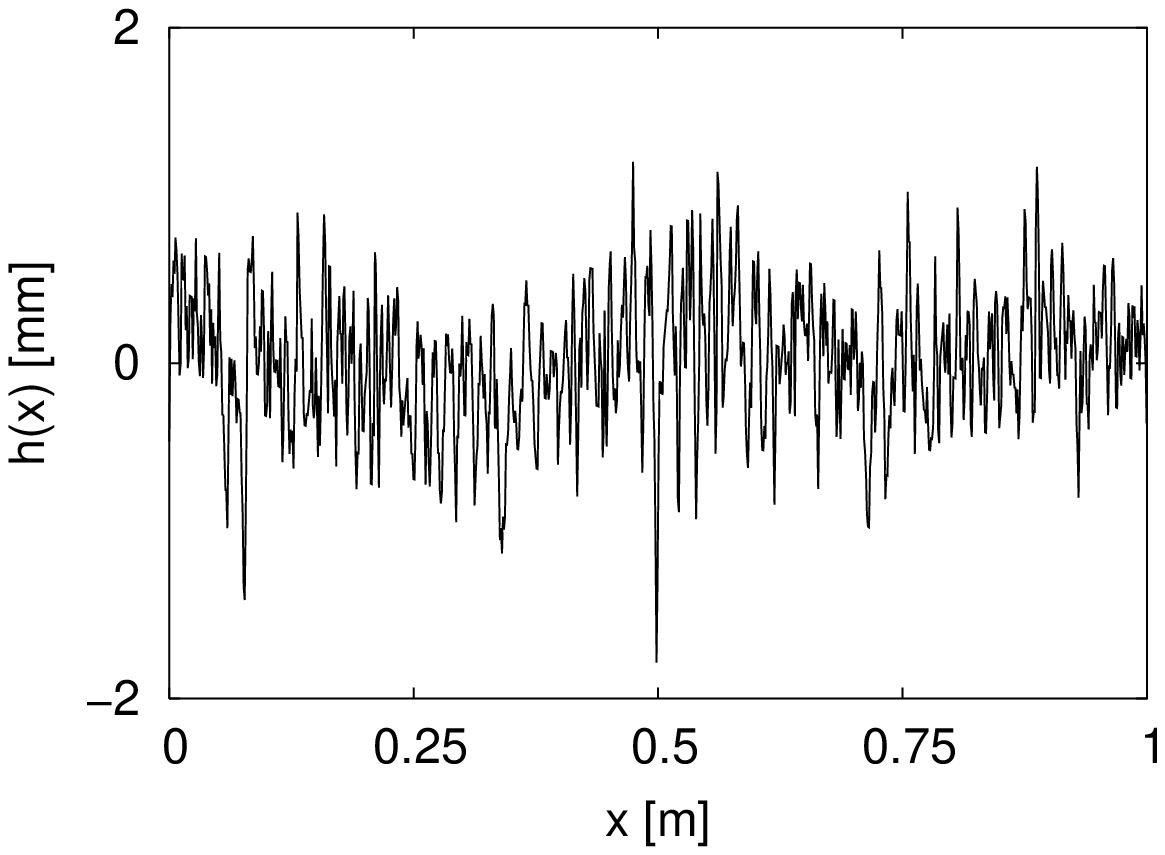}
  \begin{picture}(0,0)\put(8,29){\makebox{\small (b)}}\end{picture}%
  \includegraphics[width=\breite]{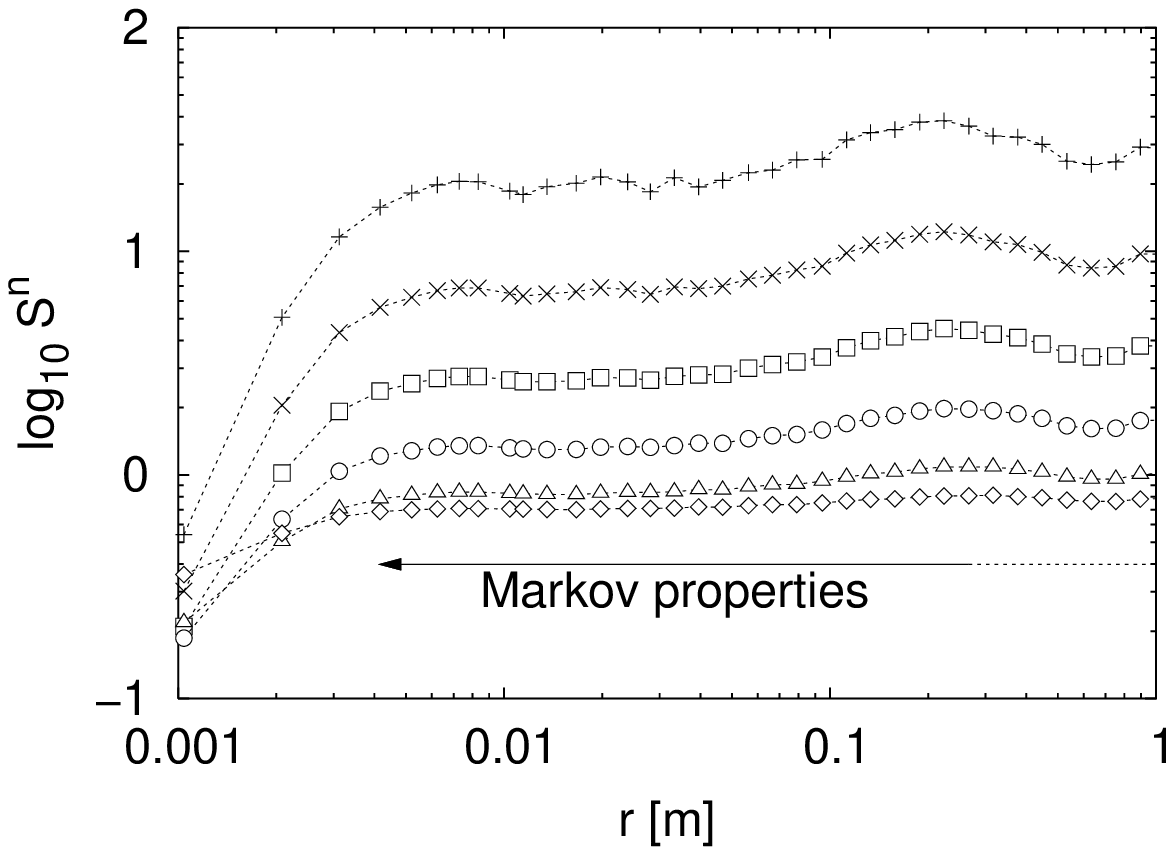}
  \caption{%
    Measurement data (a) and structure functions (b) from a road surface
    without scaling properties (Road~5). The pavement is smooth asphalt. See
    also figs.~\ref{fig:data_road_scaling} and \ref{fig:Sn_road_scaling}.}
  \label{fig:data_road_no_scaling}
\end{figure}

To complete the set of examples, we present two surfaces without scaling
properties. The first one is a smooth asphalt road (Road~5), shown in
fig.~\ref{fig:data_road_no_scaling}. No power law can be detected for the
$S^n(r)$ but a generalized scaling is observed in
fig.~\ref{fig:data_road_no_scaling_ess}(a). The range of values of $S^3$,
however, is relatively small.

\begin{figure}[htbp]
    \setlength{\breite}{0.5\linewidth}%
    \begin{picture}(0,0)%
      \put(0,33){\makebox{\small (a)}}%
      \put(0, 2){\balken{0.4\breite}{200\un{\mu m}}}%
    \end{picture}%
    \begin{picture}(0,0)%
      \put(30,33){\makebox{\small (b)}}%
      \put(30, 2){\balken{0.7143\breite}{100\un{\mu m}}}%
    \end{picture}%
    \raisebox{2.5ex}{%
      \includegraphics[width=\breite]{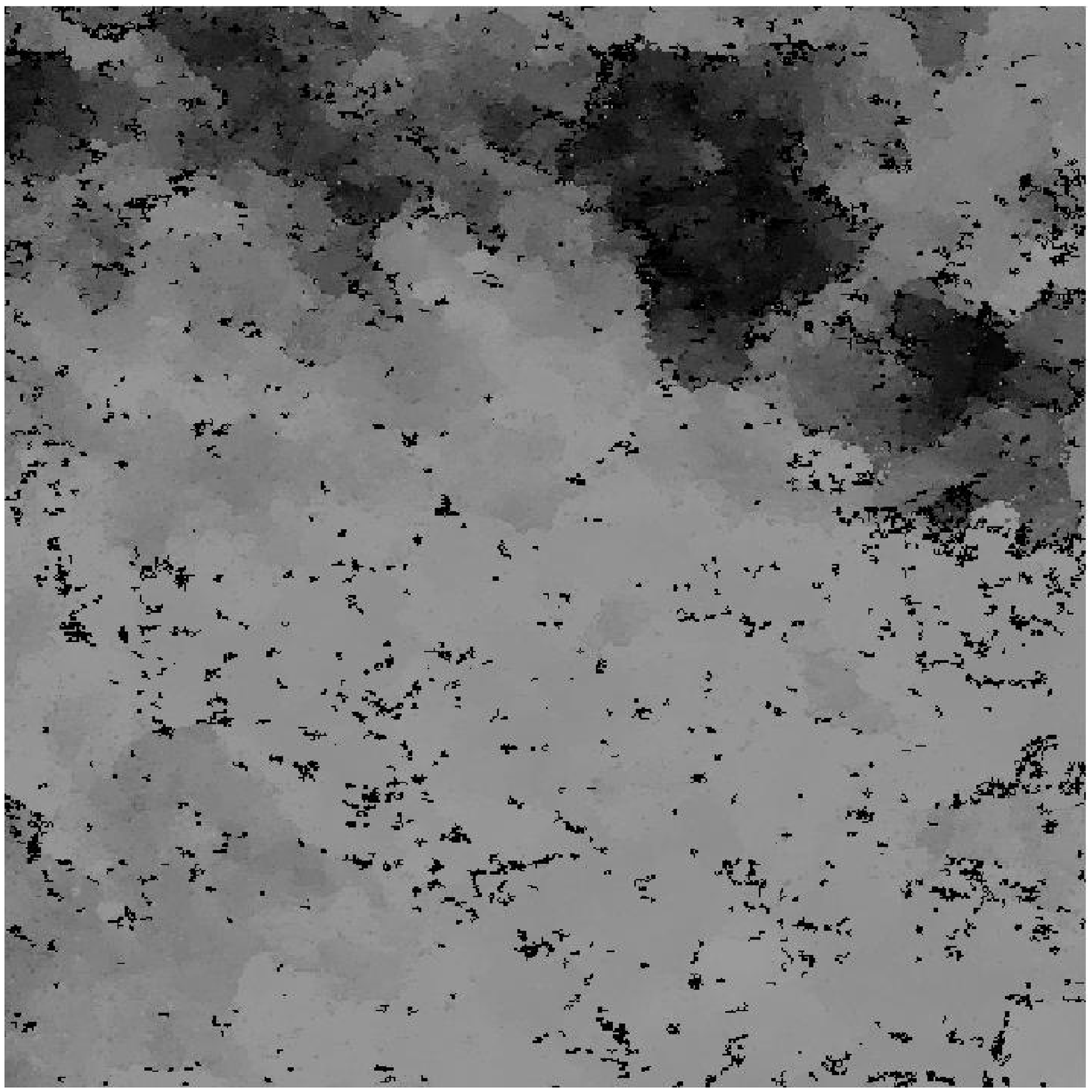}%
      \includegraphics[width=1.028\breite]{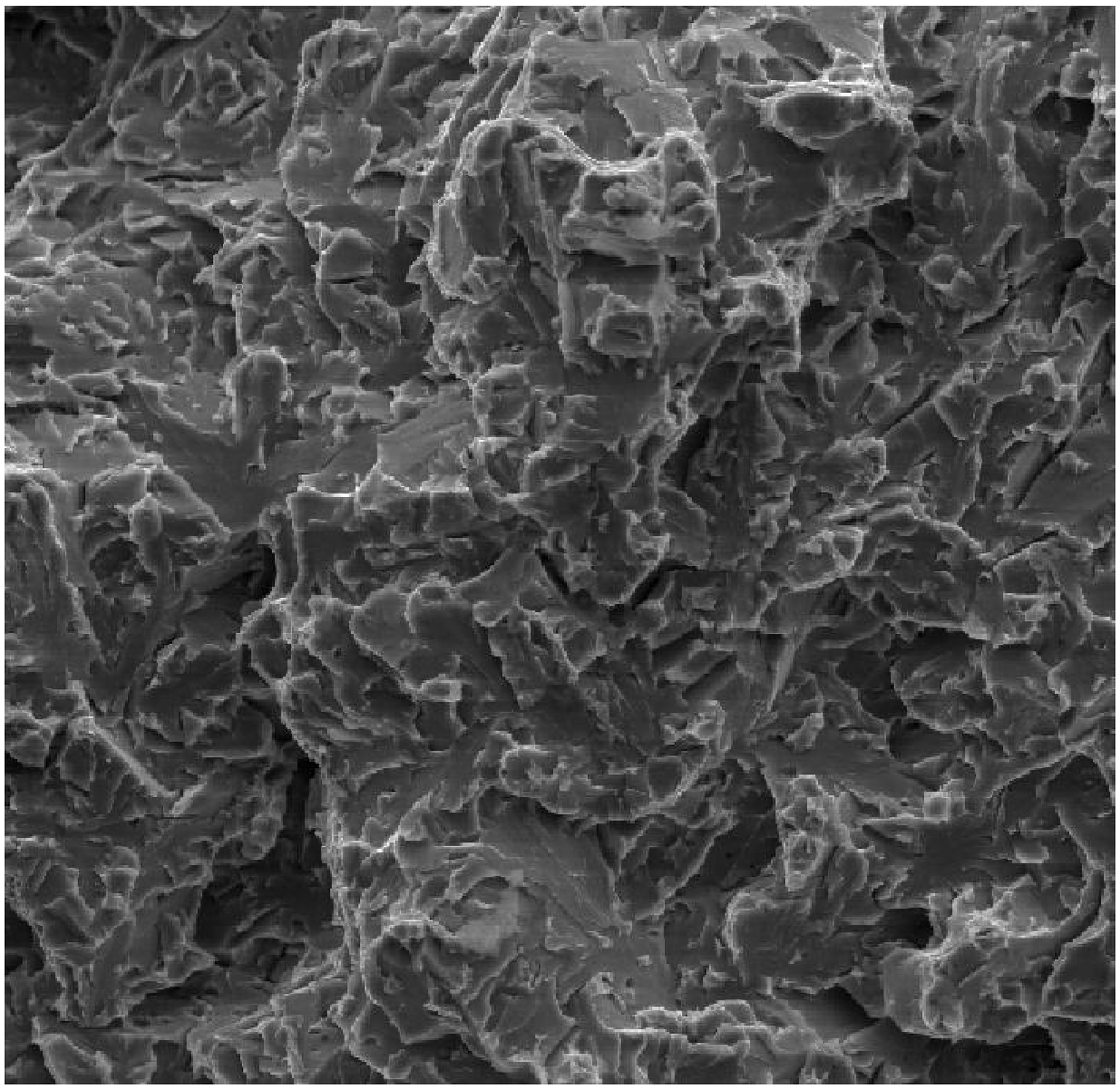}%
    }%
  \caption{%
    Measurement data from a steel crack surface (Crack).  (a) CLSM image, side
    length 502\mum, (b) REM image, side length 140\mum. }
  \label{fig:Wendt_data}
\end{figure}

\begin{figure}[htbp]\centering
  \includegraphics[width=0.8\linewidth]{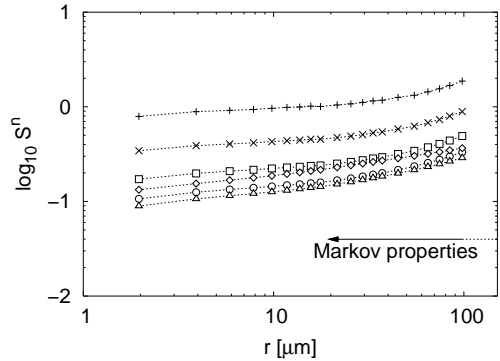}  
  \caption{%
    Structure functions $S^n(r)$ of the CLSM images from a steel crack surface
    (Crack) on a log-log scale. The symbols correspond to orders $n$ as in
    fig.~\ref{fig:Sn_road_scaling}.  }
  \label{fig:Sn_Wendt}
\end{figure}

\begin{figure}[htbp]\centering
  \setlength{\breite}{0.8\linewidth}%
  \begin{picture}(0,0)\put(8,29){\makebox{\small Road~5 (a)}}\end{picture}%
  \includegraphics[width=\breite]{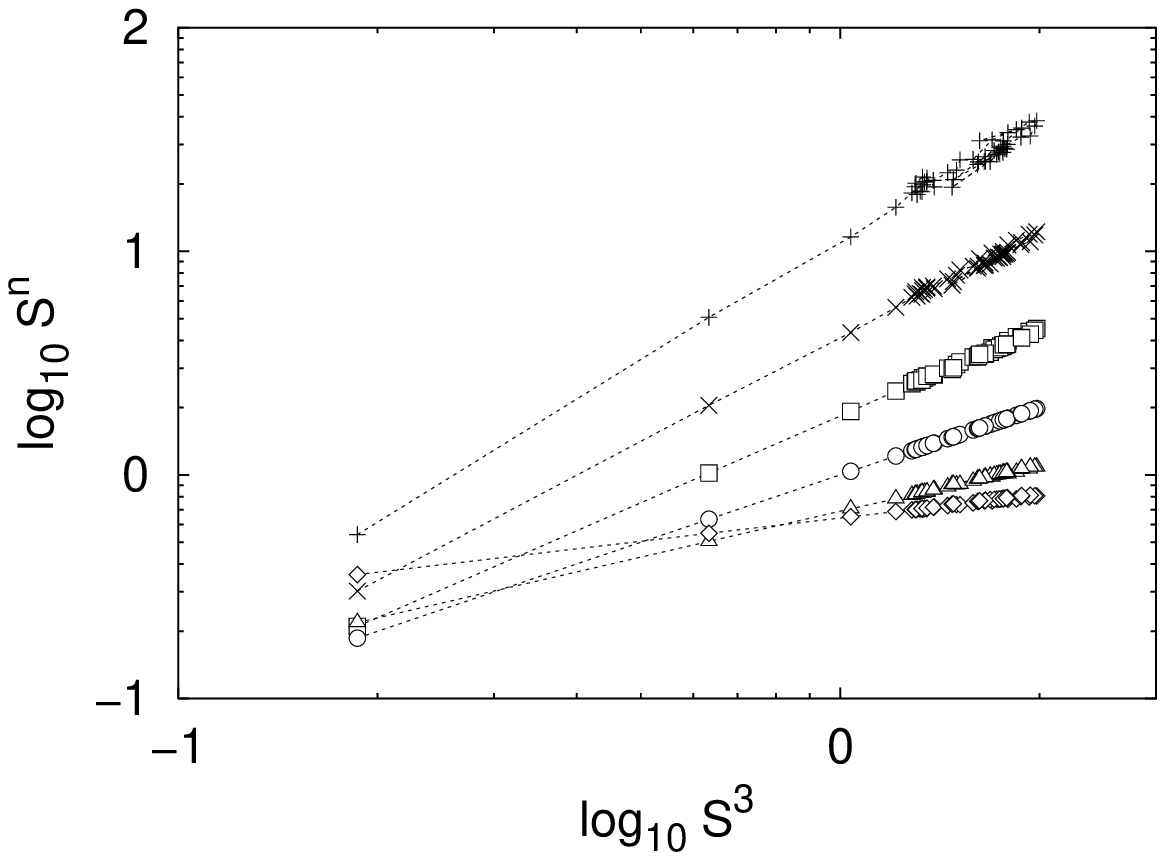}
  \begin{picture}(0,0)\put(8,29){\makebox{\small Crack (b)}}\end{picture}%
  \includegraphics[width=\breite]{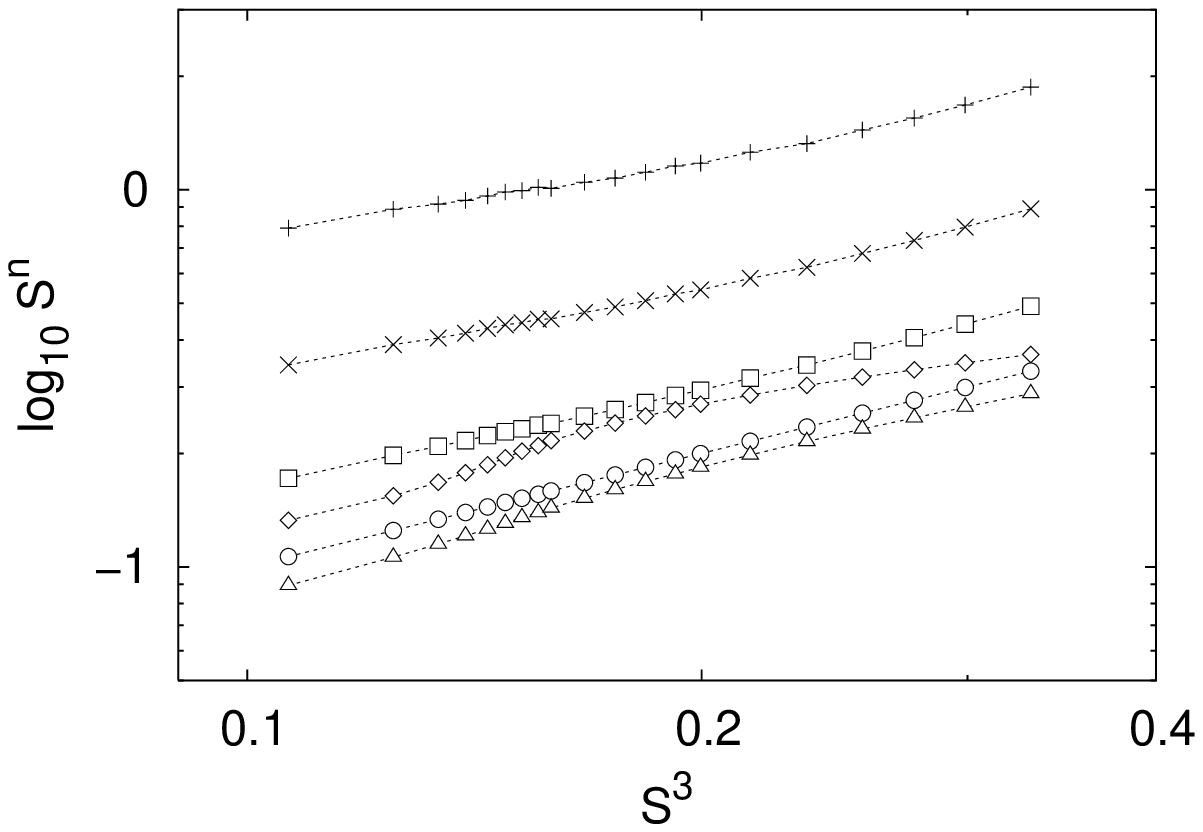}
  \caption{%
    Generalized scaling properties of the surfaces shown in
    figs.~\ref{fig:data_road_no_scaling} (Road~5 (a)) and \ref{fig:Wendt_data}
    (Crack (b)). Compare also with
    figs.~\ref{fig:data_road_scaling_ess} and \ref{fig:data_gold_ess}. }
  \label{fig:data_road_no_scaling_ess}
\end{figure}

The second example lacking a scaling regime is the steel fracture surface
(Crack).  One of the three CLSM images under investigation is shown in
fig.~\ref{fig:Wendt_data}(a).  Figure \ref{fig:Wendt_data}(b) presents an
additional REM image at a higher resolution, which gives an impression of the
surface morphology.
For the structure functions in fig.~\ref{fig:Sn_Wendt} no
scaling properties are found, and the dependences of $S^n(r)$ on $S^3(r)$ in
fig.~\ref{fig:data_road_no_scaling_ess}(b) also deviate from proper power laws.
It should be noted that in general scaling properties not only depend on the
respective data set but also on the analysis procedure.  Using other measures
than $h_r(x)$, in \cite{Wendt2002a} scaling regimes of those measures have
been found, and scaling exponents could be obtained.

\subsection{Conclusions on scaling analysis}
\label{sec:scaling_conclusions}

To conclude the scaling analysis of our examples, we have chosen surfaces with
a range of different scaling properties from good scaling over comparably wide
ranges, such as for Road~1 and Au, to the absence of scaling, such as for
Road~5 and Crack. The generalized scaling analysis, analogous to ESS
\cite{Benzi1993}, leads to the same scaling exponents as the dependence of the
structure functions $S^n(r)$ on the scale $r$, with some minor deviations.

\section{Markov properties}
\label{sec:markov_props}

As outlined in sections \ref{sec:intro} and \ref{sec:Markov_theory}, we want to
describe the evolution of the height increments $h_r(x)$ in the scale variable
$r$ as realizations of a Markov process with the help of a Fokker-Planck
equation. Consequently, the first step in the analysis procedure has to be the
verification of the Markov properties of $h_r(x)$ as a stochastic variable in
$r$.

For a Markov process the defining feature is that the $n$-scale conditional
probability distributions are equal to the single conditional probabilities,
according to eq.~(\ref{eq:markov_straight}).  With the given amount of data
points the verification of this condition is only possible for three different
scales. Additionally the scales $r$ are limited by the available profile
length.
For the sake of simplicity we will always take $r_3-r_2=r_2-r_1=\Delta r$.
Thus we can test the validity of eq.~(\ref{eq:markov_straight}) in the form
\begin{equation}
  \label{eq:markov_simple}
  p(h_1,r_1|\,h_2, r_1\!+\!\Delta r)=
  p(h_1,r_1|\,h_2, r_1\!+\!\Delta r; h_3=0, r_1\!+\!2\Delta r)\,.
\end{equation}
Note that in eq.~(\ref{eq:markov_simple}) we take $h_3=0$ to restrict the
number of free parameters in the pdf with double conditions.

Three procedures were applied to find out if Markov properties exist
for our data.  From the results of all three tests we will find a minimal
length scale $l_M$ for which this is the case. The meaning of this so-called
Markov length will be discussed below. In the following we will demonstrate
the methods using the example of the Au surface.

\subsection{Testing procedures}

The most straightforward way to verify eq.~(\ref{eq:markov_simple}) is the
visual comparison of both sides, i.e., the pdf with single and double
conditions. This is illustrated in fig.~\ref{fig:Schimmel_markov} for two
different scale separations $\Delta r=17\un{nm}$ and 35\un{nm}. In each case a
contour plot of single and double conditional probabilities $p(h_1,r_1 |
h_2,r_2)$ and $p(h_1,r_1 | h_2,r_2; h_3\!\!=\!\!0,r_3)$ is presented in the top
panel of (a) and (b), respectively.
Below two one-dimensional cuts at fixed values of $h_2 \approx \pm
\sigma_\infty$ are shown, representing directly
$p(h_1,r_1|h_2\!\!=\!\!\pm\sigma_\infty,r_2;h_3\!\!=\!\!0,r_3)$.
It can be seen that in panel (a), for the smaller value of $\Delta r$, the
single and double conditional probability are different. This becomes clear
from the crossing solid and broken contour lines of the contour plot as well
as from the differing lines and symbols of the one-dimensional plots below.
Panel (b), for $\Delta r=35$\un{nm}, shows good correspondence of both
conditional pdf. We take this finding as a strong hint that for this scale
separation $\Delta r$ eq.~(\ref{eq:markov_straight}) is valid and Markov
properties exist. Following this procedure for all accessible values of
$\Delta r$, the presence of Markov properties was examined. For this surface
Markov properties were found for scale distances from $(25\pm 5)\un{nm}$
upwards.

\setlength{\breite}{0.66\linewidth}
\begin{figure}[htbp]\centering
  \begin{picture}(0,0)\put(0,49){(a)}\end{picture}%
  \includegraphics[width=\breite]{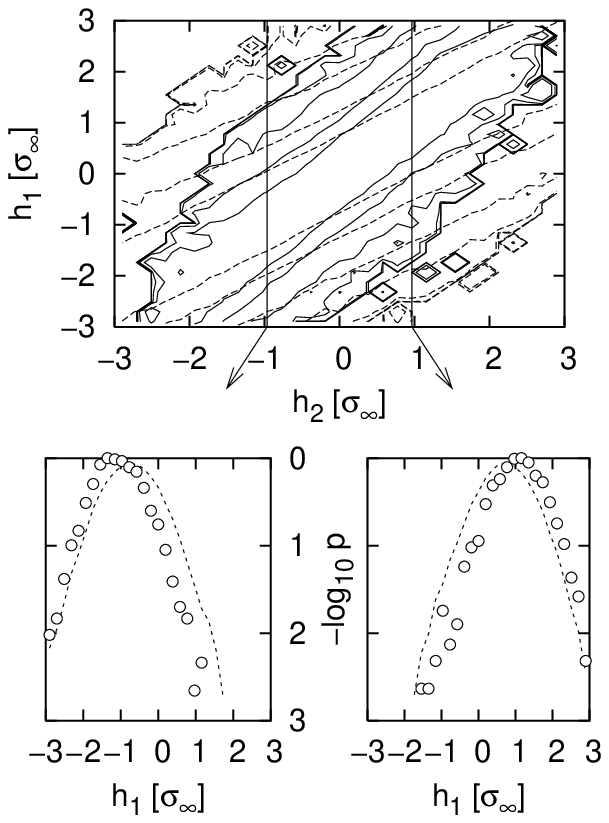}\\
  \begin{picture}(0,0)\put(0,49){(b)}\end{picture}%
  \includegraphics[width=\breite]{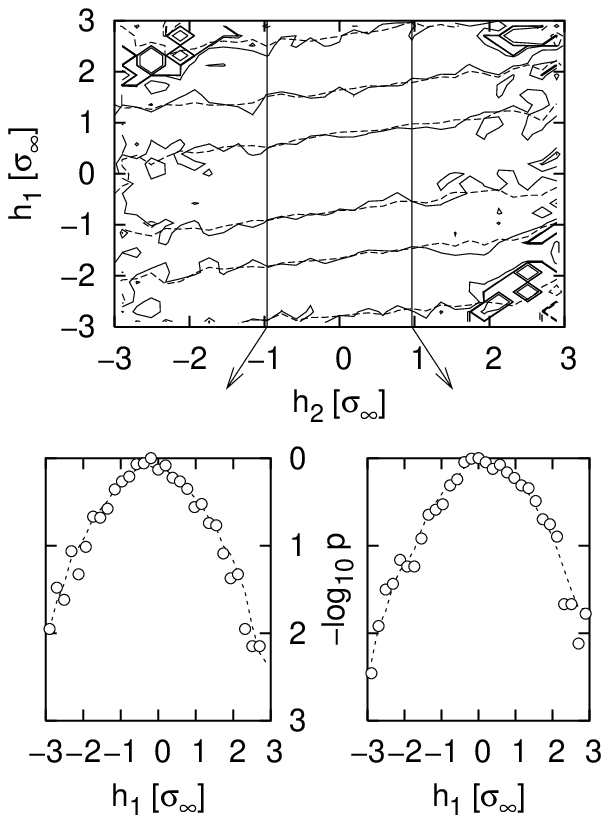}
  \caption{%
    Test for Markov properties of Au film data for two different scale
    separations $\Delta r=14$\un{nm} (a) and $35$\un{nm} (b), where
    \mbox{$\Delta r=r_3-r_2=r_2-r_1$} (see text).  In both cases
    $r_2=169\un{nm}$.
    In each case a contour plot of conditional probabilities $p(h_1,r_1 |
    h_2,r_2)$ (dashed lines) and $p(h_1,r_1 | h_2,r_2;h_3\!\!=\!\!0,r_3 )$
    (solid lines) is shown in the top panel.  Contour levels differ by a factor
    of 10, with an additional level at $p=0.3$.  Below the top panels in each
    case, two one-dimensional cuts
    at $h_2 \approx \pm \sigma_\infty$ are shown with $p(h_1,r_1 | h_2,r_2)$ as
    dashed lines and $p(h_1,r_1 | h_2,r_2;h_3\!\!=\!\!0,r_3 )$ as circles.  }
  \label{fig:Schimmel_markov}
\end{figure}

The validity of eq.~(\ref{eq:markov_simple}) can also be be quantified
mathematically using statistical tests. An approach via the well-known
$\chi^2$ measure has been presented in \cite{Friedrich1998b}, whereas in
\cite{Renner2001} the Wilcoxon test has been used. Next, we give a brief
introduction to this procedure, which will be used here, too. More detailed
discussions of this test can be found in
\cite{Bronstein1991,Renner2001,Renner2001Diss}. For this procedure, we
introduce the notation of two stochastic variables $x_i,i=1,\ldots,n$ and
$y_j,j=1,\ldots,m$ which represent the two samples from which both conditional
pdf of eq.~(\ref{eq:markov_straight}) are estimated, i.e.\
\begin{eqnarray}
  \label{eq:wilcox_xy}
  x(h_{r_2}, r_1, r_2)           &=& h_{r_1}|_{h_{r_2}}   \nonumber\\
  y(h_{r_2}, h_{r_3}, r_1, r_2, r_3) &=& h_{r_1}|_{h_{r_2}; h_{r_3}} \;.
\end{eqnarray}
Here $\cdot|_{h_x}$ denotes the conditioning.
All events of both samples are sorted together in ascending order into one
sequence, according to their value. Now the total number of so called
inversions is counted, where the number of inversions for a single event $y_j$
is just the number of events of the other sample which have a smaller value
$x_i < y_j$. If eq.~(\ref{eq:markov_simple}) holds and $n,m\geq 25$, the
total number of inversions $Q$ is Gaussian distributed with 
\begin{eqnarray}
  \label{eq:Q_mean_sigma}
  \langle Q\rangle &=& nm/2 \quad\mathrm{and} \nonumber \\
  \sigma_Q         &=& \sqrt{nm(n+m+1)/12}\; .
\end{eqnarray}
We normalize $Q$ with respect to its standard deviation and consider the
absolute value
\begin{equation}
  \label{eq:wilcox_t}
  t = | Q - \langle Q\rangle | / \sigma_Q \; .
\end{equation}
For its expectation value it is easy to show that $\langle
t\rangle=\sqrt{2/\pi}$ (still provided that (\ref{eq:markov_simple}) is valid),
where here the average $\langle\cdot\rangle$ is performed over $h_2$. If a
larger value of $\langle t\rangle$ is measured for a specific combination of
$r$ and $\Delta r$, we conclude that eq.~(\ref{eq:markov_simple}) is not
fulfilled and thus Markov properties do not exist. A practical problem with
the Wilcoxon test is that all events $x_i,y_j$ have to be statistically
independent. This means that the intervals of subsequent height increments
$h_r$ have to be separated by the largest scale involved. Thus the number of
available data is dramatically reduced. 

\setlength{\breite}{0.8\linewidth}
\begin{figure}[htbp]
  \centering
  \includegraphics[width=\breite]{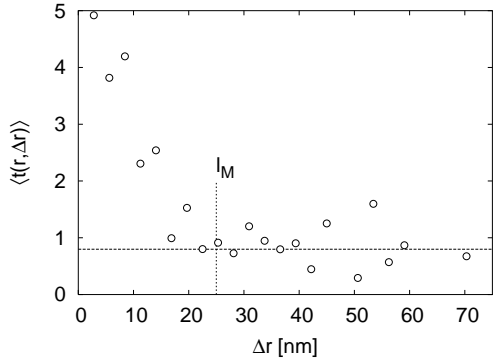}
  \caption{%
    Wilcoxon test for the Au surface. The scale $r$ is 28\un{nm}. The
    theroretically expected value $\langle t\rangle=\sqrt{2/\pi}$ is marked
    with a horizontal line, the Markov length $l_M=(25\pm5)\un{nm}$ with a
    vertical line.  }
  \label{fig:Schimmel_wit}
\end{figure}

In fig.~\ref{fig:Schimmel_wit} we present for the Au surface measured values
of $\langle t(r,\Delta r)\rangle$ at a scale $r=28\un{nm}$. The Markov length
$l_M$ is marked where $\langle t(r,\Delta r)\rangle$ has approached its
theroretical value $\sqrt{2/\pi}$.

Another method to show the validity of condition (\ref{eq:markov_simple}) is
the investigation of the well-known necessary condition for a Markovian
process, the validity of the Chapman-Kolmogorov equation \cite{Risken1984}
\begin{equation}
  \label{eq:CKE}
  p(h_1, r_1|h_3,r_3) = 
  \int_{-\infty}^{+\infty} p(h_1, r_1|h_2,r_2)p(h_2, r_2|h_3,r_3)\mbox{d}h_2
  \;.
\end{equation}
We use this equation as a method to investigate the Markov properties of our
data.  This procedure was used for example in
\cite{Friedrich1997a,Friedrich1997b,Ragwitz2001,Jafari2003} for the
verification of Markov properties. It also served to show for the first time
the existence of a Markov length in \cite{Friedrich1998b}.
The conditional probabilities in eq.~(\ref{eq:CKE}) are directly
estimated from the measured profiles. In
fig.~\ref{fig:Schimmel_markov_chapman} both sides of the Chapman-Kolmogorov
equation are compared for two different values of $\Delta r$.  In an analogous
way to fig.~\ref{fig:Schimmel_markov}, for each $\Delta r$ the two conditional
probabilities are presented together in a contour plot as well as in two cuts
at fixed values of $h_2$. While for the smaller value $\Delta r=14$\un{nm}
both the contour lines and the cuts at fixed $h_2$ clearly differ, we find
a good correspondence for the larger value $\Delta r=35$\un{nm}.

\setlength{\breite}{0.66\linewidth}
\begin{figure}[htbp]
  \centering
  \begin{picture}(0,0)\put(0,49){(a)}\end{picture}%
  \includegraphics[width=\breite]{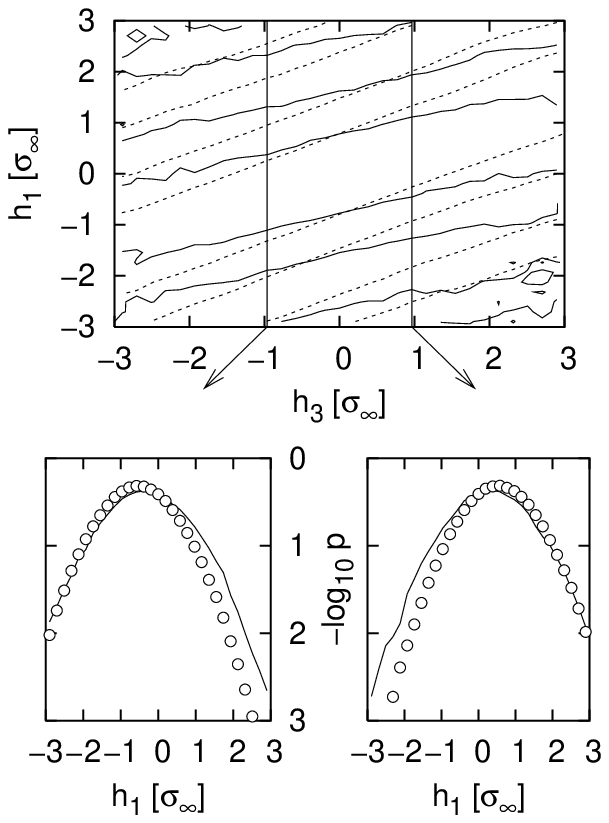}
  \begin{picture}(0,0)\put(0,49){(b)}\end{picture}%
  \includegraphics[width=\breite]{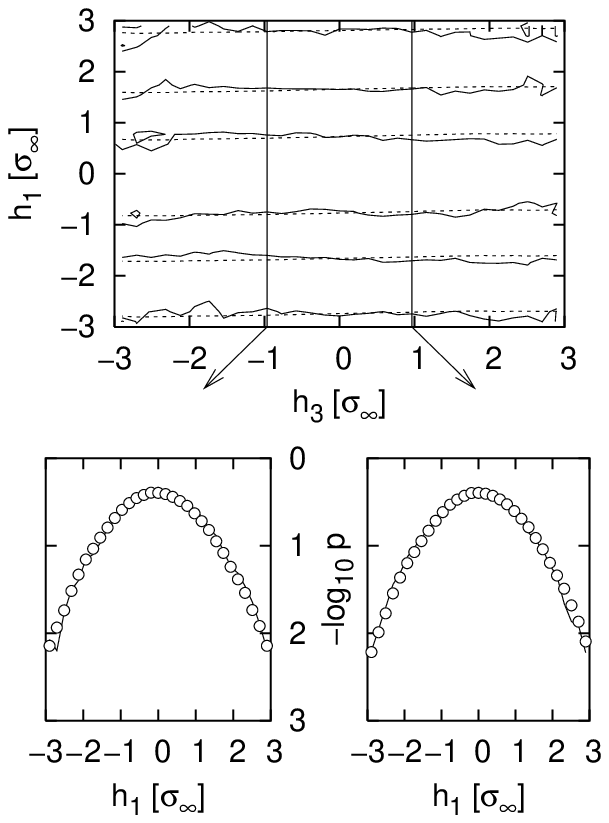}
  \caption{%
    A check of the Chapman-Kolmogorov equation (\ref{eq:CKE}) for the Au
    surface for two different scale separations $\Delta r=14$\un{nm} (a) and
    35\un{nm} (b). In both cases $r_2=169$\un{nm}.  The plots are organized in
    the same way as fig.~\ref{fig:Schimmel_markov}.  The pdf representing the
    left side of (\ref{eq:CKE}) are shown with solid lines, the integrated pdf
    of the right side of (\ref{eq:CKE}) as dashed lines and circles.  }
  \label{fig:Schimmel_markov_chapman}
\end{figure}

A third method which we did not use here but which is reported in the
literature is based on the description of the stochastic process by a Langevin
equation. With this knowledge of the Langevin equation (\ref{eq:Langevin}) the
noise can be reconstructed and analyzed with respect to its correlation
\cite{Siefert2003,Marcq2001}.

\subsection{Conclusions on Markov properties}
\label{sec:markov_conclusions}

The results of the methods described above were combined to determine
whether Markov properties of the height increment $h_r(x)$ in the scale
variable are present for our surface measurements. 
We found Markov properties for all the selected examples of surface measurement
data. 
\begin{table}[htbp]
  \newcommand{\mm}{\un{mm}}
  \begin{center}
    \begin{tabular}{cc|cc|cc}
      Surface & $l_M$ & Surface & $l_M$ & Surface & $l_M$ \\
      \hline
      Road~1  & 33\mm & Road~2  & 10\mm & Road~3  & 4.2\mm \\
      Road~4  & 17\mm & Road~5  & 4.2\mm& Au      & 25\un{nm}\\
      Crack   & 20\mum&         &       &         &        
    \end{tabular}
    \caption{Markov lengths $l_M$ for all the surfaces presented.}
    \label{tab:l_M}
  \end{center}
\end{table}
It is also common to all examples that these Markov properties are not
universal for all scale separations $\Delta r$ but there exists a lower
threshold which we call the Markov length $l_M$. It was determined in each
case by systematic application of the three testing procedures for all
accessible length scales $r$ and scale separations $\Delta r$. 
The resulting values are listed in table~\ref{tab:l_M}. The presence of
Markov properties only for values of $\Delta r$ above a certain threshold has
also been found for stochastic data generated by a large variety of processes
and especially occurs in turbulent velocities
\cite{Friedrich1998a,Friedrich1998b,Friedrich2000b,Renner2001,Ghasemi2003}.

The meaning of this Markov length $l_M$ may be seen in comparison with a mean
free path length of a Brownian motion. Only above this mean free path is a
stochastic process description valid. For smaller scales there must be some
coherence which prohibits a description of the structure by a Markov process.
If for example the description of a surface structure requires a second order
derivative in space, a Langevin equation description (\ref{eq:Langevin})
becomes impossible. In this case a higher dimensional Langevin equation (at
least two variables) is needed. It may be interesting to note that the Markov
length we found for the Au surface of about 30\un{nm} coincides quite well with
the size of the largest grain structures we see in fig.~\ref{fig:data_gold}.
Thus the Au surface may be thought of as a composition of grains (coherent
structures) by a stochastic Markov process.

In the case of Road~2 with its strong periodicity at 0.2\un{m} the Markov
properties end slightly above this length scale. It seems evident that here
the Markov property is destroyed by the periodicity. While some of the other
surfaces also have periodicities, these are never as sharp as for Road~2. An
upper limit for Markov properties could not be found for any of the other
surfaces.
 
Another interesting finding can be seen from figs.~\ref{fig:Sn_road_scaling},
\ref{fig:Sn_gold}, \ref{fig:data_road_no_scaling}, and \ref{fig:Sn_Wendt}.
There is no connection between the scaling range and the range where Markov
properties hold. Regimes of scaling and Markov properties are found to be
distinct, overlapping or covering, depending on the surface.  Data sets which
fulfill the Markov property do not in all cases show a scaling regime at all.
Also, on the other hand, scaling features seem not to imply Markov properties,
which has been indicated previously for some numerically generated data in
\cite{Friedrich1998a}. While there is always an upper limit of the scaling
regime, we found only one surface for which the Markov properties possess an
upper limit.

\section{Estimation of drift and diffusion coefficients}
\label{sec:Dk_estimation}

As a next step we want to concentrate on extracting the concrete form of the
stochastic process, if the Markov properties are fulfilled.
As mentioned in section~\ref{sec:Markov_theory} our analysis is based on the
estimation of Kramers-Moyal coefficients.
The procedure we use to obtain the drift ($D^{(1)}$) and diffusion coefficient
($D^{(2)}$) for the Fokker-Planck equation (\ref{eq:FPE1}) was already outlined
by Kolmogorov \cite{Kolmogorov1931}, see also \cite{Risken1984,Renner2001}.
First, the conditional moments $M^{(k)}(h_r,r,\Delta r)$ for finite step sizes
$\Delta r$ are estimated from the data via the moments of the conditional
probabilities.
This is done by application of the definition in eq.~(\ref{eq:Mk_def}),
which is recalled here:
%
\begin{equation}
   \label{eq:Mk_2}
   M^{(k)}(h_r,r,\Delta r) =
     \frac{r}{k!\Delta r}
     \int_{\scriptscriptstyle-\infty}^{\scriptscriptstyle+\infty}
     (\tilde{h}-h_r)^k \; p(\tilde{h},r-\Delta r | h_r,r) \;  d\tilde{h}
\end{equation}
The conditional probabilities in the integral are obtained by counting events
in the measurement data as shown already in section~\ref{sec:markov_props}.
Here, one fundamental difficulty of the method arises: For reliable estimates
of conditional probabilities we need a sufficient number of events even for
rare combinations of $\tilde{h}, h_r$. Consequently, a large amount of data
points is needed.
This problem becomes even more important if one takes into account that a large
range in $r$ should be considered. The number of statistically independent
intervals $h_r$ is limited by the length of the given data set and decreases
with increasing $r$.

In a second step, the coefficients $D^{(k)}(h_r,r)$ are obtained from the
limit of $M^{(k)}(h_r,r,\Delta_r)$ when $\Delta r$ approaches zero (see
definition in eq.~(\ref{eq:Dk_def})).
For fixed values of $r$ and $h_r$ a straight line is fitted to the sequence of
$M^{(k)}(h_r,r,\Delta r)$ depending on $\Delta r$ and extrapolated against
$\Delta r=0$. 
The linear dependence corresponds to the lowest order term when the $\Delta
r$-dependence of $M^{(k)}(h_r,r,\Delta r)$ is expanded into a Taylor series
for a given Fokker-Planck equation \cite{Friedrich2002, Siefert2003}.
Our interpretation is that this way of estimating the $D^{(k)}$ is the most
advanced one, and also performs better than first parameterizing the $M^{k}$
and then estimating the limit $\Delta r\rightarrow 0$ for this
parameterization, as previously suggested in \cite{Renner2001,Renner2001Diss}.

There have been suggestions to fit functions to $M^{(k)}$ other than a straight
line, especially for the estimation of $D^{(2)}$, see \cite{Renner2001Diss}.
Furthermore it has been proposed to use particular terms of the above-mentioned
expansion to directly estimate $D^{(k)}(h_r, r)$ without
extrapolation \cite{Ragwitz2001}.
On the other hand, in \cite{Sura2002} it becomes clear that there can be
manifold dependences of $M^{(k)}$ on $\Delta r$ which in general are not known
for a measured data set. Consequently, one may state that there is still a
demand to improve the estimation of $D^{(k)}$. At the present time we
suggest to show the quality of the estimated $D^{(k)}$ by verification of the
resulting Fokker-Planck equation, once its drift and diffusion coefficients
have been estimated.
However, for our data neither nonlinear fitting functions nor correction terms
applied to the $M^{(k)}$ resulted in improvements of the estimated $D^{(k)}$.

A crucial point in our estimation procedure is the range of $\Delta r$ where
the fit can be performed. Only those $\Delta r$ can be used where Markov
properties were found in the scale domain. In section \ref{sec:markov_props} we
showed that for our data Markov properties are given for $\Delta r$ larger than
the Markov length $l_M$ (see table \ref{tab:l_M}).  In order to reduce
uncertainty, a large range of $\Delta r$ as the basis of the extrapolation is
desirable.  From eq.~(\ref{eq:Mk_2}) it can be seen, however, that $\Delta r$
must be smaller than $r$.  As a compromise between accuracy and extending the
scale $r$ to smaller values, in many cases an extrapolation range of $l_M \leq
\Delta r \leq 2\,l_M$ was used (cf.\ table \ref{tab:l_extrapol}).  This
procedure is shown in fig.~\ref{fig:drh_ex} for Road~1.

\begin{table}[htbp]
  \centering
  \newcommand{\mm}{\un{mm}}
  \newcommand{\lmax}{\ensuremath{l_{\mathit{max}}}}
    \begin{tabular}{ccc|ccc}
      Surface& $l_M$ & \lmax& Surface& $l_M$ & \lmax \\
      \hline
      Road~1 & 33\mm & 67\mm& Road~2 & 10\mm & 21\mm \\
      Road~3 & 4.2\mm& 19\mm& Road~4 & 17\mm & 25\mm \\ 
      Road~5 & 4.2\mm&8.3\mm& Au   &25\un{nm}&84\un{nm}\\
      Crack  & 20\mum&44\mum&        &       &       
    \end{tabular}
    \caption{Extrapolation ranges for all the presented surfaces. Listed are
      the smallest ($l_M$) and largest ($l_\mathit{max}$) values of $\Delta r$
      used for extrapolation of the $D^{(k)}(h_r, r)$.  }
    \label{tab:l_extrapol}
\end{table}

\setlength{\breite}{0.8\linewidth}
\begin{figure}[htbp]
  \centering
  \begin{picture}(0,0)\put(0,30){\makebox{\small (a)}}\end{picture}%
  \includegraphics[width=\breite]{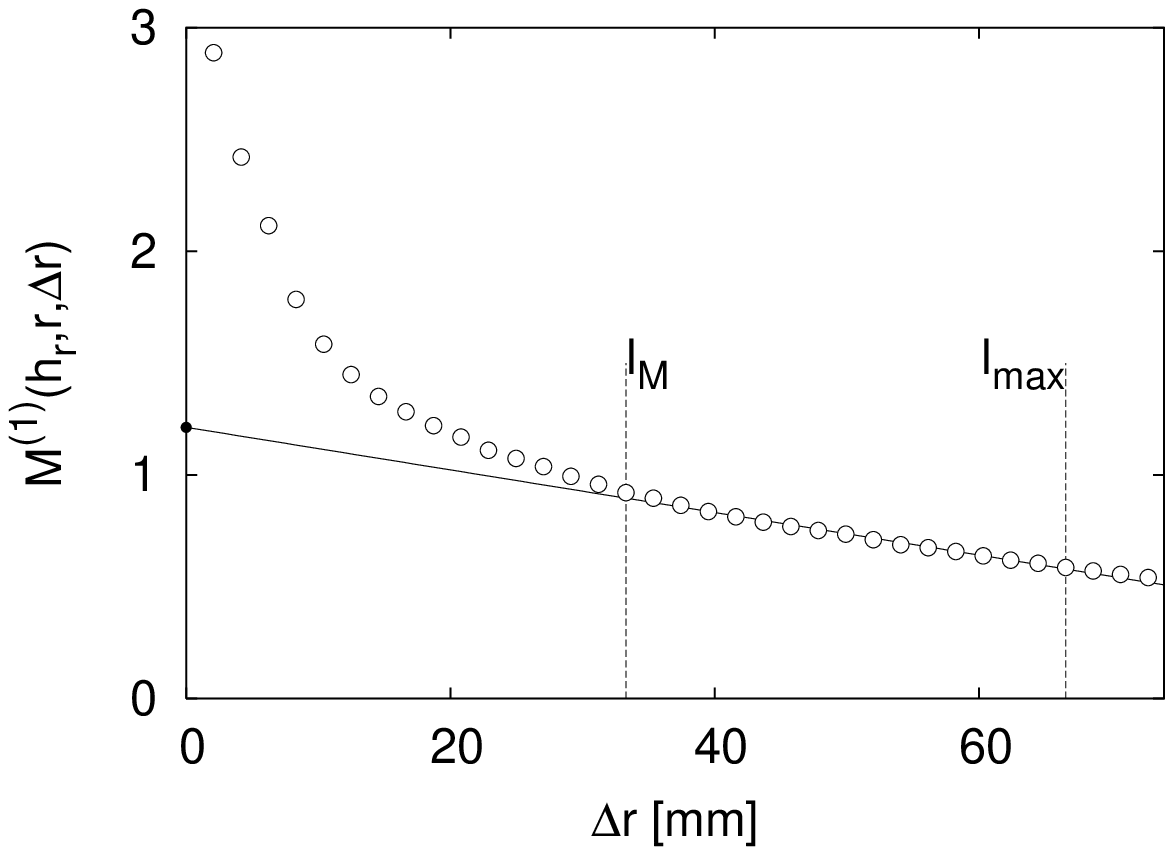}
  \begin{picture}(0,0)\put(0,30){\makebox{\small (b)}}\end{picture}%
  \includegraphics[width=\breite]{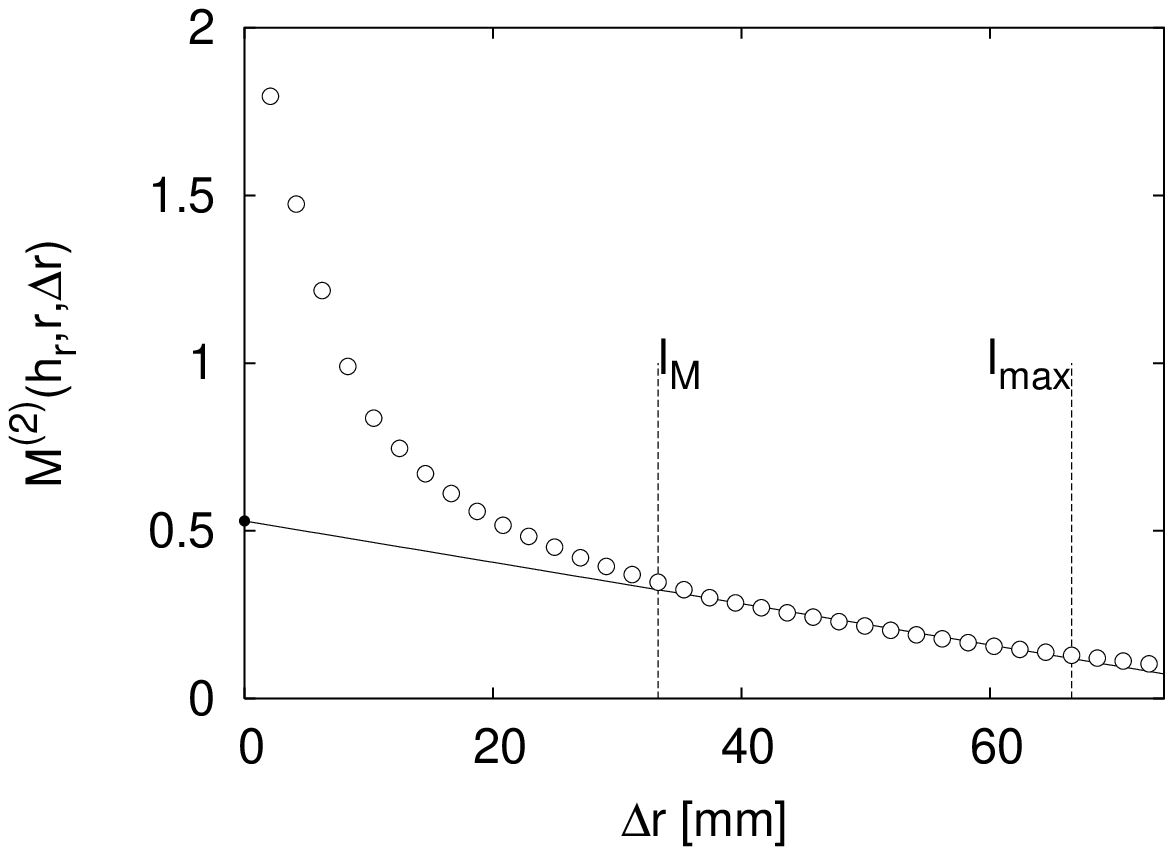}
  \caption[Extrapolation procedure for $D^{(1)}$.]{%
    Extrapolation procedure for $D^{(1)}$ (a) and $D^{(2)}$ (b), illustrated
    for surface Road~1.  Length scale $r$ is 108\un{mm}, $h_r$ is
    $-\sigma_\infty$. Values of $M^{(1)}$ and $M^{(2)}$ inside the range
    marked by broken lines were used for the extrapolation. The results
    $D^{(1)}(h_r,r)$ and $D^{(2)}(h_r,r)$ are marked with filled circles on
    the vertical axis.  }
  \label{fig:drh_ex}
\end{figure}

\subsection{Estimation results}
\label{sec:Dk_results}

Following the procedure outlined above, $D^{(1)}(h_r,r)$ and $D^{(2)}(h_r,r)$
were derived for the measurement data presented in section
\ref{sec:measurement_data}, with the exception of Road~5 (see below).
For the road surfaces, estimations were performed for length scales $r$
separated by ten measurement steps or 10.4\un{mm}, respectively, to reach a
sufficient density over the range where the coefficients were accessible.
Figures~\ref{fig:D1_road_scaling} and \ref{fig:D2_road_scaling} show
estimations of the drift coefficients $D^{(1)}$ and the diffusion coefficients
$D^{(2)}$ for the road surfaces, each performed for one fixed length scale
$r$.  The error bars are estimated from the errors of $M^{(k)}(h_r,r,\Delta
r)$ via the number of statistically independent events contributing to each
value, assuming that each bin of $p(h_1,r_1|h_0,r_0)$ containing $N$ events
has an intrinsic uncertainty of $\pm\sqrt{N}$.
Additionally, values of $D^{(4)}$ are added to the plots of $D^{(2)}$ which
have been estimated in the same way. Thus it can be seen that in all cases
$D^{(4)}$ is small compared to $D^{(2)}$, except for Road~4, and in most cases
its statistical errors are larger than the values themselves. Negative values
are not shown because the vertical axes start at zero. As $M^{(4)}$ is positive
by definition, the occurence of negative values of $D^{(4)}$ results from the
limit $\Delta r\rightarrow 0$ and should be only due to the statistical errors
involved.
Even if there is no evidence that $D^{(4)}$ is identically zero, the presented
values give a hint that its influence in the Kramers-Moyal expansion
(\ref{eq:KME}) is rather small and the assumption of a Fokker-Planck equation
(\ref{eq:FPE1}) is justifiable, with the possible exception of Road~4. 

\setlength{\breite}{0.8\linewidth}
\begin{figure}[htbp]
  \centering
  \begin{picture}(0,0)\put(35,28){\makebox{\small Road~1}}\end{picture}%
  \includegraphics[width=\breite]{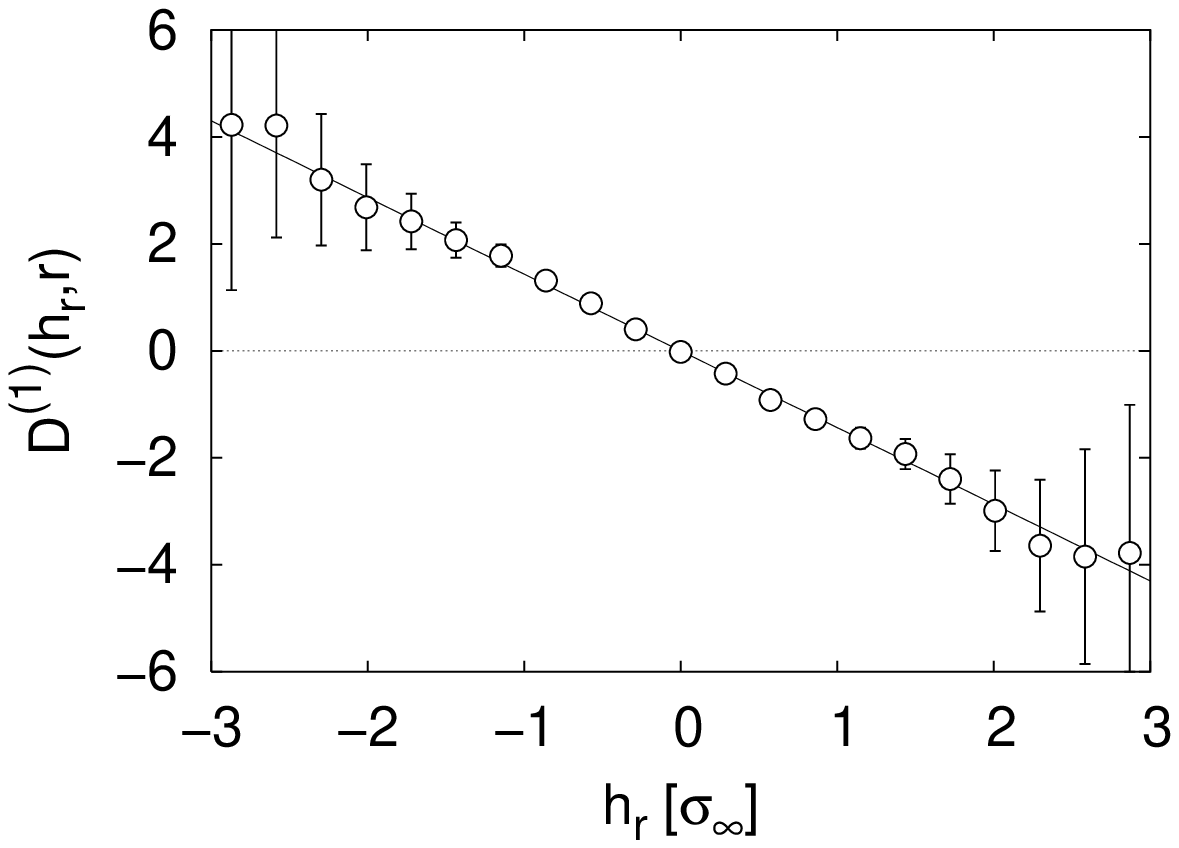}
  \begin{picture}(0,0)\put(35,28){\makebox{\small Road~2}}\end{picture}%
  \includegraphics[width=\breite]{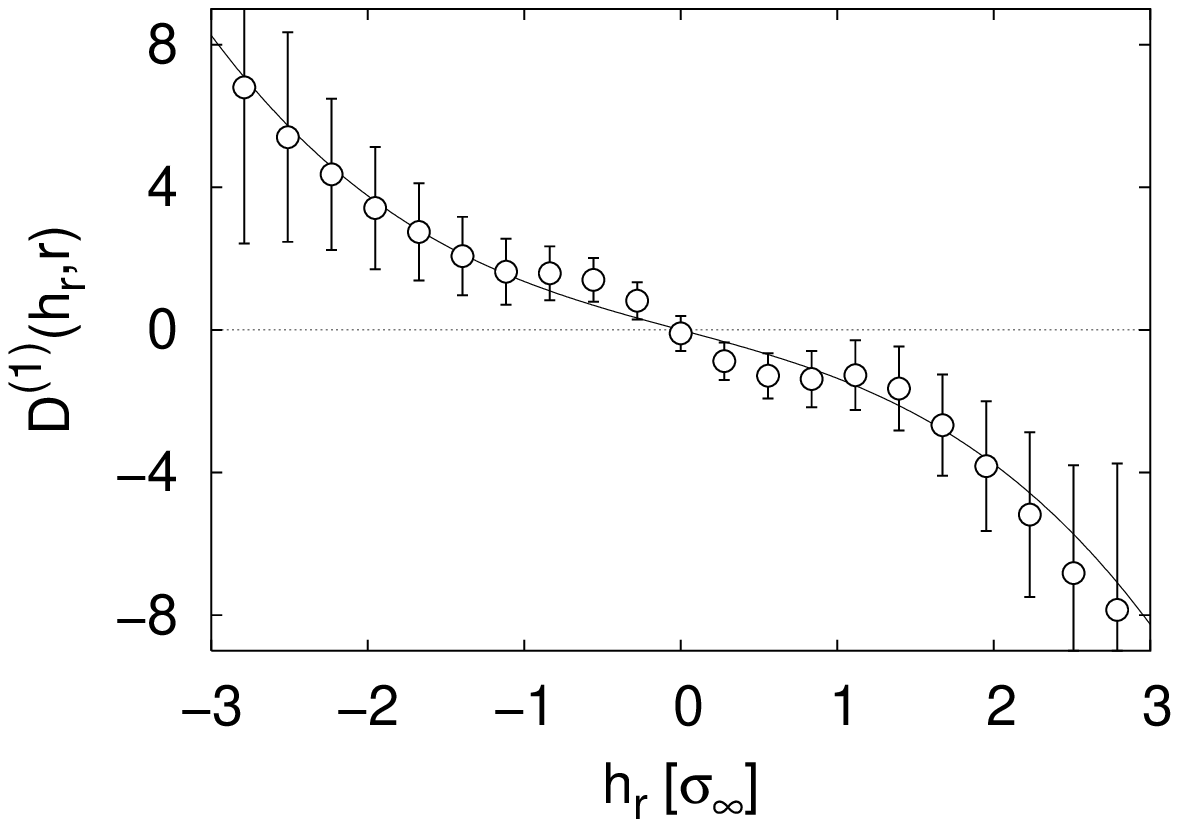}
  \begin{picture}(0,0)\put(35,28){\makebox{\small Road~3}}\end{picture}%
  \includegraphics[width=\breite]{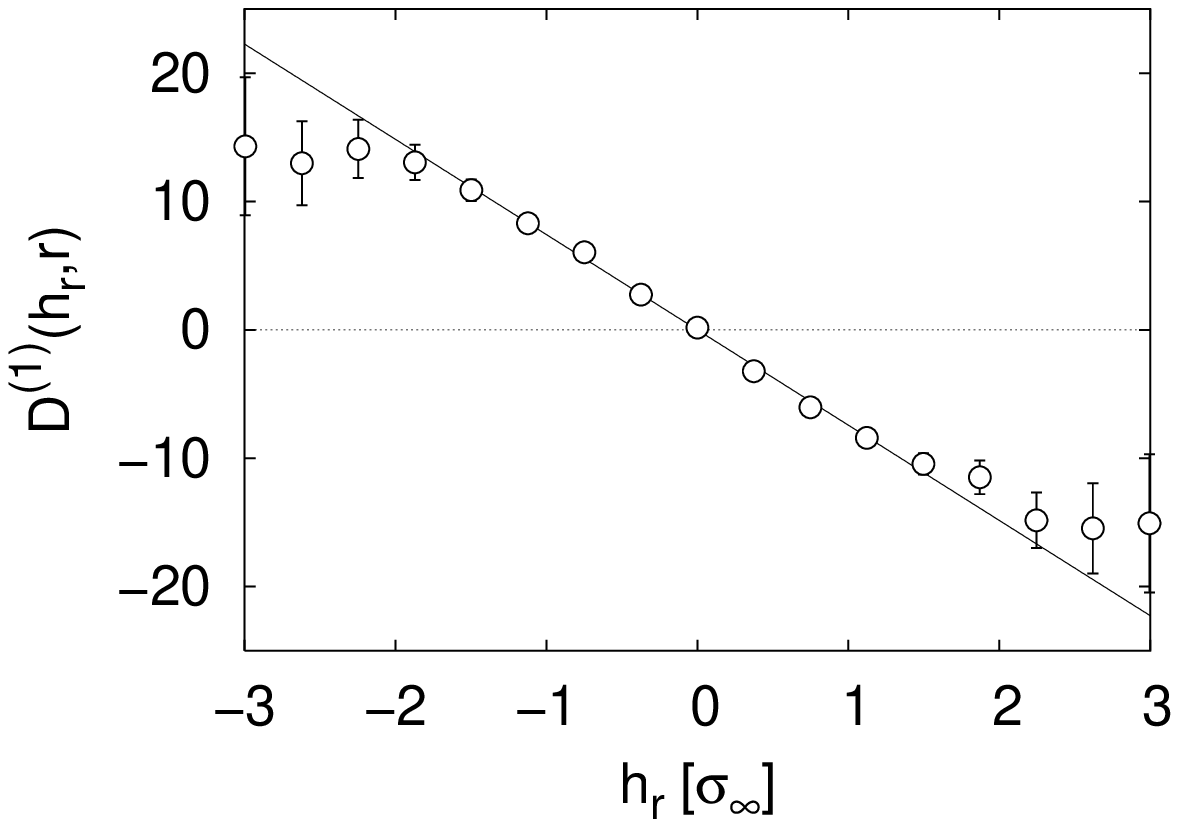}
  \begin{picture}(0,0)\put(35,28){\makebox{\small Road~4}}\end{picture}%
  \includegraphics[width=\breite]{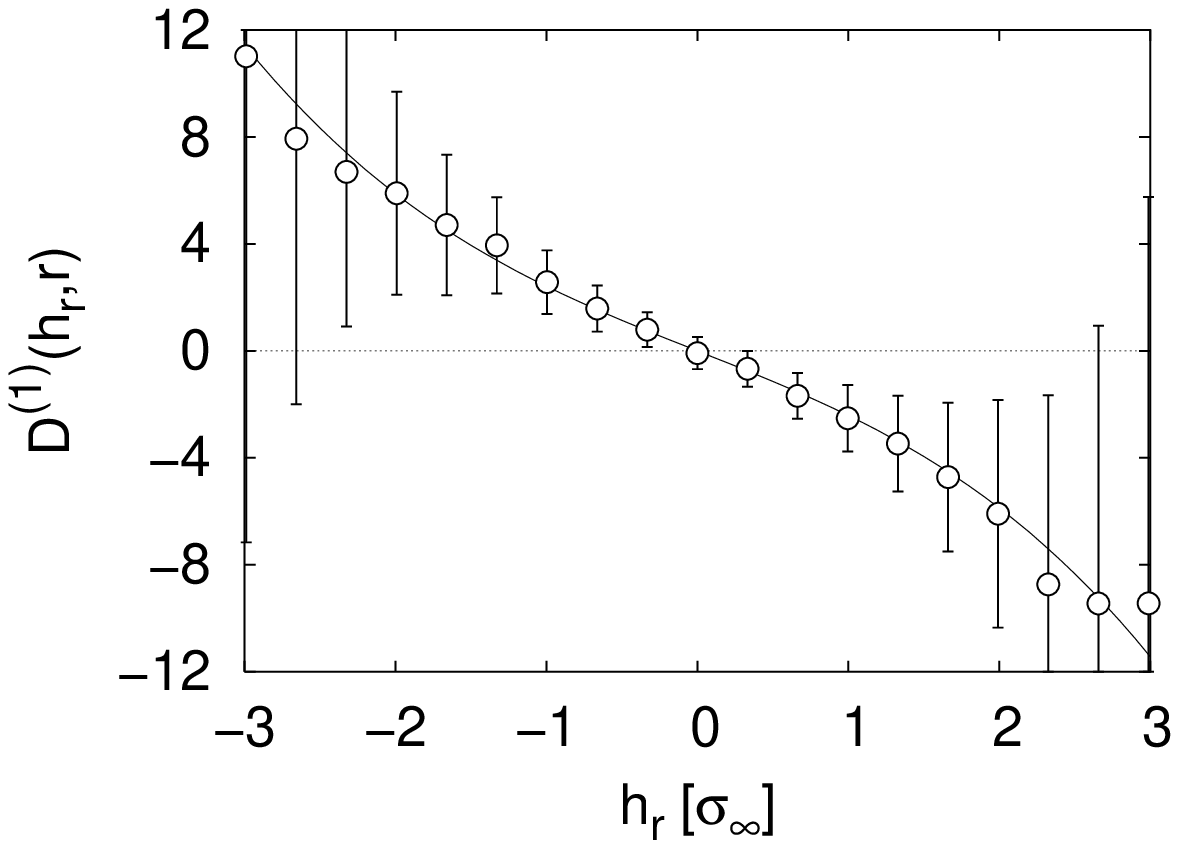}
  \caption[Estimated drift coefficient $D^{(1)}(h_r, r)$]%
  {Estimated drift coefficients $D^{(1)}(h_r, r)$ of the Fokker-Planck
    equation for the road surfaces shown in fig.~\ref{fig:data_road_scaling}.
    Scales $r$ are 108\un{mm} (Road~1), 114\un{mm} (Road~2), 94\un{mm}
    (Road~3), and 104\un{mm} (Road~4).  Parameterizations are shown as lines.}
  \label{fig:D1_road_scaling}
\end{figure}

\begin{figure}[htbp]
  \centering
  \begin{picture}(0,0)\put(23,28){\makebox{\small Road~1}}\end{picture}%
  \includegraphics[width=\breite]{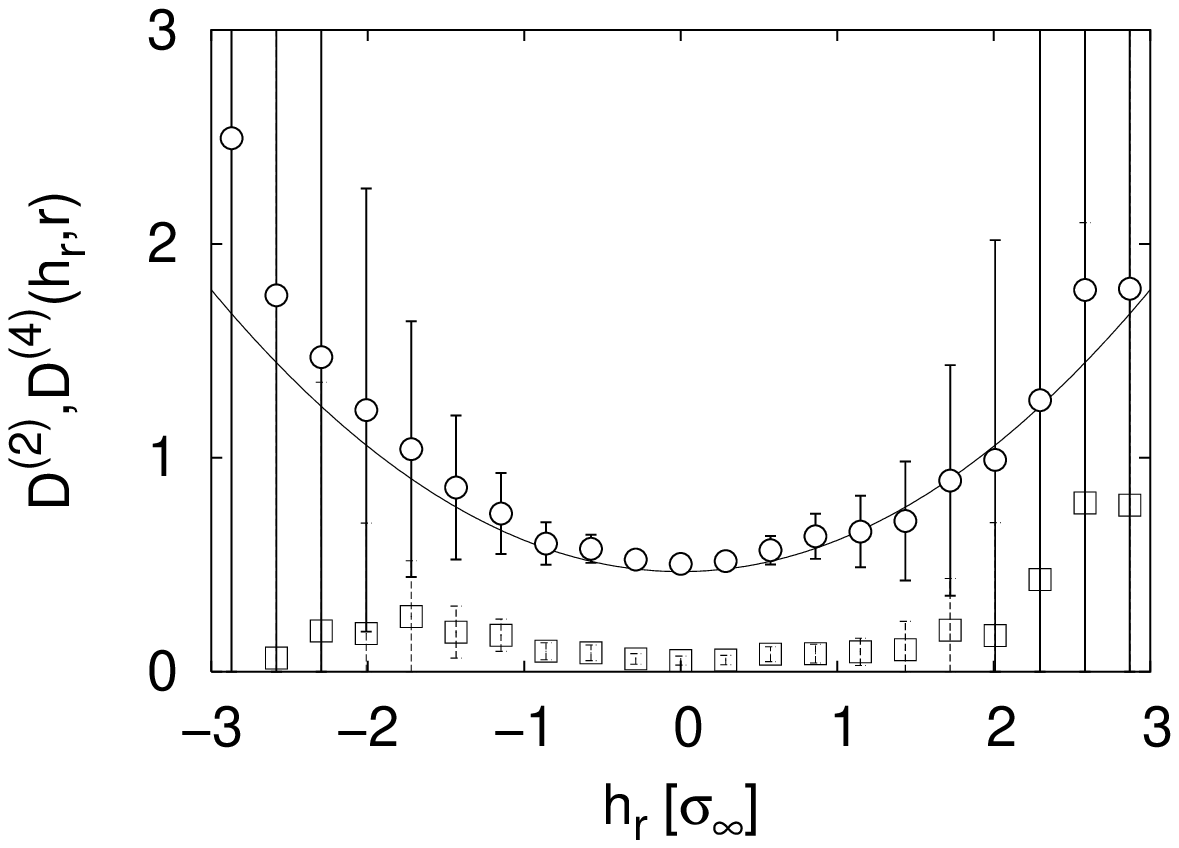}
  \begin{picture}(0,0)\put(23,28){\makebox{\small Road~2}}\end{picture}%
  \includegraphics[width=\breite]{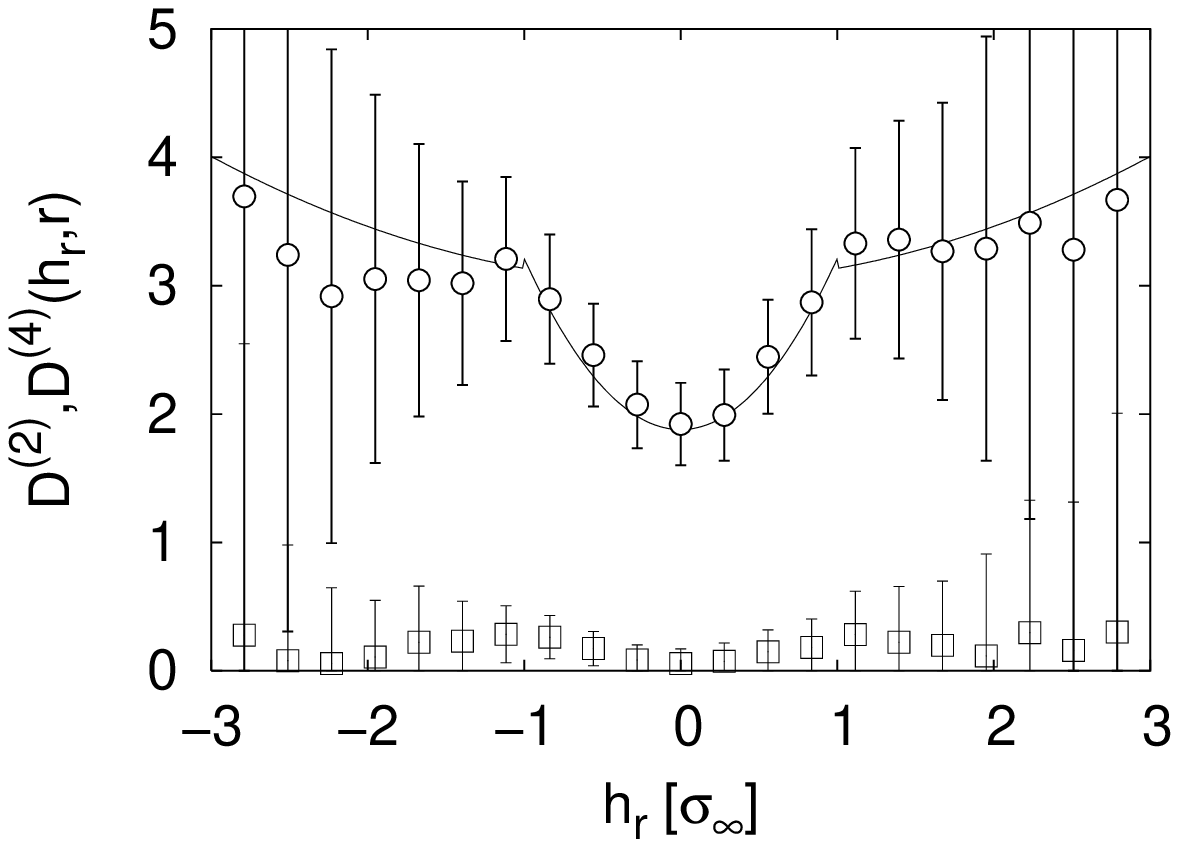}
  \begin{picture}(0,0)\put(23,28){\makebox{\small Road~3}}\end{picture}%
  \includegraphics[width=\breite]{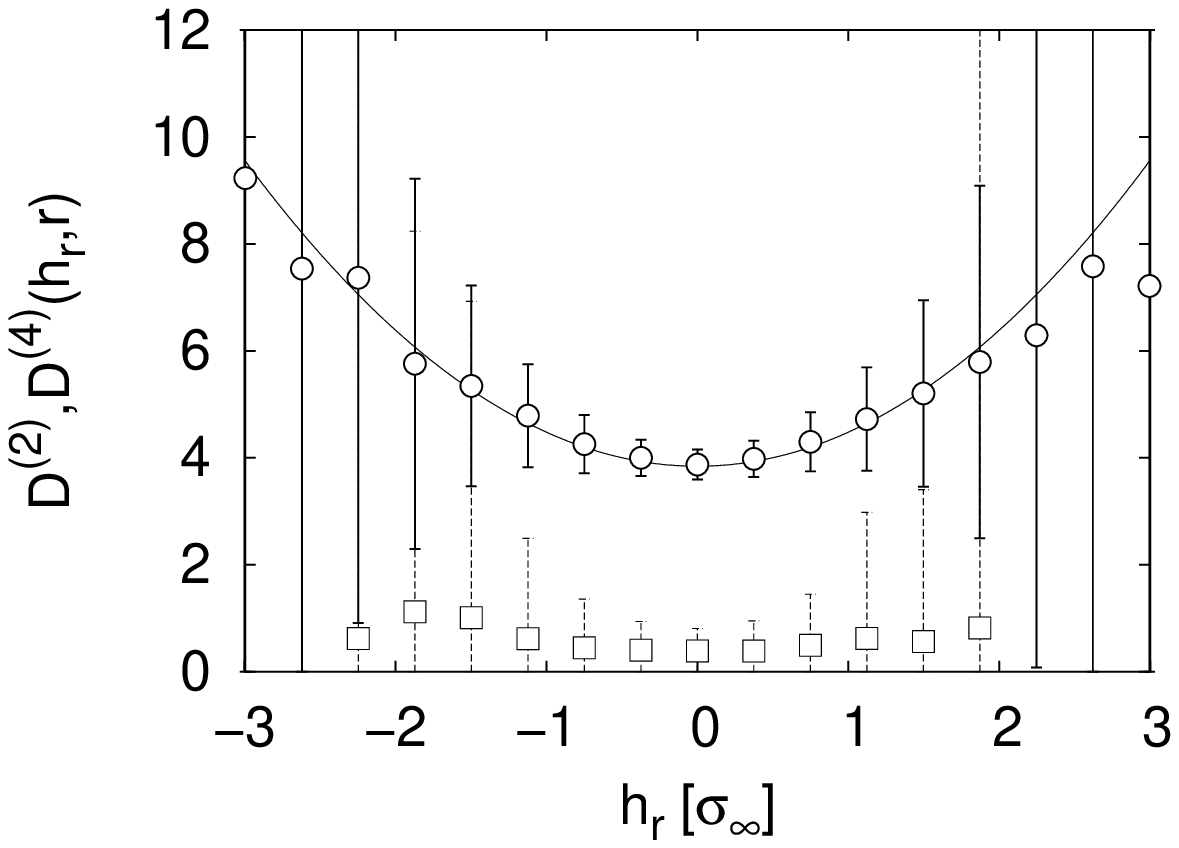}
  \begin{picture}(0,0)\put(23,28){\makebox{\small Road~4}}\end{picture}%
  \includegraphics[width=\breite]{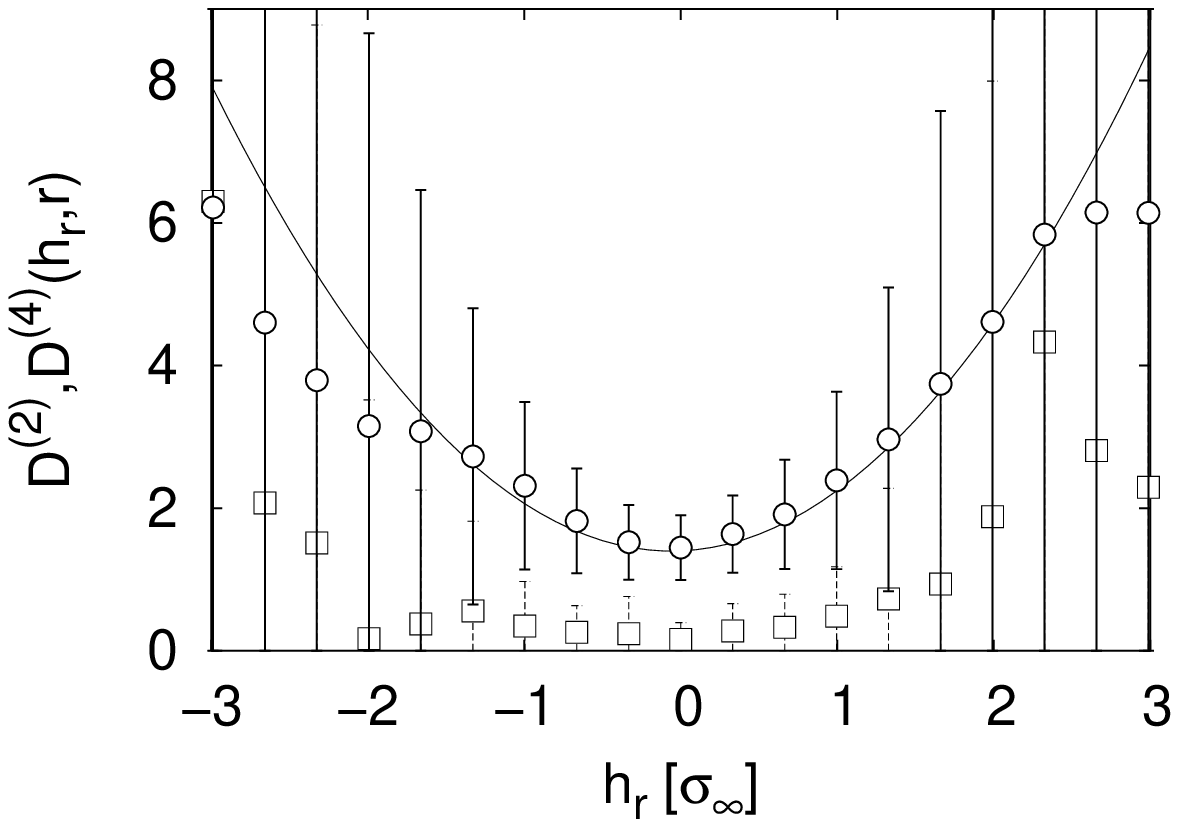}
  \caption[Estimated drift coefficient $D^{(1)}(h_r, r)$]%
  {Estimated diffusion coefficients $D^{(2)}(h_r, r)$ (circles) of the
    Fokker-Planck equation for road surfaces shown in
    fig.~\ref{fig:data_road_scaling}. Additionally the fourth Kramers-Moyal
    coefficients $D^{(4)}(h_r, r)$ are shown as squares.  Scales $r$ are as in
    fig.~\ref{fig:D1_road_scaling}.  Parameterizations are shown as lines.}
  \label{fig:D2_road_scaling}
\end{figure}

\begin{figure}[htbp]
  \centering
  \begin{picture}(0,0)\put(40,28){\makebox{\small (a)}}\end{picture}%
  \includegraphics[width=\breite]{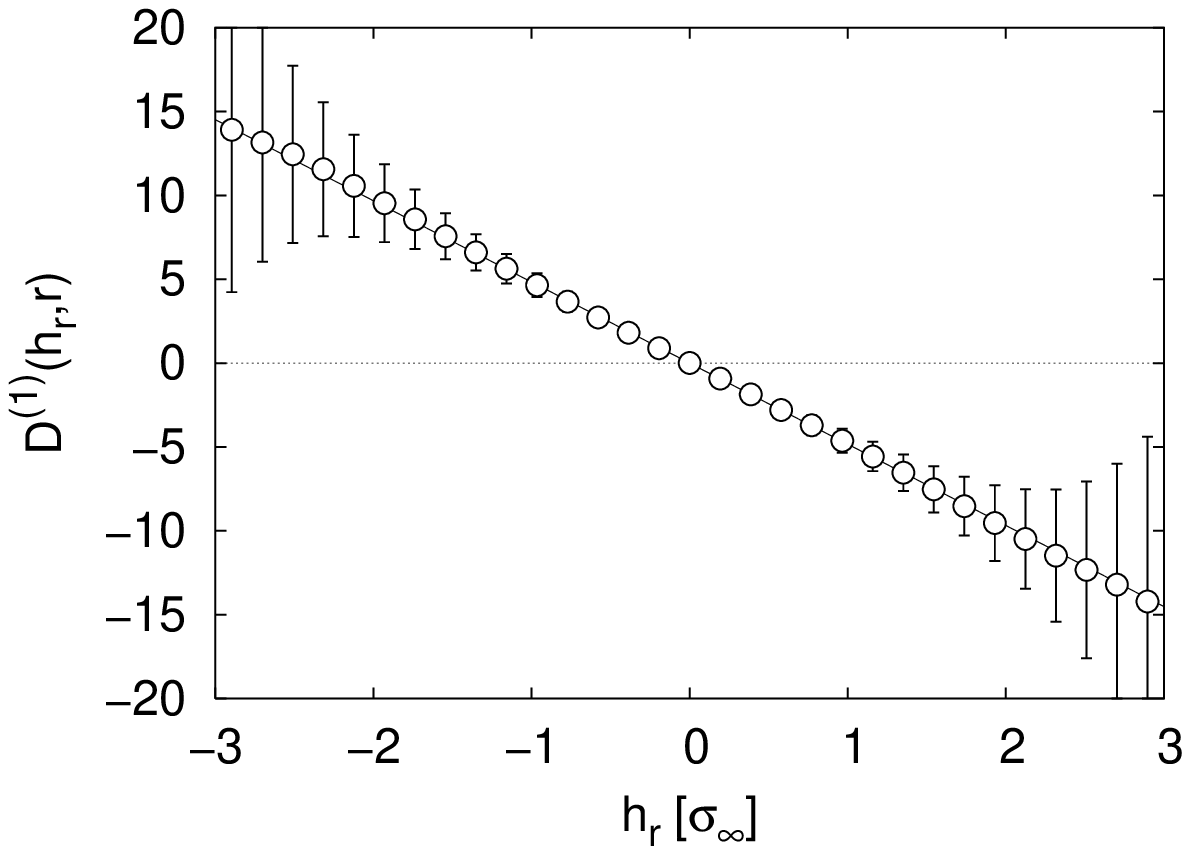}
  \begin{picture}(0,0)\put(22,28){\makebox{\small (b)}}\end{picture}%
  \includegraphics[width=\breite]{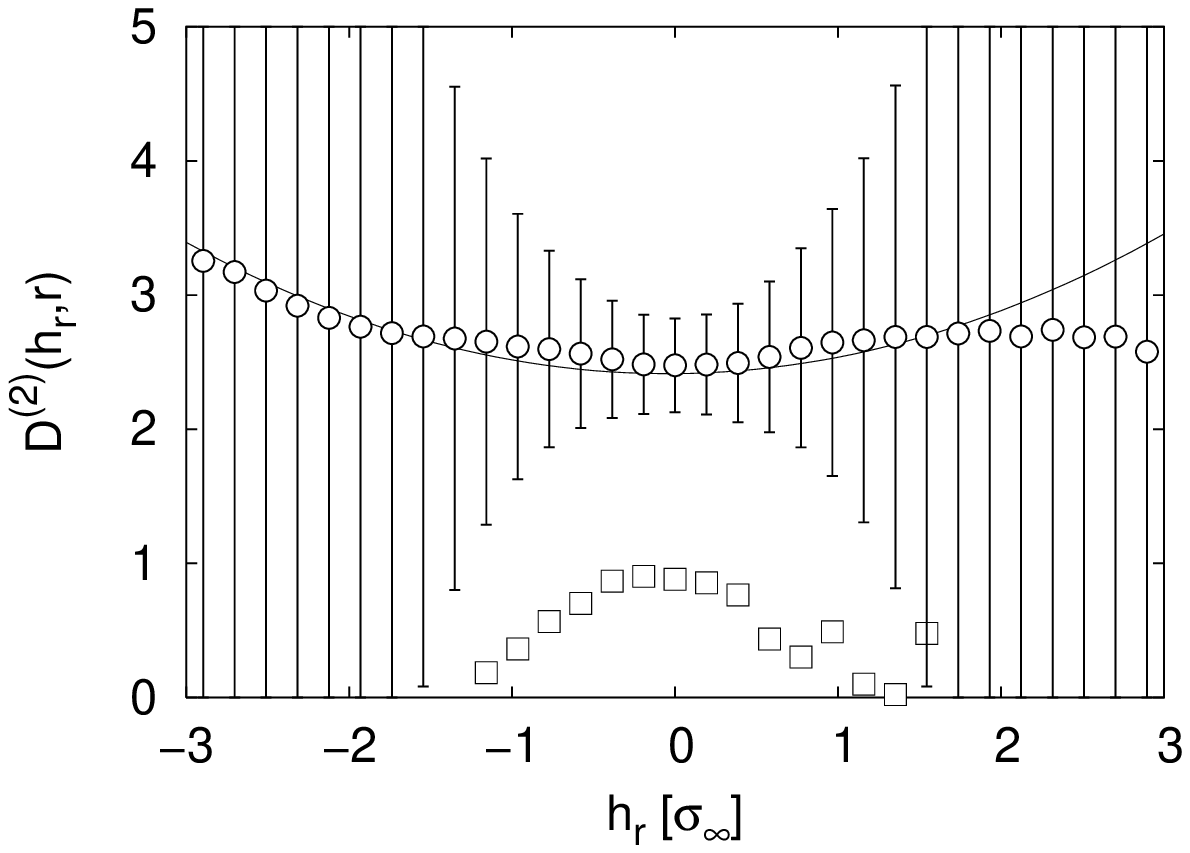}
  \caption{%
    Estimated drift (a) and diffusion (b) coefficient of the Au surface for
    $r=169$\un{nm}. Estimates of $D^{(4)}$ are added as squares in (b).}
  \label{fig:Schimmel_D12}
\end{figure}

Estimated drift and diffusion coefficients $D^{(1)}$ and $D^{(2)}$ for the Au
surface are shown in fig.~\ref{fig:Schimmel_D12} for $r=169$\un{nm}. Again,
$D^{(4)}$ was added to the plot of $D^{(2)}$, in this case without error bars
to enhance clarity. Errors of $D^{(4)}$ are in this case always much larger
than the values themselves and would cover the values of $D^{(2)}$ as well as
their errors. Also the error bars of $D^{(2)}$ appear to be quite large for
the Au surface. The data here are measured as two-dimensional images, thus
the number of statistically independent $h_r(x)$ decreases quadratically with
increasing $r$, resulting in rather large error estimates. For the calculation
of $D^{(k)}$ nevertheless all accessible $h_r(x)$ were used.
As the regime of Markov properties starts at $\Delta r=25\un{nm}$, the range
$25\un{nm} \leq \Delta r \leq 84\un{nm}$ was used as basis for the
extrapolation (see table~\ref{tab:l_extrapol}). For $r<84\un{nm}$ the upper
limit was reduced in order to derive the coefficients also for smaller scales
$r$ (compare also section \ref{sec:Dk_estimation}). In this way the drift and
diffusion coefficients of the Au film could be worked out from 281 down to
56\un{nm}. 

\begin{figure}[htbp]
  \centering
  \begin{picture}(0,0)\put(40,28){\makebox{\small (a)}}\end{picture}%
  \includegraphics[width=\breite]{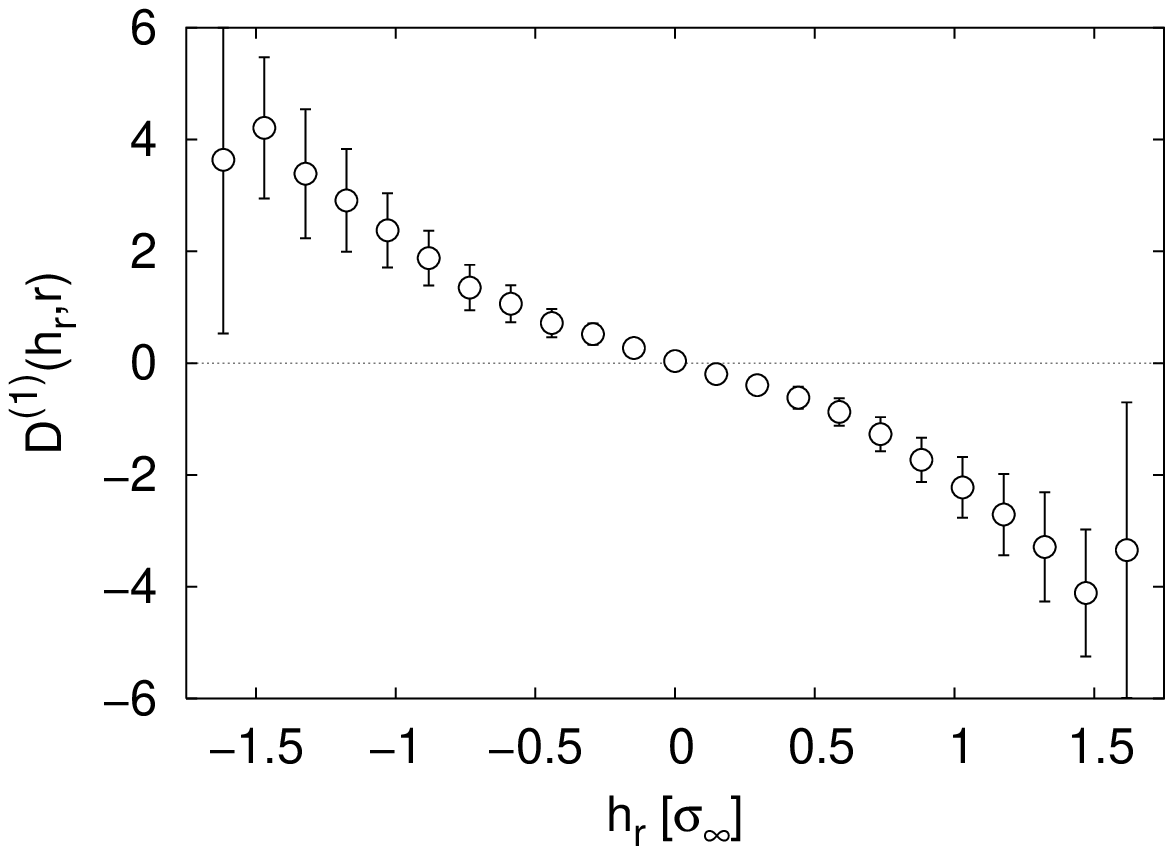}
  \begin{picture}(0,0)\put(22,28){\makebox{\small (b)}}\end{picture}%
  \includegraphics[width=\breite]{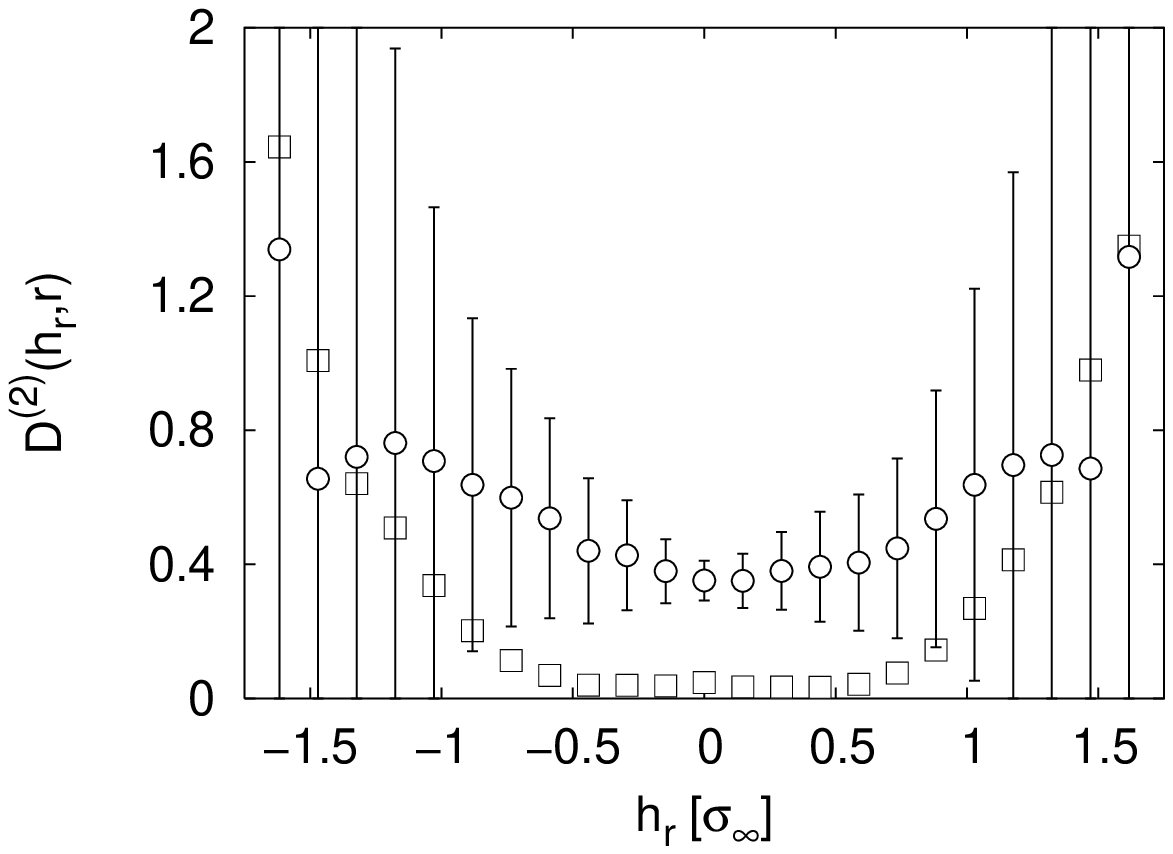}
  \caption{%
    Estimated Kramers-Moyal coefficients 
    for surface Crack for a length scale $r=49$\mum. (a)
    $D^{(1)}(h_r,r)$, (b) $D^{(2)}(h_r,r)$ and $D^{(4)}(h_r,r)$.}
  \label{fig:Wendt_D1234}
\end{figure}

In the same way as for the other surfaces, estimations of the Kramers-Moyal
coefficients were performed for the steel crack. The results are shown in
fig.~\ref{fig:Wendt_D1234}. Again, the estimates for $D^{(4)}$ are also
presented, which are of the same order of magnitude as $D^{(2)}$ for
higher values of $h_r$ ($|h_r|>0.5\sigma_\infty$).

\begin{figure}[htbp]
  \centering
  \begin{picture}(0,0)\put(40,28){\makebox{\small (a)}}\end{picture}%
  \includegraphics[width=\breite]{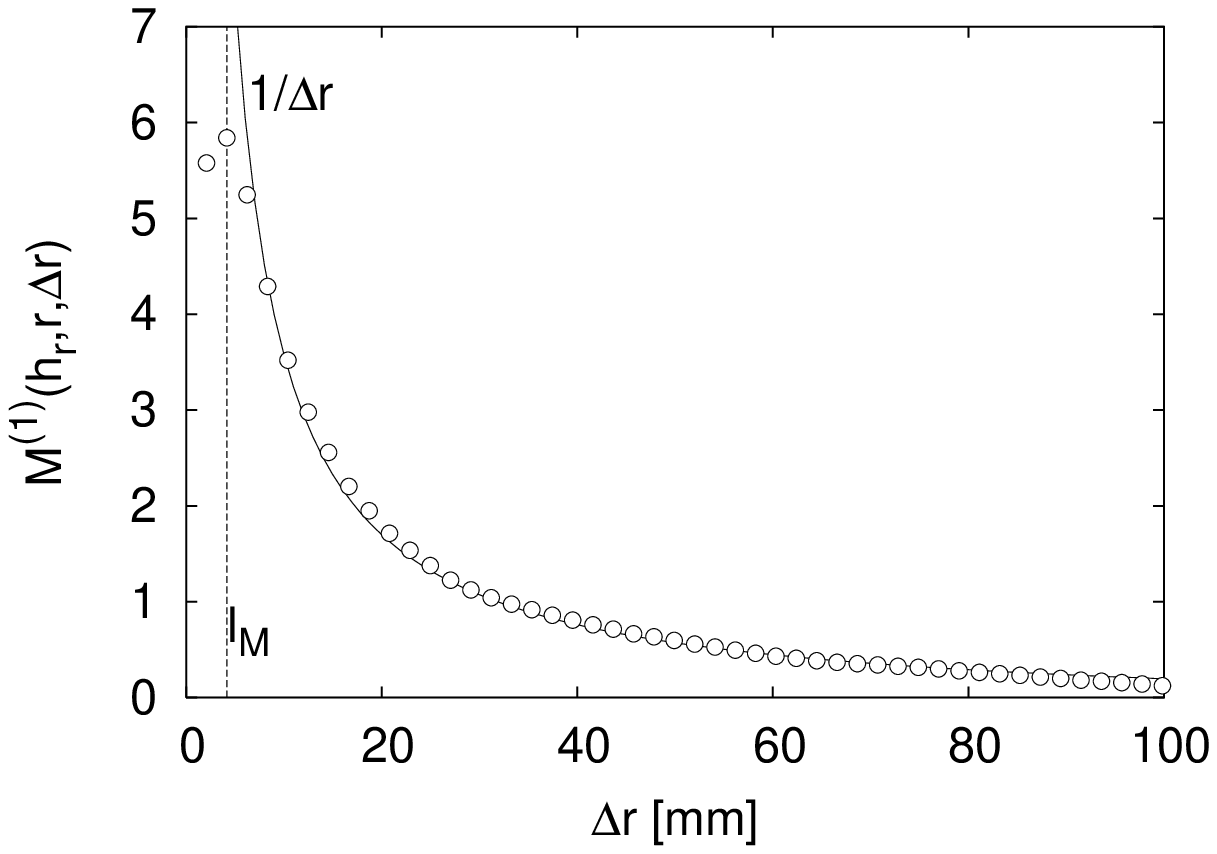}
  \begin{picture}(0,0)\put(40,28){\makebox{\small (b)}}\end{picture}%
  \includegraphics[width=\breite]{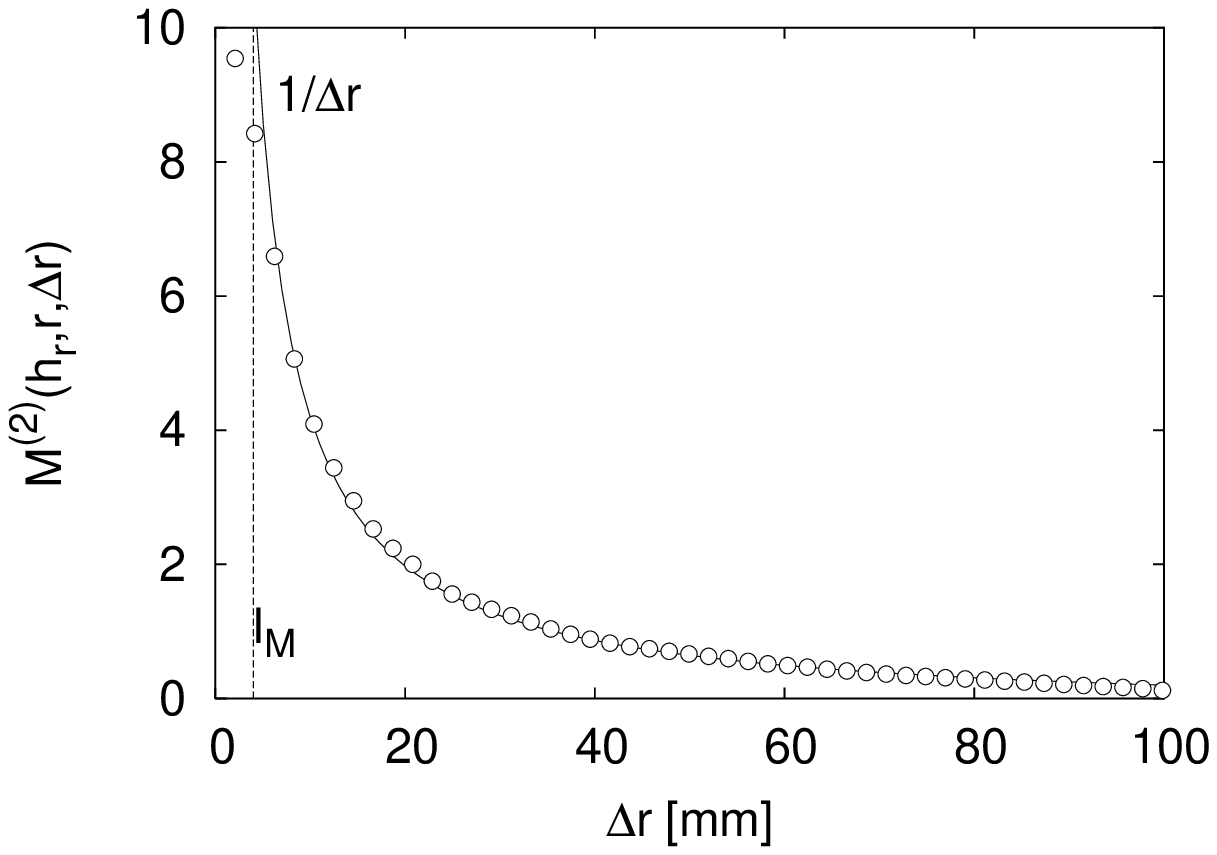}
  \caption[Extrapolation procedure for $D^{(1)}$.]{%
    Uncorrelated noise in the case of surface Road~5.  Dependence of (a)
    $M^{(1)}(h_r,r,\Delta r)$ and (b) $M^{(2)}(h_r,r,\Delta r)$ on $\Delta r$
    for $r=104\un{mm}$ and $h_r=-0.6\sigma_\infty$. The Markov length $l_M$ is
    marked with a dashed line. For illustration a function proportional to
    $1/\Delta r$ is fitted to the $M^{(k)}$. 
  }
  \label{fig:Dk_road_no_scaling_ex}
\end{figure}

For the surface Road~5 (cf.\ fig.~\ref{fig:data_road_no_scaling}) drift and
diffusion coefficients could not be estimated. The reason can be seen in
fig.~\ref{fig:Dk_road_no_scaling_ex}. The diagram shows the dependence of
$M^{(1)}(h_r,r,\Delta_r)$ and $M^{(2)}(h_r,r,\Delta r)$ on $\Delta r$ for fixed
$r$ and $h_r$, in this case 104\un{mm} and $-0.6\sigma_\infty$. For $\Delta
r>l_M$ it can be seen that $M^{(1)}$ and $M^{(2)}$ behave like $1/\Delta r$.
This behaviour can be explained by the presence of some additional uncorrelated
noise, where additional means independent of the stochastic process. A similar
behaviour was found for financial market data \cite{Renner2001b}. In this case
the integral in eq.~(\ref{eq:Mk_2}) will tend to a constant for small $\Delta
r$, independent of the value of $\Delta r$. Because we divide the integral by
$\Delta r$, the $M^{(k)}$ will then diverge as $\Delta r$ approaches zero.
Note that within the same mathematical framework the presence of uncorrelated
noise can be quantitatively determined \cite{Siefert2003}.

\subsection{Conclusions on the estimation of drift and diffusion coefficients}

Estimations of the drift and diffusion coefficients $D^{(1)}(h_r,r)$ and
$D^{(2)}(h_r,r)$ have been performed for all the surfaces introduced in section
\ref{sec:measurement_data}. An exception is Road~5, where the stochastic
process in the scale variable, while still Markovian, appears to be dominated
by additional uncorrelated noise. From eqs.~(\ref{eq:Dk_def}) and
(\ref{eq:Mk_def}) it can be seen that this leads to diverging Kramers-Moyal
coefficients $D^{(k)}$, as is the case for Road~5.

As mentioned above, the magnitude of the fourth Kra\-mers-Moyal coefficient
$D^{(4)}$ is of particular importance. If $D^{(4)}$ can be taken as zero, the
whole scale dependent complexity can be described by a Fokker-Planck equation.
Otherwise, if $D^{(4)}$ is not zero, an infinite set of $D^{(k)}$ is necessary.
In terms of a Langevin equation (\ref{eq:Langevin}), for $D^{(4)}\neq 0$ no
Gaussian noise is present. This case is related to unsteady stochastic
processes \cite{Honerkamp1990}. As we see from the topographies in
figs.~\ref{fig:data_road_scaling} and \ref{fig:Wendt_data}, jumps are more
likely to be present for the Road~4 and Crack surfaces than for the remaining
ones. This impression is consistent with the result that here we find
$D^{(4)}\neq 0$.
As a consequence, in these cases the Fokker-Planck equation with a drift and
diffusion coefficient is not sufficient to describe the stochastic process in
the scale variable, because the higher coefficients cannot be neglected. The
reconstruction of conditional probabilities (cf.\ section
\ref{sec:veri_coeff}) failed for these surfaces.

The range of scales where the drift and diffusion coefficients could be
estimated varies for the different surfaces, depending on the Markov length on
one side and on the length of the measured profiles on the other side. In
the case of Road~2 an additional upper limit for the Markov properties was
caused by the influence of a strong periodicity of the pavement.

\section{Verification of the estimated Fokker-Planck equations}
\label{sec:veri_coeff}

In the previous section methods to estimate the Kramers-Moyal coefficients were
discussed. We found that this estimation is not trivial. 
To prove the quality of the estimated $D^{(k)}$
we now want to verify the corresponding Fokker-Planck equations.

\subsection{Parameterization of Drift and Diffusion Coefficients}
\label{sec:Dk_parameterization}

With the estimations of the drift and diffusion coefficient from section
\ref{sec:Dk_estimation} for each surface a Fokker-Planck equation
(\ref{eq:FPE1}) is defined which should describe the corresponding process. For
the verification of these coefficients it is additionally desirable to generate
parameterizations which define $D^{(1)}(h_r, r)$ and $D^{(2)}(h_r, r)$ not only
at discrete values but at arbitrary points in the $(h_r, r)$-plane.

Such parameterizations have already been shown in
figs.~\ref{fig:D1_road_scaling}, \ref{fig:D2_road_scaling},
\ref{fig:Schimmel_D12}, and \ref{fig:Wendt_D1234}, as lines together with the
estimated discrete values. For $D^{(1)}$ it can be seen that for all surfaces
a straight line with negative slope was used, with additional cubic terms for
Road~2, Road~4, and Crack.
The diffusion coefficients were in all cases parameterized as parabolic
functions. The special shape of the diffusion coefficient for Road~2 was
parameterized as one inner and one outer parabola for small and larger values
of $h_r$, respectively (compare with fig.~\ref{fig:D2_road_scaling}).
We would like to note that both the drift and diffusion coefficients of the
cobblestone road presented in \cite{Waechter2003_plus_preprint} are best
fitted by piecewise linear functions with steeper slopes for larger $h_r$.

It is easy to verify that with a linear $D^{(1)}$ and a constant $D^{(2)}$ the
Fokker-Planck equation~(\ref{eq:FPE1}) describes a Gaussian process, while
with a parabolic $D^{(2)}$ the distributions become non-Gaussian, also called
intermittent or heavy tailed.
For the Au surface it can be seen in fig.~\ref{fig:Schimmel_D12} that
$D^{(2)}$ has only a weak quadratic dependence on $h_r$ and possibly could
also be interpreted as constant (we nevertheless kept the small quadratic term
because it is confirmed by the verification procedure below). If $D^{(2)}$ is
constant in $h_r$ the type of noise in the corresponding Langevin equation
(\ref{eq:Langevin}) is no longer multiplicative but additive, which results in
Gaussian noise in the process. Thus the statistics of $h_r$ in $r$ will always
stay Gaussian, and all moments $\langle h_r^n\rangle$ with $n>2$ can be
expressed by the first and second one. As a further consequence, the scaling
exponents $\xi_n$ (see section~\ref{sec:scaling}) are obtained by $\langle
h_r^n\rangle\sim \langle h_r^2\rangle^{n/2}$ as $\xi^n = \frac{n}{2}\xi_2$.
This linear dependence on $n$ denotes self-affinity
rather than multi-affinity and is confirmed by the scaling analysis in
section~\ref{sec:scaling_with}.

\subsection{Reconstruction of empirical pdf}

Next, we want to actually evaluate the precision of our results. Therefore we
return to eq.~(\ref{eq:FPE1}). Knowing $D^{(1)}$ and $D^{(2)}$ it should be
possible to calculate the pdf of $h_r$ with the corresponding Fokker-Planck
equation.
Equation~(\ref{eq:FPE1}) can be integrated over $h_0$ and is then valid also
for the unconditional pdf:
\begin{eqnarray}
   \label{eq:FPE2}
   \lefteqn{-r\,\frac{\partial}{\partial r}\;p(h_r,r)\,=} \\
   && \left\{ -\frac{\partial}{\partial h_r} D^{(1)}(h_r,r)
       + \frac{\partial^2}{\partial h_r^2} D^{(2)}(h_r,r)
     \right\} \, p(h_r,r) \nonumber
\end{eqnarray}
Now at the largest scale $r_0$ where the drift and diffusion coefficients could
be worked out the empirical pdf is parameterized
and used as the initial condition for a numerical solution of
eq.~(\ref{eq:FPE2}). For several values of $r$ the reconstructed pdf is
compared to the respective empirical pdf, as shown below in this section. If
our Fokker-Planck equation successfully reproduces these single scale pdf,
the structure functions $\langle h_r^n\rangle$ can also easily be obtained.

A second verification is the reconstruction of the conditional pdf by a
numerical solution of Fokker-Planck equation (\ref{eq:FPE1}) for the
conditional pdf. Reconstructing the conditional pdf this way is much more
sensitive to deviations in $D^{(1)}$ and $D^{(2)}$.  This becomes evident by
the fact that the conditional pdf (and not the unconditional pdf of
figs.~\ref{fig:road_scaling_veri1} and \ref{fig:Schimmel_veri1}) determine
$D^{(1)}$ and $D^{(2)}$ and thus the stochastic process, see
eqs.~(\ref{eq:Dk_def}) and (\ref{eq:Mk_def}).
The knowledge of the conditional pdf also gives access to the
complete $n$-scale joint pdf (eq.~\ref{eq:markov_straight}).
Here again the difference from the
multiscaling analysis becomes clear, which analyses higher moments $\langle
h_r^n\rangle = \int h_r^n\cdot p(h_r)\,\mathrm{d} h_r$ of $h_r$, and does not
depend on the conditional pdf. It is easy to show that there are many different
stochastic processes which lead to the same single scale pdf $p(h_r)$.

For both verification procedures we use a technique which is mentioned in
\cite{Risken1984} and has already been used in
\cite{Renner2001,Waechter2003_plus_preprint}. An approximative solution of the
Fokker-Planck eq.~(\ref{eq:FPE1}) for infinitesimally small steps $\Delta r$
over which $D^{(k)}$ can be taken as constant in $r$, is known
\cite{Risken1984}
\begin{eqnarray}
  \label{eq:FPE_approx_sol}
  \lefteqn{ p(h_1, r-\Delta r| h_0,r) =  
      \frac{1}{2\sqrt{\pi D^{(2)}(h_r,r)\Delta r}}  } \nonumber\\
      &&\times\exp{\left(-\frac{(h_1-h_0-D^{(1)}(h_0,r)\Delta r)^2}
                       {4D^{(2)}(h_r,r)\Delta r}\right)}\,. 
\end{eqnarray}
A necessary condition for a Markov process is the validity of the
Chapman-Kolmogorov equation (\ref{eq:CKE}) \cite{Risken1984},
which allows to combine two conditional pdf with adjacent intervals in $r$
into one conditional pdf spanning the sum of both intervals. By an iterative
application of these two relations we are able to obtain conditional
probabilities \mbox{$p(h_i, r_0-i\Delta r| h_0,r_0)$} spanning large intervals
in the scale $r$, given that for all involved scales $r_i$ the drift and
diffusion coefficients are known.

In the following the results of this verification procedure are shown for
those surfaces where the drift and diffusion coefficients of the Fokker-Planck
equation could be obtained.

\subsection{Verification results}
\label{sec:veri_results}

The results of the reconstruction of the unconditional pdf for the road
surfaces with scaling properties are presented in
fig.~\ref{fig:road_scaling_veri1}. The pdf of Road~1 show at smaller scales a
peak around 5\un{\sigma_\infty} which is not reproduced by our Fokker-Planck
equation because in this regime of $h_r$ $D^{(1)}$ and $D^{(2)}$ could not be
estimated with sufficient precision. Here it has to be noted that according to
eq.~(\ref{eq:sigma_inf}) $\sigma_\infty>\sigma_r$ for any $r$, and thus
5\un{\sigma_\infty} is a large value for a pdf, denoting quite rare events (the
$r$-dependence of $\sigma_r$ has been presented by $S^2(r)=\sigma_r^2$, see
section \ref{sec:scaling}).  The magnitudes of the estimated drift and
diffusion coefficients had to be adjusted by a factor of 0.65 to give optimal
results in the reconstruction.
For Road~2 it is likely that the correspondence between the emprical and
reconstructed pdf could be improved by a more advanced parameterization of the
nontrivial shape especially of the estimated drift coefficient (see
fig.~\ref{fig:D1_road_scaling}). Here, the estimated drift and diffusion
coefficients could be used without adjustment.
The reconstructed pdf for Road~3 are in perfect agreement with the empirical
ones. A substantial adjustment factor of 0.20 for $D^{(1)}$ and 0.26 for
$D^{(2)}$ was necessary to achieve the best result.

\setlength{\breite}{0.66\linewidth}
\begin{figure}[htbp]
  \centering
  \begin{picture}(0,0)\put(8,31){\makebox{\small Road~1}}\end{picture}%
  \includegraphics[width=\breite]{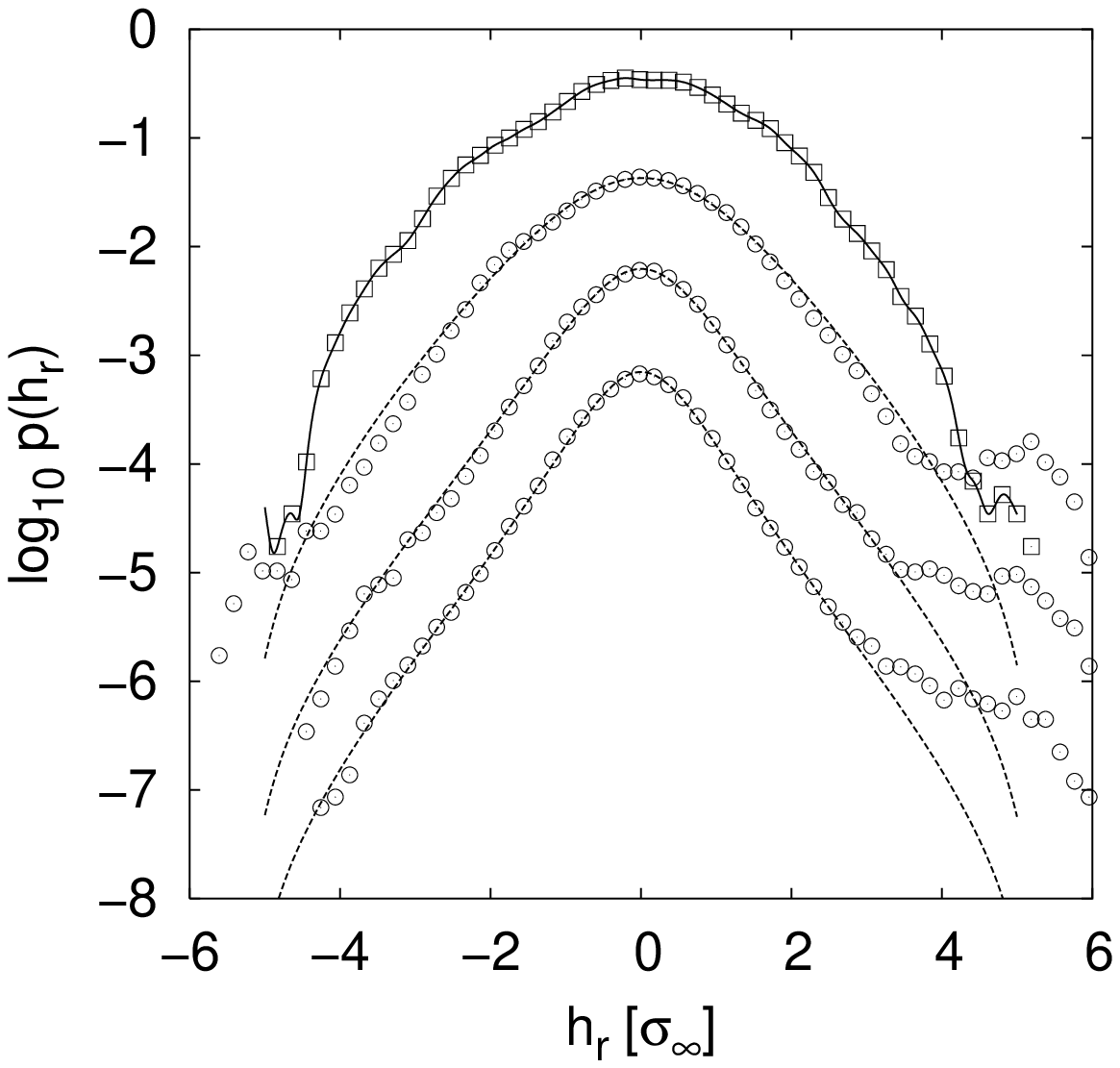}
  \begin{picture}(0,0)\put(8,31){\makebox{\small Road~2}}\end{picture}%
  \includegraphics[width=\breite]{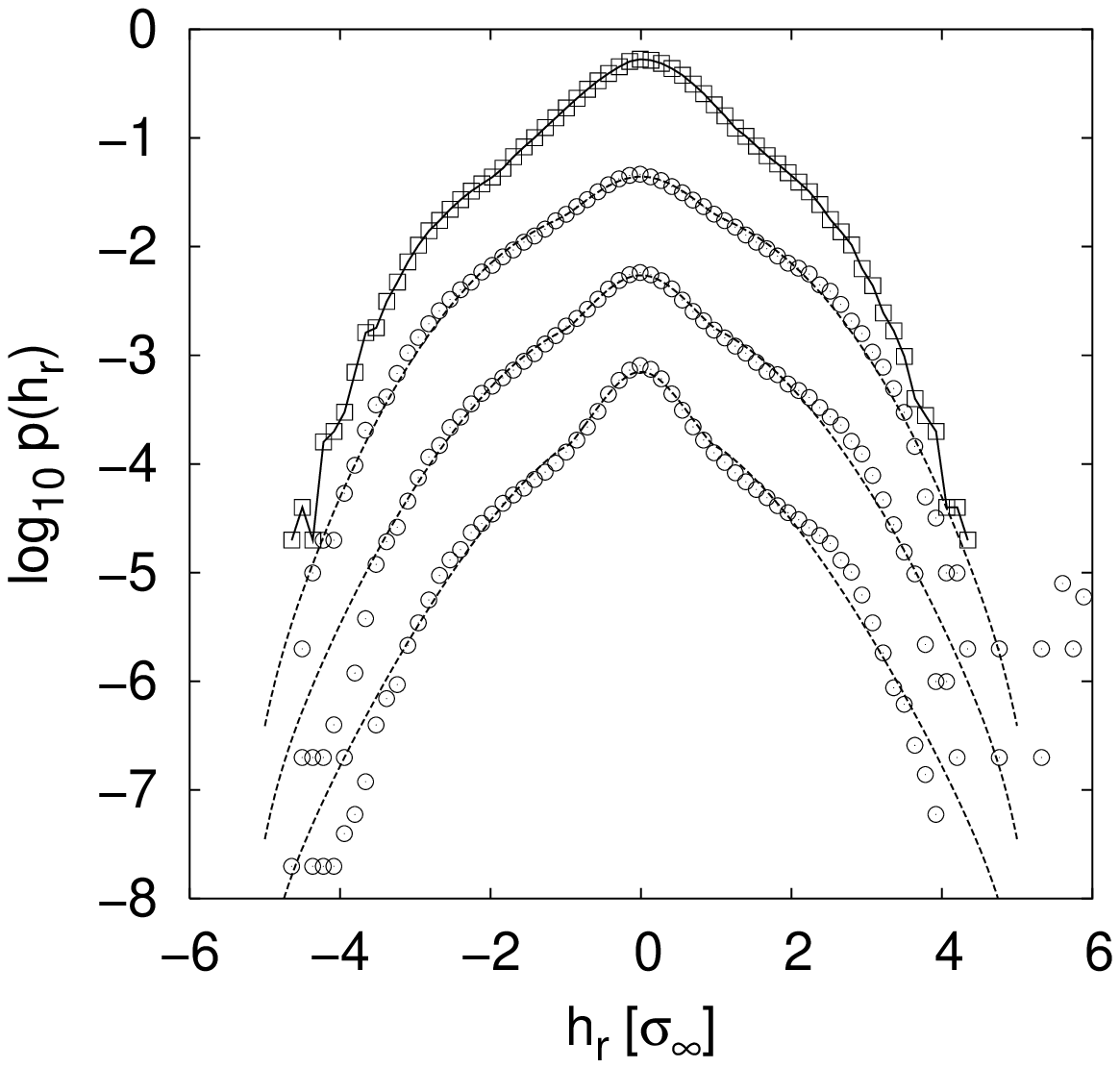}
  \begin{picture}(0,0)\put(8,31){\makebox{\small Road~3}}\end{picture}%
  \includegraphics[width=\breite]{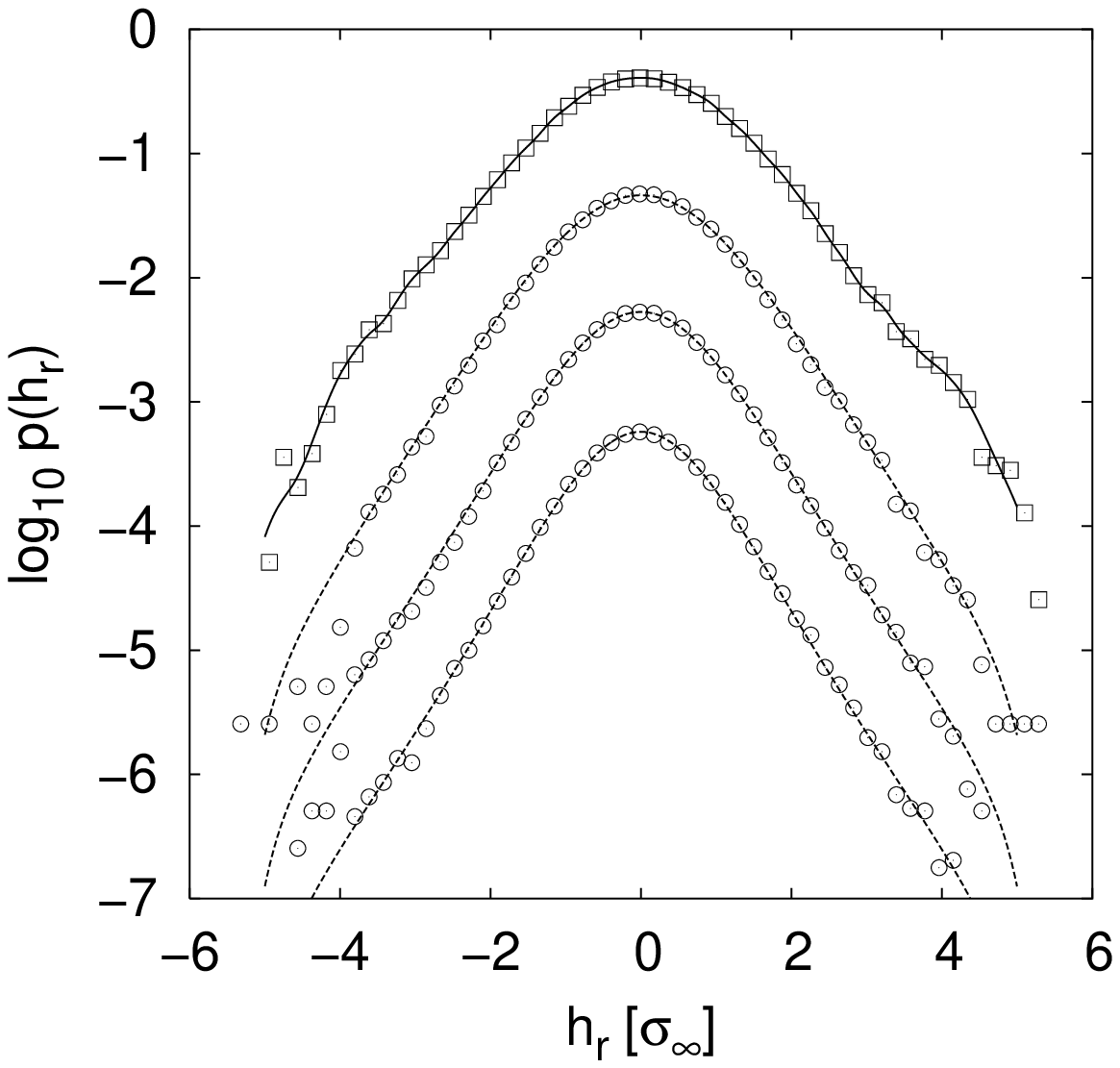}
  \caption{%
    Numerical solution of the Fokker-Planck equation (\ref{eq:FPE2}) compared
    to the empirical pdf (symbols) for road surfaces with scaling properties.
    For each surface, the topmost solid line corresponds to an empirical pdf
    parameterized at the largest scale, and the dashed lines to the
    reconstructed pdf.  Scales are (from top to bottom)
    for Road~1: 316, 158, 79, 66\un{mm}, %
    for Road~2: 158, 79, 47, 20\un{mm},  %
    for Road~3: 188, 95, 47, 24\un{mm}.  %
    Pdf are shifted in the vertical direction for clarity of presentation.  }
  \label{fig:road_scaling_veri1}
\end{figure}

Reconstructed conditional pdf are shown in fig.~\ref{fig:road_scaling_veri2}
for the road surfaces. While there are deviations for larger values of
$h_0,h_1$, the overall agreement between the empirical and reconstructed pdf is
good. Especially the rather complicated shape of the conditional pdf of Road~2
appears to be well modelled by our coefficients $D^{(1)},D^{(2)}$. As mentioned
above, an improved parameterization of $D^{(1)}$ may lead to even better
results. The magnitudes of $D^{(1)},D^{(2)}$ were adjusted by the same
factors as for the unconditional pdf above.

\setlength{\breite}{0.64\linewidth}
\begin{figure}[htbp]
  \centering
  \begin{picture}(0,0)\put(0,42){\makebox{\small Road~1}}\end{picture}%
  \includegraphics[width=\breite]%
    {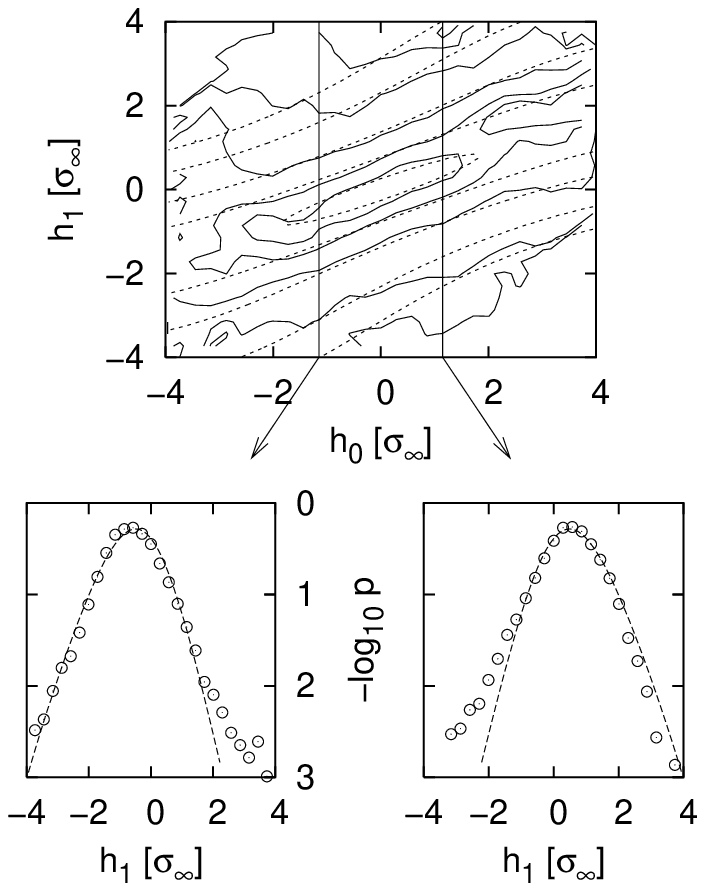}
  \begin{picture}(0,0)\put(0,42){\makebox{\small Road~2}}\end{picture}%
  \includegraphics[width=\breite]%
    {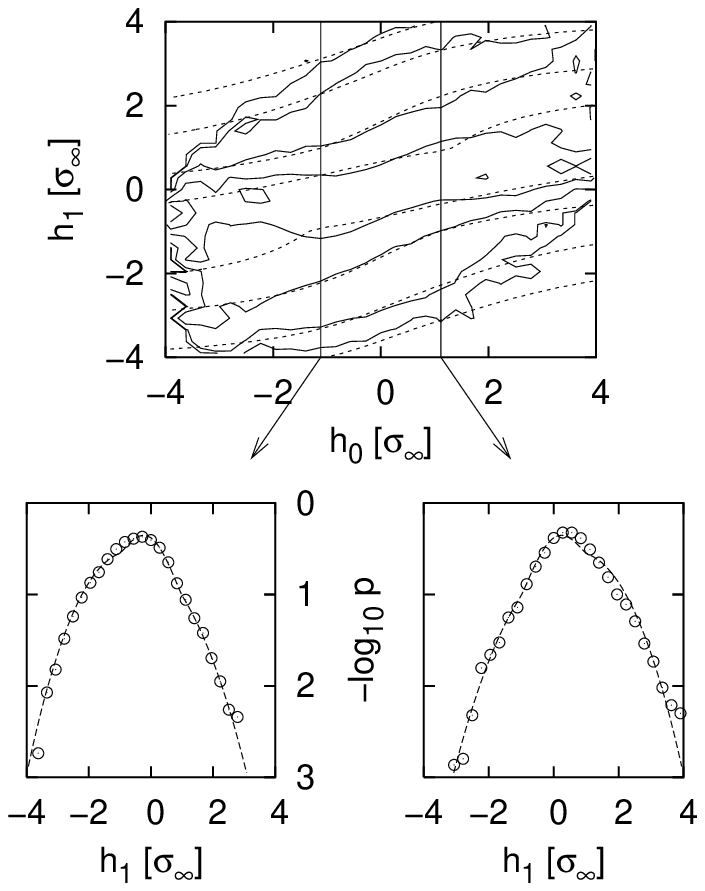}
  \begin{picture}(0,0)\put(0,42){\makebox{\small Road~3}}\end{picture}%
  \includegraphics[width=\breite]%
    {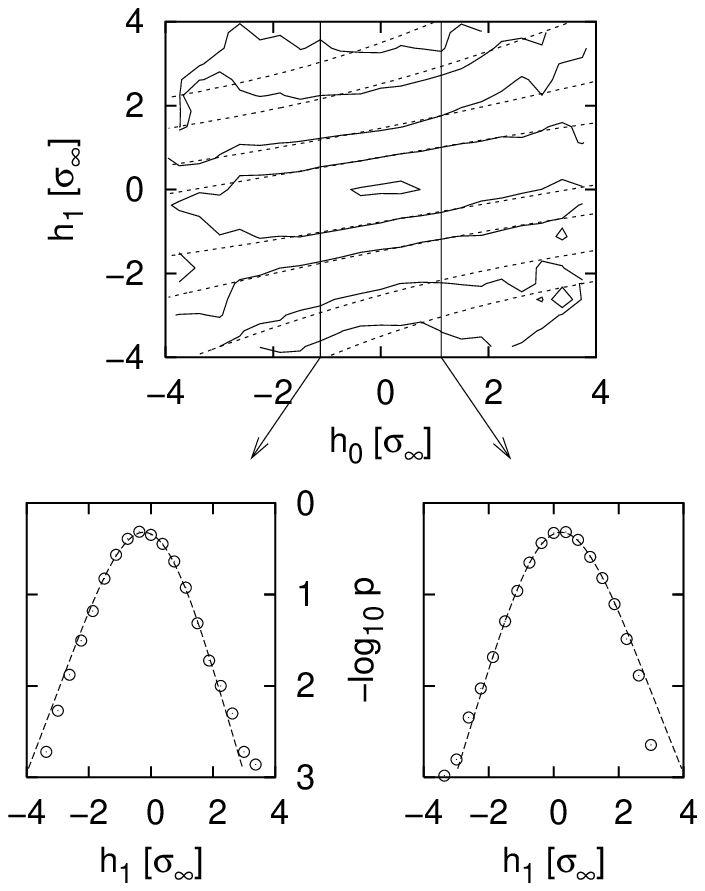}
  \caption[Numerical solution of Fokker-Planck equation]%
  {Numerical solution of the Fokker-Planck equation (\ref{eq:FPE1}) compared to
    the empirical pdf for road surfaces with scaling properties. Similar to
    fig.~\ref{fig:Schimmel_markov} in each case a contour plot of empirical
    (solid lines) and reconstructed pdf (broken lines) is shown on top, with
    contour levels as in fig.~\ref{fig:Schimmel_markov}. Below two cuts at
    $h_0\approx\pm\sigma_\infty$ are located. Here, empirical pdf are plotted
    as symbols.
    Scales are %
    $r_0=304\un{mm}$, $r_1=158\un{mm}$ (Road~1), %
    $r_0=158\un{mm}$, $r_1=112\un{mm}$ (Road~2), %
    and $r_0=188\un{mm}$, $r_1=92\un{mm}$ (Road~3). %
    }
  \label{fig:road_scaling_veri2}
\end{figure}

\setlength{\breite}{0.64\linewidth}
\begin{figure}[htbp]
  \centering
  \begin{picture}(0,0)\put(8,30){\makebox{\small Au (a)}}\end{picture}%
  \includegraphics[width=\breite]{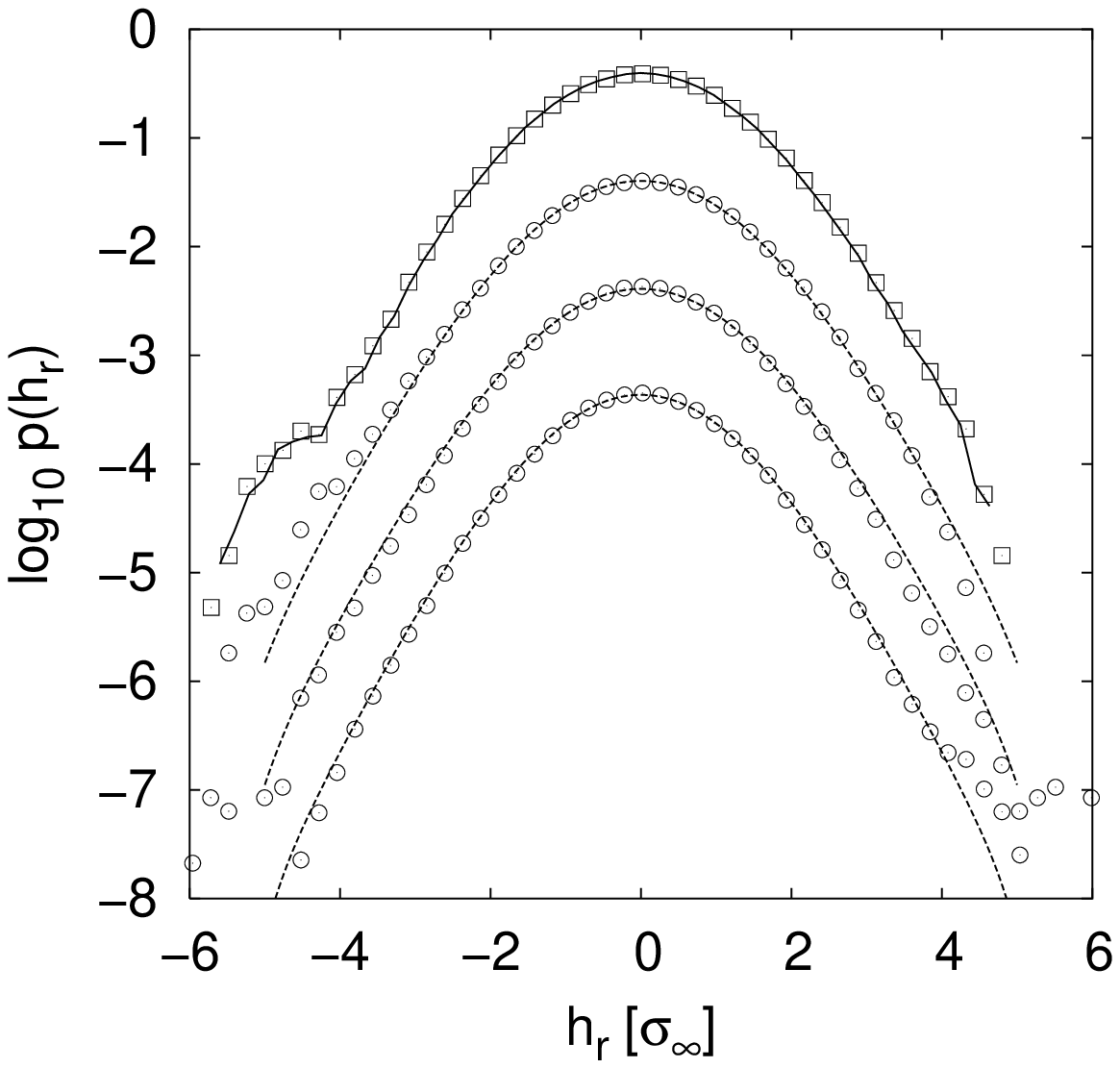}
  \begin{picture}(0,0)\put(0,42){\makebox{\small Au (b)}}\end{picture}%
  \includegraphics[width=\breite]%
    {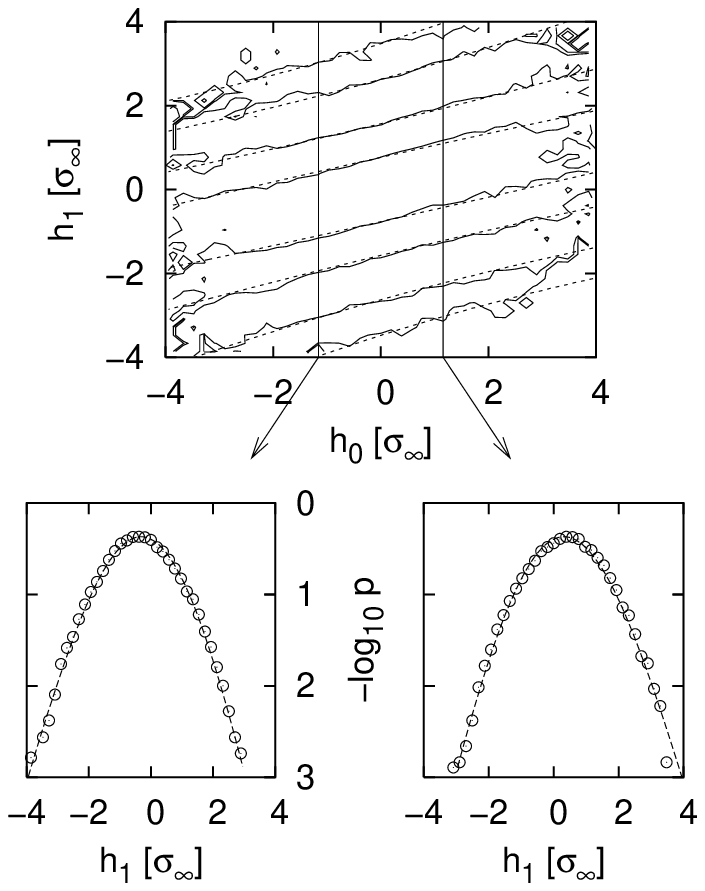}
  \begin{picture}(0,0)\put(0,42){\makebox{\small Au (c)}}\end{picture}%
  \includegraphics[width=\breite]%
    {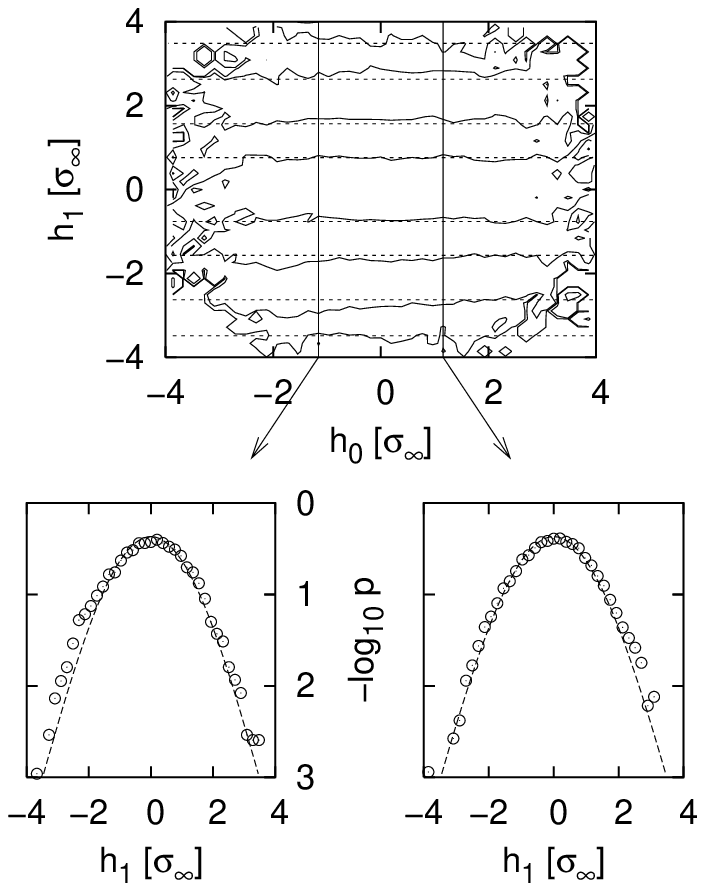}
  \caption[Au: Numerical solutions of Fokker-Planck equations]{%
    Numerical solutions of the Fokker-Planck equations (\ref{eq:FPE1}) and
    (\ref{eq:FPE2}) compared to the empirical pdf for the Au surface. \\
    (a) Results of the integrated equation (\ref{eq:FPE2}) presented as in
    fig.~\ref{fig:road_scaling_veri1}.  Scales $r$ are 281, 246, 148, and
    56\un{nm} (from top to bottom). \\
    (b), (c) Numerical solution of equation (\ref{eq:FPE1}) for the
    conditional pdf compared to the empirical pdf at scales $r_0=183\un{nm}$,
    $r_1=155\un{nm}$ (b) and $r_0=281\un{nm}$, $r_1=56\un{nm}$ (c).
    The organisation of the diagram is as in fig.~\ref{fig:road_scaling_veri2}.
  }
  \label{fig:Schimmel_veri1}
\end{figure}

In the case of the Au film the drift and diffusion coefficients could be worked
out from 281 down to 56\un{nm}, see section \ref{sec:Dk_estimation}. In
contrast to this regime, the range of correlation between scales is only about
40\un{nm}, i.e., height increments on scales which are separated by at least
40\un{nm} are uncorrelated.
Nevertheless, both verification procedures outlined in section
\ref{sec:veri_coeff} gave good results over the whole range from 281 to
56\un{nm} as shown in fig.~\ref{fig:Schimmel_veri1}. Here the estimated
$D^{(1)}$ and $D^{(2)}(h_r,r)$ were multiplied by factors 1.3 and 2.2,
respectively.

\subsection{Discussion of the verification procedure}

For the verification of the drift and diffusion coefficients estimated in
section~\ref{sec:Dk_estimation} numerical solutions of the Fokker-Planck
equations (\ref{eq:FPE1}) and (\ref{eq:FPE2}) have been performed using these
estimations. The reconstructed pdf have been compared to the empirical ones to
validate the descripition of the data sets as realizations of stochastic
processes obeying the corresponding Fokker-Planck equation.

Good results were obtained for most surfaces where the drift and diffusion
coefficients could be derived. In the case of Road~4 and Crack we found that
the higher Kramers-Moyal coefficients $D^{(3)}$ and $D^{(4)}$ were
significantly different from zero, and the empirical pdf could not be
reproduced with a Fokker-Planck equation (which only uses $D^{(1)}$ and
$D^{(2)}$). 

It may be surprising that the correspondence between the empirical and
reconstructed pdf seems better for the conditional rather than for the
unconditional pdf in some cases (compare figs.~\ref{fig:road_scaling_veri1} and
\ref{fig:road_scaling_veri2}). One reason may be that in
fig.~\ref{fig:road_scaling_veri2} it is clear that the empirical pdf are not
precisely defined for combinations of large $h_0$ and $h_1$. The eye
concentrates on the central regions of the contour plots where the uncertainty
of the empirical pdf is reduced, as well as deviations due to possible
inaccuracies and uncertainties of our drift and diffusion coefficients. This
effect is also confirmed by our mathematical framework where all steps in the
procedure are based on the estimation and evaluation of the conditional (not
the unconditional) pdf.

The reconstruction procedure allows also to adjust the estimated coefficients
in order to improve the above-men\-tioned description, thus compensating for a
number of uncertainties in the estimation process.
While the functional form of $D^{(1)}$ and $D^{(2)}$ found in
section~\ref{sec:Dk_estimation} for all surfaces could be confirmed, in most
cases the magnitudes of the estimated values had to be adjusted to give
satisfactory results in this reconstruction procedure.  
We found this effect also when analysing turbulent velocities and financial
data.
One reason may be the uncertainties of the estimation procedure. 
%
A second source of deviations may be that the dependence of
$M^{(k)}(h_r,r,\Delta r)$ on $\Delta r$ is not always purely linear in the
extrapolation range (see section~\ref{sec:Dk_estimation}). Thus fitting a
straight line and extrapolating against $\Delta r=0$ may lead to coefficients
$D^{(1)}$ and $D^{(2)}$ which still have the correct functional form in $h_r$
but incorrect magnitudes. As mentioned in section~\ref{sec:Dk_estimation}, in
our case no general improvements could be achieved by the use of nonlinear
(i.e.\ polynomial) fitting functions or higher order terms of the corresponding
Taylor expansion. It is possible that the range of $\Delta r<l_M$ where no
Markov properties are given is in most cases large enough that approximations
for small $\Delta r$ are inaccurate. We would like to note that there are also
data sets which did not require any adjustment of the estimated coefficients,
see Road~2 and \cite{Waechter2003_plus_preprint}.
A last remark concerns the latest results in the case of Road~1, see
fig.~\ref{fig:drh_ex}.  If the fraction of $M^{(k)}(\Delta r)$ which is
proportional to $1/\Delta r$ is substracted before performing the
extrapolation, the resulting $D^{(k)}$ are substantially improved in
their magnitudes. This may be a way to correct the extrapolation of the
$D^{(k)}$ in cases where uncorrelated noise is involved.

In any case, whether an adjustment of $D^{(1)}$ and $D^{(2)}$ was needed or
not, for the presented surfaces a Fokker-Planck equation was found which
reproduces the conditional pdf. Together with the verification of the Markov
property (\ref{eq:markov_straight}) thus a complete description of the
$n$-scale joint pdf is given, which was the aim of our work.

\section{Conclusions}
\label{sec:conclusions}

For the analysis and characterization of surface roughness we have presented a
new approach and applied it to different examples of rough surfaces. The
objective of the method is the estimation of a Fokker-Planck equation
(\ref{eq:FPE1}) which describes the statistics of the height increment $h_r(x)$
in the scale variable $r$. A complete characterization of the corresponding
stochastic process in the sense of multiscale conditional probabilities is the
result.

The application to different examples of surface measurement data showed that
this approach cannot serve as a universal tool for any surface, as it is also
the case for other methods like those based on self- and multi-affinity. With
given conditions, namely the Markov property and a vanishing fourth order
Kramers-Moyal coefficient (cf.\ section~\ref{sec:Markov_theory}), 
a comprehensive characterization of a single surface is obtained. The
features of the scaling analysis are included, and beyond that a deeper
insight in the complexity of roughness is achieved.
As shown in \cite{Jafari2003} such knowledge about a surface allows the
numerical generation of surface structures which should have the same
complexity. This may be of high interest for many research fields based on
numerical modelling.

The precise estimation of the magnitudes of the drift and diffusion
coefficients for surface measurement data still remains an open problem. While
for other applications a number of approaches have been developed
\cite{Renner2001,Friedrich2002,Ragwitz2001,Sura2002} in any case a verification
of the estimated Fokker-Planck equation is necessary and may lead to
significant adjustments, as it is the case for some of our data sets.

\begin{acknowledgement}
  
  We enjoyed helpful and stimulating discussions with R.\ Friedrich, A.\ 
  Kouzmitchev and M.\ Haase. Financial support by the Volks\-wa\-gen
  Foundation is kindly acknowledged.

\end{acknowledgement}

\bibliography{obf_overview,mw}
\bibliographystyle{unsrt}

\end{document}